\documentclass[twoside,11pt]{article}

\usepackage{blindtext}

\usepackage[utf8]{inputenc}
\usepackage{setspace}
\usepackage{placeins}

\usepackage{amsmath, amssymb, amsthm}

\usepackage{graphicx}
\usepackage{caption}
\usepackage{subcaption}
\usepackage{multirow}

\usepackage[ruled,vlined]{algorithm2e}
\usepackage{forest}

\usepackage[numbers]{natbib}

\newcommand{\rmdo}{\mathrm{do}}
\newcommand{\param}{\mathrm{pm}}
\newcommand{\np}{\mathrm{np}}
\newcommand{\pert}{\mathrm{pert}}

\newcommand{\p}{\mathrm{p}}
\newtheorem{assumption}{Assumption}

\newcommand{\code}{\url{https://github.com/bohanwu2000/npp}}
\usepackage{jmlr2e}
\usepackage{aliascnt}

\newaliascnt{proposition}{theorem}
\newtheorem{proposition}[proposition]{Proposition}
\aliascntresetthe{proposition}
\newaliascnt{lemma}{theorem}
\newtheorem{lemma}[lemma]{Lemma}
\aliascntresetthe{lemma}
\newaliascnt{remark}{theorem}
\newtheorem{remark}[remark]{Remark}
\aliascntresetthe{remark}

\input{macros}

\usepackage{lastpage}
\ShortHeadings{Nonparametric Perturbations with Generalized Bayes}{Wu et al. }
\firstpageno{1}

\usepackage{cleveref}
\crefname{assumption}{Assumption}{Assumptions}
\crefname{proposition}{Proposition}{Propositions}
\Crefname{proposition}{Proposition}{Propositions}
\crefname{theorem}{Theorem}{Theorems}
\Crefname{theorem}{Theorem}{Theorems}
\crefname{equation}{eq.}{eqs.}
\Crefname{equation}{Eq.}{Eqs.}
\Crefname{algocf}{Algorithm}{Algorithms}
\crefname{algocf}{algorithm}{algorithms} 

\begin{document}

\title{Adaptive Nonparametric Perturbations of Parametric Models with Generalized Bayes}

\author{
\name Bohan Wu\thanks{Equal contribution} \email bw2766@columbia.edu \\
\addr Department of Statistics\\
Columbia University\\
New York, NY, USA
\AND
\name Eli N. Weinstein\footnotemark[1] \email enawe@dtu.dk \\
\addr Department of Chemistry\\
Technical University of Denmark\\
Kgs. Lyngby, Denmark
\AND
\name Sohrab Salehi \email sohrab.salehi@mskcc.org \\
\addr Computational Oncology, Department of Epidemiology and Biostatistics\\
Memorial Sloan Kettering Cancer Center\\
New York, NY, USA
\AND
\name Yixin Wang \email yixinw@umich.edu \\
\addr Department of Statistics\\
University of Michigan\\
Ann Arbor, MI, USA
\AND
\name David M. Blei \email david.blei@columbia.edu \\
\addr Department of Computer Science and Department of Statistics\\
Columbia University\\
New York, NY, USA
}

%\editor{Pierre Alquier}
\editor{}

\maketitle

\begin{abstract}
Parametric Bayesian modeling offers a powerful and flexible toolbox for machine learning. Yet the model, however detailed, may still be wrong, and this can make inferences untrustworthy. In this paper we introduce a new class of semiparametric corrections for parametric Bayesian models, when the target of inference is a functional of the true data distribution.
Our starting point is a fully Bayesian modeling approach, which explicitly accounts for the possibility that the parametric model is wrong.
Asymptotic analysis shows that this approach is both robust to model misspecification and data efficient, achieving fast convergence when the parametric model is close to true.
However, the fully Bayesian approach is limited in its practical usefulness by the challenges of conducting inference and computing a Bayes factor for a nonparametric model.
We therefore propose a novel model correction based on generalized Bayes, which entirely avoids the need to compute a nonparametric Bayes factor, but preserves the robustness and efficiency of the fully Bayesian approach.
We demonstrate our method by estimating causal effects of gene expression from single cell RNA sequencing data. Overall, we offer a new efficient approach to robust Bayesian inference with parametric models.
\end{abstract}


\section{Introduction}
% Idea #1
Parametric Bayesian modeling offers a flexible toolbox for incorporating scientific knowledge and hypotheses into probabilistic machine learning.
Doing so has many benefits, such as increased data efficiency and interpretability.
Yet it can also present risks, as the model remains only an approximation of reality.
How can we ensure inferences are trustworthy even when the model might be wrong?
% Thanks to increasingly automated and easy-to-use software for Bayesian analysis, the problem of model misspecification has only grown increasingly pressing.
% Experts develop, criticize and refine models based on the datasets they have at hand.
% Then they release a convenient software package, which users apply  off-the-shelf to draw inferences from their own datasets.
% But users often do not have the time or knowledge to check and refine these models, to make sure they accurately describe their data.
% As a result, models are likely to be applied to data they describe poorly.

In this paper, we study this problem in the context of functional inference.
While standard posterior inference focuses on the distribution of model parameters, functional inference concerns the estimation and uncertainty quantification of functionals of the population distribution. % From the perspective of de Finetti's theorem, Bayesian inference can be viewed as inference on the population distribution through a prior on random probability measures. 
It can leverage flexible machine learning models, which learn complex population distributions. Developing accurate, calibrated and robust inference of functional estimands is fundamental to research areas such as causal inference \citep{Rose2011,Hill2011,Castillo2015}. 
However, developing Bayesian methods for functional inference remains an active challenge \citep{Lyddon2018,Ray2020BvM}.

%Our proposed method is straightforward to implement, as it does not require any change to how the model's posterior is computed, instead only needing access to samples from the predictive.

In this article, we study the problem of reliable and efficient functional inference based on a parametric Bayesian model, even when the parametric model might be wrong. We are specifically concerned with the possibility that the model likelihood is misspecified.
Concretely, assume the data is i.i.d. from some true underlying distribution, $x_{1:n} \iid \p_0$. A parametric Bayesian model of the data is specified by a prior over an unknown parameter $\pi(\theta)$, together with a likelihood $\p_\theta$,
\begin{equation} \label{eqn:parametric}
\begin{aligned}
\theta &\sim \pi(\theta),\\
x_{1:n} &\iid \p_\theta.
\end{aligned}
\end{equation}
The problem is that the model may not be able to describe $\p_0$. The induced prior over the distribution of the data, $\pi(\p_\theta)$, only has support on a subset of all possible distributions, $\mathcal{M}_\param = \{\p_\theta : \theta \in \Theta\} \subset \mathcal{P}$.
If the likelihood is misspecified, such that $\p_0 \notin \mathcal{M}_\param$, inferences drawn from the posterior may be unreliable, as no amount of data can overwhelm such an overconfident prior.

A fully Bayesian solution is to adjust the model to account for this possibility. 
From the perspective of de Finetti's theorem, Bayesian inference can be viewed as inference of the data distribution through a prior on probability measures.
Following this perspective, rather than insist the true data distribution matches the likelihood distribution exactly, we allow for distributional perturbations, so the data is generated from an alternative distribution, $x_{1:n} \iid \tilde \rmp$, instead of $x_{1:n} \iid \p_\theta$.
We place a prior on $\tilde \rmp$ that is centered at $\rmp_\theta$ but has sufficient support that $\tilde \rmp$ can be any distribution in $\mathcal{P}$. 
The resulting \emph{nonparametrically perturbed parametric} (NPP) Bayesian model encodes the belief that the likelihood is ``probably roughly true.''

This basic idea of combining  a parametric model and a nonparametric component has a long history in Bayesian statistics. 
It can, however, be challenging to implement in large-scale machine learning settings, as it requires posterior inference over a nonparametric model.
In this paper, we aim to develop a practical alternative.

%In practice, the statistician who uses a particular parametric likelihood $\p_\theta$ to analyze their data usually does not do so because they actually believe that their chosen likelihood perfectly describes the truth.
%They do so because the likelihood offers a convenient simplification of reality, often one grounded in specific domain knowledge.
%That is to say, they believe the likelihood is ``probably roughly true.''

%The problem is how to encode such a mature opinion into a Bayesian model.
%In this paper, we study a general approach.
%Rather than insist the distribution of the data matches the likelihood exactly, we allow for perturbations, so the data is generated as $x \sim \p_\theta + \delta$ instead of as $x \sim \p_\theta$.
%We place a prior on $\delta$ that is centered at $0$ but has support everywhere, so $\p_\theta + \delta$ can be any distribution in $\mathcal{P}$.
%The resulting nonparametrically perturbed parametric (NPP) Bayesian model encodes the belief that the likelihood is ``probably roughly true.''

First, we elucidate the key theoretical advantages of NPP models. We show that, with an appropriate choice of prior, NPP models can achieve robustness without sacrificing asymptotic efficiency. 
We then develop a generalized Bayes approach that imitates the NPP posterior, retaining its nice statistical properties but with more tractable computation.
This \textit{generalized NPP} (gNPP) posterior does not require any change to how the parametric model's posterior is computed, instead only needing access to samples from the predictive distribution. This efficiency is supported by asymptotic results. We demonstrate the method in simulations and with an application to gene expression in cancer.

\subsection{Related Work}

Robustness and efficiency are core questions in functional estimation. Perhaps the most well-established route to achieving these properties is to employ semiparametric corrections, such as influence functions, doubly robust estimators, and Neyman orthogonality~\citep{Kennedy2022-ml}.
We show how to achieve similar properties via a very different methodological strategy, which extends beyond the context of causal inference.
Specifically, consider a doubly robust estimator that uses a parametric outcome model and a nonparametric propensity model. 
When the outcome model is well-specified, the estimator is efficient: it converges to the estimand, the treatment effect, at a parametric rate. 
When the outcome model is misspecified, the estimator is robust: it still converges to the true effect, albeit at a slower rate \citep{Bang2005,Antonelli2020}.
We show that NPP models offer the same guarantees, but are more general: they are not specific to causal functionals, and they do not require a propensity model or another influence function-based correction.
%Indeed, NPP models carry these guarantees to any functional, not just causal functionals, without any functional-specific derivations.
We engineer gNPP posteriors to inherit these properties at reduced computational cost.
Note, though, that NPP and gNPP modeling is not in conflict with influence function-based methods, and in \Cref{sect-theory} we explain how the two methods can be combined.

Our work sits in the broader context of robust Bayesian learning. We estimate functionals of the true data distribution $\p_0$, following previous work on robust Bayesian modeling such as \citet{Lyddon2019,Pompe2021-nj,Dellaporta2022}. Recent work has focused on achieving robust and efficient functional estimation by adapting semiparametric corrections to the Bayesian context \citep{Antonelli2020,Walker2023-pn,Yiu2023Semiparametric}.
Inspired by this progress, we aim for similar guarantees via alternative and more general methods.

Other robust Bayes approaches consider a different problem, in which the target of inference may not be identified as a functional of $\p_0$~\citep{DiFinetti1961,Wang2017robustness,Wang2018robustBayesianmodeling,Miller2018,Bhatia2024-wi}.
For example, $\p_0$ may be corrupted by a population of outliers. The classical $\epsilon$-contamination model posits $\p_0$ is distorted to $(1-\epsilon) \p_0 + \epsilon q$ \citep{Huber2009-eu}.
Our use of \textit{robust} concerns reliable inference in the face of possible model misspecification, rather than data contamination.

%One powerful way of achieving robust and efficient functional estimates is to employ semiparametric corrections such as influence functions, doubly robust estimators, and Neyman orthogonality~\citep{Antonelli2020,Walker2023-pn,Yiu2023Semiparametric}. 
%We achieve similar theoretical guarantees using a very different methodological strategy.
%Our approach is not in conflict with influence function-based methods, and in \Cref{sect-theory} we explain how the two methods can be combined.

%The robustness and efficiency of gNPP models bear analogy to the properties of doubly robust estimators used for causal inference.
%Specifically, consider a doubly robust estimator that uses a parametric outcome model and a nonparametric propensity model. When the outcome model is well-specified, the estimator will converge at a parametric rate. When the outcome model is misspecified, the estimator will still converge to the true effect, but at a slower rate \citep{Bang2005,Antonelli2020}.
%A gNPP model with a parametric outcome model perturbed by a nonparametric outcome model offers the same guarantees, even without a propensity model.
%But the gNPP framework is not restricted to causal functionals. 

Our work builds on the long literature on nonparametric perturbations of parametric Bayesian models. In this paper, we are interested in perturbation techniques for generic parametric models $\p_\theta$, so we focus on reviewing these methods. 

\citet{antoniak1974mixtures} consider nonparametric perturbations with a Dirichlet process, in the form of mixture of Dirichlet processes. \citet{Berger2001} consider nonparametric perturbations with a Polya tree, but in practice their method is limited to low-dimensional Euclidean data.
\citet{Lyddon2018} consider perturbations using a Dirichlet process.
We prove their approach sacrifices data efficiency, and propose an alternative that does not. 
%\citet{Amin2021-yk} develop nonparametric perturbations that retain data efficiency, but their method is restricted in scope to discrete sequence data.
\citet{Miller2019-ha} consider perturbations based on Dirichlet process mixture models, but they do not provide theoretical analysis. Another prominent class of examples of combining a parametric component with a nonparametric model is semiparametric Bayesian regression that involves a finite-dimensional regression parameter alongside a nonparametric function. The area has a large body of work, see e.g. \citet{Blight1975,OHagan1978,Kottas2001,Berry2002,Kowal2024} and the references therein.

Another approach that is used to perturb a parametric model is to create a mixture of different copies of the model, e.g. replacing a Gaussian with a Dirichlet process mixture of Gaussians~\citep{Lo1984,Escobar1995}.
One could consider similarly replacing a parametric model with a Dirichlet process mixture of that model.
But this approach grows rapidly in theoretical and computational difficulty as the complexity of the parametric model grows. 

Our posterior approximation builds on generalized Bayesian posteriors, which replace the model likelihood with an alternative loss~\citep{Jiang2008-zw,Zhang2006-sq,Bissiri2016,Jewson2018-mw,Shao2018-ef,Miller2021,Knoblauch2022-tl}. 
Following \citet{Weinstein2023}, we construct a generalized Bayes factor between a parametric and nonparametric model by using a statistical divergence. 
Unlike in \citet{Weinstein2023}, we only need samples from the parametric model's posterior to estimate this generalized Bayes factor.

\section{Nonparametric Perturbations of Bayesian Models} \label{sect-method}
\subsection{Target of inference} \label{inference-target}

% 1. Introduce target of inference.
% 2. Here's the challenge: misspecified.
Suppose we observe i.i.d. data $x_{1:n} \iid \rmp_0$. Our target estimand is the output of a functional applied to the data generating distribution, $\psi(\p_0)$, where $\psi: \mathcal{P} \to \R^s$.
Here are some examples:
\begin{itemize}
    \item \textbf{Summary statistics.} The target of inference could be the mean, in which case the functional is $\psi(\p_0) = \int x\, \p_0(x) \di x$. Likewise we could infer the variance, etc.
    \item \textbf{Loss minimizers.} The target of inference could be the minimizer of a loss or utility, $\psi(\p_0) = \argmin_\alpha \int \ell(x\s \alpha) \p_0(x) \di x$.  The loss can even be specified in terms of the parametric model $\p_\theta$, e.g. if $\ell(x\s \alpha) = -\log \p_\alpha(x)$ then the target of inference corresponds to the maximum likelihood estimate (MLE) from infinite data. This is equivalent to the KL minimizer, $\psi(\p_0) = \argmin_{\theta} \kl{\p_0 \| \p_\theta}$.
    \item \textbf{Causal effects.} The target of inference could be a functional of $\p_0$ describing the effect of an intervention~\citep{Kennedy2022-ml}. For example, consider data $x = (a, y, w)$ consisting of a treatment $a \in \{0, 1\}$, outcome $y$, and confounder $w$. 
    The average treatment effect is, 
\begin{equation*}
\begin{aligned}
    \psi(\p_0) &= \EE_{\p_0}[Y\s \rmdo(a=1)] - \EE_{\p_0}[Y\s \rmdo(a=0)]\\ &= \int \int y\, [\p_0(y \mid a=1, w) - \p_0(y \mid a=0, w)] \di y\, \p_0(w) \di w.
\end{aligned}
\end{equation*}
    % Other causal functionals include conditional average treatment effects and the effects of stochastic interventions, and may involve instrumental variables, proxies, or other types of variables used to correct for confounding~\citep{Kennedy2022-ml}.
\end{itemize}

The inference challenge is that if the parametric model is misspecified, the posterior will always put zero mass on the true data distribution, and so we cannot learn the true functional, $\psi(\p_0)$.

\subsection{Nonparametric Perturbations of Parametric Models}
We will study nonparametric perturbations of parametric Bayesian models that use a specific type of prior. In particular, we consider NPP models that mix the parametric Bayesian model with a perturbed distribution,
\begin{equation} \label{npp}
\begin{aligned}
    \theta &\sim \pi(\theta),\quad b \sim \mathrm{Bernoulli}(\eta), \\
    &\begin{cases}
\mathrm{p} = \mathrm{p}_\theta & \text{if } b = 1 \\
\mathrm{p} \sim \pi_{\pert}(\p \mid \p_\theta) & \text{if } b = 0
    \end{cases}, \\
x_{1:n} &\iid \p. 
\end{aligned}
\end{equation}
In this model, the entire dataset $x_{1:n}$ is either drawn from the parametric model ($b=0$) or from an alternative ($b=1$).
There are many ways to perturb. 
%If not, the data comes from a distribution $\p$ drawn from a nonparametric model $\pi_{\pert}(\rmp \mid \rmp_\theta)$, which samples perturbations of $\p_\theta$.
%  model could be a Dirichlet process, $\pi_{\pert}(\p \mid \p_\theta) = \textsc{dp}(\rmp_\theta, 1/h)$, where $1/h$ is the concentration. Or, the perturbation model
For example, $\pi_{\pert}(\rmp \mid \rmp_\theta)$ could be a Dirichlet process mixture model \citep{Miller2019-ha},
\begin{equation} \label{eqn:dpmm_npp}
\Sigma \sim g(\Sigma), \quad \sum_{k=1}^\infty w_k \delta_{\mu_k} \sim \textsc{dp}(\rmp_\theta, \alpha), \quad \rmp = \sum_{k=1}^\infty w_k \mathrm{Normal}(\mu_k, \Sigma). 
\end{equation}
%Note as the concentration parameter $\alpha \to \infty$ and scale parameter $\tau \to 0$, the perturbation reduces to zero, i.e. $\p \sim \pi_{\pert}(\rmp \mid \rmp_\theta) \to \delta_{\p_\theta}$.
Other possibilities are Polya trees~\citep{Berger2001} and Gaussian process density models~\citep{Adams2009-sa}.
We assume the perturbation model is nonparametric, placing its support on any distribution over $x$, including $\p_\theta$.
We do not assume that the perturbation depends on $\p_\theta$ explicitly, i.e. $\pi_{\pert}(\p \mid \p_\theta)$ can be replaced by a generic nonparametric model $\pi_{\pert}(\p)$.

We want to use an NPP model instead of a conventional nonparametric model because parametric models are efficient for small data.  
We can think of an NPP model as a nonparametric model with a special spike-and-slab prior on the data distribution, that is centered at the parametric model.
The prior has a ``spike'' component that puts its mass just on the parametric model ($b=1$), and a ``slab'' component extending over all distributions ($b=0$). 
This prior is nonparametric but regularizes inferences towards the parametric model. 
This choice of an adaptive prior enables data-efficient learning, as we describe next.

\subsection{NPP Bayesian Posterior}\label{sec:method_robust_eff}
We now examine some of the advantages of NPP models for functional inference.
We can perform functional inference with an NPP model by computing the posterior over the target functional, $\Pi(\psi(\p) \mid x_{1:n})$. 
The NPP posterior is a mixture between a parametric and nonparametric posterior, 
\begin{equation} \label{eqn:npp_decomp}
\Pi\left(\psi \left(\p \right) \mid x_{1:n}\right) = \eta_n \Pi_{\param}\left( \psi \left(\p_\theta\right)  \mid x_{1:n} \right)  + (1 -  \eta_n)\Pi_\pert \left(  \psi \left(\p\right)  \mid x_{1:n} \right) , 
\end{equation}
where $ \Pi_{\param}\left( \psi \left(\p_\theta\right)  \mid x_{1:n} \right)$ is the posterior from the parametric model, $ \Pi_\pert\left(  \psi \left(\p\right)  \mid x_{1:n} \right)$ is the posterior from the nonparametric perturbation model, and $\eta_n := \Pi\left( b = 1 \mid x_{1:n} \right)$.
% (the posterior given $b = 1$) (the posterior given $b=0$)
This decomposition fully separates the parametric and nonparametric components.
The weight $\eta_n$ adaptively trades off between the parametric and nonparametric posterior, based on the data.

%That is, we propagate our uncertainty about the data distribution, $\Pi(\p \mid x_{1:n})$, into uncertainty about the functional $\psi(\p)$.

%and efficiency. Larger $\eta_n$ shrinks the nonparametric posterior towards the parameter posterior that achieves faster posterior convergence under limited sample size. On the other hand, a misspecified parametric model results in vanishing $\eta_n$ that shifts the inference towards the more robust nonparametric posterior. 

NPP models provide a \textit{robust} and \textit{efficient} way of learning about $\psi(\p_0)$.
To make these notions precise, we compare NPP models to two alternatives: (a) the original parametric model, given by $\theta \sim \pi(\theta), x_{1:n} \iid \p_\theta$, and (b) a nonparametric model with a generic prior, $\p \sim \pi_{\np}(\p), x_{1:n} \iid\p$, where $\pi_{\np}(\p)$ has support over all distributions on $\mathcal{X}$.
We use $\Pi_{\param}(\cdot \mid x_{1:n})$ to denote the parametric model posterior, $\Pi_{\np}(\cdot \mid x_{1:n})$ to denote the nonparametric model posterior, and $\Pi(\cdot \mid x_{1:n})$ to denote the NPP model posterior.
We will show that the parametric model is efficient but not robust, while the nonparametric model is robust but not efficient.
Formal statements and proofs of the following results are in \Cref{sect-theory}.

\paragraph{Misspecified case.} We want inferences about $\psi(\p_0)$ to be \textit{robust} to model misspecification, such that they always converge to the truth with enough data.
The NPP model is robust: \Cref{thm-NPP} shows that $\Pi\left(\psi \left(\p \right) \mid x_{1:n}\right) \to \delta_{\psi(\p_0)}$ even when $\p_0 \notin \mathcal{M}_\param$.
To see why, recall from \Cref{eqn:npp_decomp} that the NPP posterior is a mixture of a parametric and nonparametric posterior. The parametric posterior will not converge to the truth, $\Pi_{\param}(\psi(\rmp_\theta) \mid x_{1:n}) \not\to \delta_{\psi(\rmp_0)}$, but the nonparametric component will, $\Pi_{\pert}(\psi(\rmp) \mid x_{1:n}) \to \delta_{\psi(\rmp_0)}$.
The overall NPP posterior converges to the truth because the mixing weight $\eta_n$ asymptotically places all weight on the nonparametric component: if $\rmp_0 \notin \cM_\param$ then 
$\eta_n \to 0$ a.s. as $n \to \infty$ (\Cref{thm:lambda-consistency}).

To understand the behavior of the mixing weight, we can write $\eta_n/(1-\eta_n)$ in terms of a Bayes factor that compares the marginal likelihood of the data under the parametric and nonparametric models, weighted by the prior odds $\eta/(1-\eta)$,
\begin{equation} \label{Bayes-factor}
    \frac{ \eta_n}{1 -  \eta_n} = \frac{\Pi_{\param}(x_{1:n}) \eta}{\Pi_{\pert}(x_{1:n})(1 - \eta)} =: \BF_n,
\end{equation}
where
\begin{equation*}
    \Pi_{\param}(x_{1:n}) = \int \prod_{i=1}^n \p_\theta(x_i) \pi(\theta) \di \theta, 
\end{equation*}
and
\begin{equation*}
    \Pi_{\pert}(x_{1:n}) = \int \int \prod_{i=1}^n \p(x_i) \pi_{\pert}(\p|\p_\theta) \pi(\theta) \di \p \di \theta. 
\end{equation*}
When the parametric model is misspecified, the Bayes factor prefers the more expressive nonparametric model, since it can match the data distribution, while the parametric model cannot.
Hence, $\BF_n \to 0$, so the mixing weight $\eta_n \to 0$ a.s. as $n \to \infty$ (\Cref{thm:lambda-consistency}).

In summary, the NPP model is robust: $\Pi(\psi(\rmp) \mid x_{1:n}) \to \delta_{\psi(\rmp_0)}$ even when when $\p_0 \notin \cM_\param$. 
In this way, the NPP model behaves like a generic nonparametric model, for which we also expect $\Pi_{\np}(\psi(\rmp) \mid x_{1:n}) \to \delta_{\psi(\rmp_0)}$.
A parametric model, by contrast, is not robust: in general, $\Pi_{\param}(\psi(\rmp_\theta) \mid x_{1:n}) \not\to \delta_{\psi(\rmp_0)}$ when $\p_0 \notin \cM_\param$.
% The NPP model inherits its robustness from the nonparametric compone nonparametric models: $\Pi_{\np}(\psi(\rmp) \mid x_{1:n}) \to \delta_{\psi(\rmp_0)}$.

\paragraph{Well-specified case.}
We next consider the case where the parametric model is well-specified, $\p_0 \in \cM_\param$.
In this case, the NPP model converges to the truth at an efficient rate: the posterior 
$\Pi(\psi(\p) \mid x_{1:n})$ contracts at a rate $1/\sqrt{n}$ around $\psi(\rmp_0)$ when $\p_0 \in \mathcal{M}_\param$. 
To see this, return to the decomposition in \Cref{eqn:npp_decomp}. 
Both the parametric posterior $\Pi_{\param}(\psi(\rmp_\theta) \mid x_{1:n})$ and the nonparametric posterior $\Pi_{\pert}(\psi(\rmp) \mid x_{1:n})$ will converge to the truth, but the parametric posterior will converge faster, at a rate $1/\sqrt{n}$, while the nonparametric component will in general converge more slowly.
The NPP posterior inherits the rate of the parametric posterior because the mixing weight asymptotically places all weight on the parametric component: if $\rmp_0 \in \cM_\param$ then 
$\eta_n \to 1$ a.s. as $n \to \infty$ (\Cref{thm:lambda-consistency}).

To see why $\eta_n \to 1$ intuitively, consider again the Bayes factor, \Cref{Bayes-factor}. 
Both the parametric and the nonparametric model describe the data distribution, but the parametric model uses fewer parameters than the nonparametric model: a finite number, rather than infinite.
Moreover, the parametric model is nested inside the nonparametric model, since it corresponds to sampling $\p = \p_\theta$ from $\pi_\pert(\p \mid \p_\theta)$.
By Bayesian Occam's razor, the Bayes factor prefers the simpler parametric model, i.e. $\BF_n \to \infty$~\citep[][Chap. 28]{Dawid2011-kb,Hong2012-ar,MacKay2003-db}.
Hence $\eta_n \to 1$.

In summary, when the parametric model is correctly specified, $\p_0 \in \mathcal{M}_\param$, the NPP posterior $\Pi(\psi(\p) \mid x_{1:n})$ converges to the truth $\psi(\rmp_0)$ at an efficient parametric rate of $1/\sqrt{n}$.
So here, the NPP posterior behaves like a parametric Bayes posterior, for which we also expect a convergence rate of $1/\sqrt{n}$ by the Bernstein-von Mises theorem.
In a nonparametric model, however, the convergence rate is generally slower, e.g. $\Pi_{\np}(\psi(\p) \mid x_{1:n})$ may converge at a rate of $1/\sqrt[3]{n}$ or worse~\citep{Ghosal2000-pf}.

%We quantify efficiency with the rate at which the posterior converges to the truth.
% TODO: In some sense an NPP model achives the best of both worlds. But if the parametric model is misspecified, its rate will be slower. 

\paragraph{Finite samples.} A possible concern is that in practice there is always some amount of misspecification, and so we cannot ever expect a parametric rate. However, the Bartlett-Lindley effect says the Bayes factor will place more weight on the simpler model up until there is enough data to determine that model is wrong~\citep{Miller2018}.
So, heuristically, we expect the NPP posterior to converge towards the truth as quickly as the parametric posterior, up until the parametric model stops providing a good description of the data.

%When the parametric model is misspecified, the NPP model is only guaranteed to converge at a nonparametric rate. Heuristically, however, we expect the NPP posterior to converge towards the truth about as quickly as the parametric posterior, up until the parametric model stops providing a good description of the data.
%Specifically, the Bartlett-Lindley effect says the Bayes factor will place more weight on the simpler model up until there is enough data to determine that model is wrong~\citep{Miller2018}.

\subsection{NPP Generalized Bayesian Posterior}
We have seen that NPP models offer both robustness and efficiency, but this comes at a computational cost.
Computing the NPP posterior (\Cref{eqn:npp_decomp}) requires computing not only the parametric model posterior, but also (a) the mixing weight $\eta_n$, and (b) the nonparametric posterior, $ \Pi_\pert\left(  \psi \left(\p\right)  \mid x_{1:n} \right)$.
In this section we propose a new inference approach that replaces each term with alternatives that are efficient to compute and practical to implement, but also preserve robustness and efficiency.
The aim of this \textit{generalized NPP} (gNPP) approach is not to directly approximate the NPP, but rather to mimic its statistical behavior while easing computation.

% The purpose of the Bayes factor is to select a model.
% We fill in an alternative that is easier to compute.
% Overkill to get full Bayes factor.

\paragraph{Mixing weight.} The mixing weight $\eta_n$ depends on the Bayes factor $\BF_n$ (\Cref{Bayes-factor}), which compares the marginal likelihood of the parametric model to the marginal likelihood of a nonparametric alternative. Computing marginal likelihoods is often challenging, especially for nonparametric models.
We propose to replace the Bayes factor with a generalized Bayes factor, which uses divergences instead of marginal likelihoods to evaluate the parametric model~\citep{Shao2018-ef,Weinstein2023}.
Our approach is motivated by the observation that, asymptotically, the standard Bayes factor will converge to the posterior expected KL divergence, $\frac{1}{n} \log \BF_n \to -\mathbb{E}[\kl{\p_0 \|\p_\theta} \mid x_{1:n}]$ a.s., under regularity conditions \citep{Dawid2011-kb,Shao2018-ef,Miller2021,Weinstein2023}. We consider alternative divergences, easing computation while matching the behavior of the Bayes factor.

Let $\diverge(\p, \p_0)$ denote a divergence between probability distributions, which satisfies $\diverge(\p,\p_0) = 0$ when $\p = \p_0$ and $\diverge(\p, \p_0) > 0$ when $\p \neq \p_0$. Let $\diverge_n(\p, \p_0)$ denote an estimate of $\diverge(\p, \p_0)$ based on data $x_{1:n}$ i.i.d. from $\p_0$. Concretely, the divergences we consider are the Wasserstein distance, the maximum mean discrepancy and the kernelized Stein discrepancy, though other choices are also possible~\citep{Gretton2012-do,Liu2016}.
We use the divergence to construct a generalized Bayes factor (gBF),
\begin{equation} \label{gBF}
\gBF_n := \Xi\left(\frac{\mathbb{E}[\diverge_n(\p_\theta, \p_0)]}{\mathbb{E}[\diverge_n(\p_\theta, \p_0) \mid x_{1:n}]} (n+1)^{-r}\right) \frac{\eta}{1-\eta}.
\end{equation}
Here, $\mathbb{E}[\diverge_n(\p_\theta, \p_0) \mid x_{1:n}]$ is the expected value of the divergence under the posterior $\Pi_\param(\theta \mid x_{1:n})$, as appears in asymptotic Bayes factor. $\mathbb{E}[\diverge_n(\p_\theta, \p_0)]$ is the expected divergence under the prior $\pi(\theta)$, which ensures the gBF reverts to the prior when there is no data.
The rate $r > 0$ is a hyperparameter, and the function $\Xi(x) = \exp(1-1/x)x$ is monotonic and satisfies $\Xi(x)\to0$ as $x \to 0$ and $\Xi(x) \to \infty$ as $x \to\infty$.
Although \Cref{gBF} does not contain the standard Bayes factor as a special case, it is related: like the standard marginal likelihood, it is a posterior average over an estimate of the model-data mismatch.
Now, as an alternative to the NPP posterior, we propose to consider a generalized posterior that uses $\hat{\eta}_n = 1/(1 + \gBF_n^{-1})$ in place of $\eta_n$.

The aim of this generalized Bayes method is not to directly estimate the mixing weight $\eta_n$, but rather to match its statistical behavior.
When the parametric model is misspecified, $\hat{\eta}_n \to 0$, just like $\eta_n$.
The reason is that $\diverge(\p_\theta, \p_0) > 0$ for all $\theta$ when $\p_0 \notin \mathcal{M}_\param$, so $\mathbb{E}[\diverge_n(\p_\theta, \p_0) \mid x_{1:n}]$ will converge to a positive value.
On the other hand, when the parametric model is well-specified, $\hat{\eta}_n \to 1$, just like $\eta_n$.
The reason is that the posterior will concentrate at $\p_0$, so $\mathbb{E}[\diverge_n(\p_\theta, \p_0) \mid x_{1:n}] \to 0$.
So long as the rate hyperparameter $r$ is chosen small enough that $\mathbb{E}[\diverge_n(\p_\theta, \p_0) \mid x_{1:n}](n+1)^r \to 0$ also, we will have $\hat{\eta}_n \to 1$.
So, the generalized Bayes factor is asymptotically \textit{consistent} in detecting misspecification.

In addition to achieving the same asymptotic limits as the original Bayes factor, the generalized Bayes factor also approaches those limits at a similar rate, thanks to the choice of function $\Xi(x)$ (\Cref{sec-gbf-scaling}).
Before we observe data (i.e. for $n=0$), both the Bayes factor and the generalized Bayes factor coincide with the prior odds ratio, $\gBF_0 = \BF_0 = \eta/(1-\eta)$.
In the generalized Bayes factor, this is thanks to the inclusion of the prior divergence, $\mathbb{E}[\diverge_n(\p_\theta, \p_0)]$.

However, the key advantage of the generalized Bayes factor is computation: it requires only the ability to draw samples from the prior and posterior of the parametric model, to approximate $\mathbb{E}[\diverge_n(\p_\theta, \p_0)]$ and $\mathbb{E}[\diverge_n(\p_\theta, \p_0) \mid x_{1:n}]$. The marginal likelihood of the parametric model does not appear. The nonparametric model does not appear at all.

As a result, an important difference with the standard Bayes factor is that the standard approach compares the parametric model's fit to the nonparametric model's fit.
The generalized Bayes factor instead focuses on just checking the parametric model's fit, and does not directly trade off against an alternative.
In other words, the gNPP treats the alternative as a backup in case the parametric model goes wrong.

Indeed, the generalized Bayes factor can also be understood as a form of predictive check on the parametric model~\citep{Gelman1996,Moran2023-pa,Li2022-yd}. Specifically, the term $\mathbb{E}[\diverge_n(\p_\theta, \p_0) \mid x_{1:n}]$ compares the posterior over the parametric model's predictive distribution, $\Pi_\param(\p_\theta \mid x_{1:n})$, to data, using $\diverge_n$ as measure of model-data discrepancy.

\paragraph{Choice of divergence and rate.} We consider several types of divergence, which prioritize different kinds of mismatch between the model and data. All of the following divergences can be approximated by Monte Carlo methods. 
\begin{itemize}
\item \textbf{Wasserstein distance.} The Wasserstein measures the difference between two distributions as the cost of transforming one into the other by transporting probability mass. The $p$-Wasserstein distance is typically defined as 
\begin{align}
    \W_p^p(\rmp, \rmq) :=  \inf_{\pi \in \Pi(\rmp, \rmq)} \mathbb{E}_{\pi}[\|X - Y\|_2^p], 
\end{align}
where $\Pi(\rmp, \rmq)$ represents the set of all couplings between $\rmp$ and $\rmq$, i.e. the set of all joint distributions with $\rmp$ and $\rmq$ as marginals, and $\|\cdot\|_2$ denotes the Euclidean metric on $\R^\kappa$. We can estimate the distance between the model and the data using $m$ samples drawn from the model. We draw the same number of samples as data points, $m = n$, by default. Computing the Wasserstein distance between two empirical distributions reduces to solving a linear program, which can be done in \( O(n^3) \) time using interior-point methods \citep{pele2009fast}.  In practice, the Euclidean cost with $p\in\{1,2\}$ is most commonly used. The case $W_1$ admits the convenient Kantorovich--Rubinstein dual formulation in terms of $1$-Lipschitz test functions, while $W_2$ enjoys favorable geometric properties \citep{Villani2009}. In high dimensions, however, the empirical $p$-Wasserstein distances suffer from the curse of dimensionality, with a slow convergence rate of $O(n^{-1/\kappa})$ even for compactly supported measures. In contrast, alternatives such as the sliced Wasserstein distance and smoothed Wasserstein distance can achieve dimension-free $O(n^{-1/2})$ rates under the same conditions \citep{CheNilRig25OT}. For large-scale problems, entropically-regularized optimal transport \citep{Cuturi2013,Peyr2019} offers scalable approximations that can be computed with iteration complexity that does not depend on $\kappa$ \citep{Carlier2022}, and the associated Sinkhorn divergence \citep{genevay2018learning,genevay2019sample} comes with improved statistical convergence rates, $o(n^{-1/2})$ for sub-Gaussian measures and $O(n^{-1})$ for compactly supported measures \citep{del2023improved}.

\item \textbf{Maximum mean discrepancy (MMD).} The maximum mean discrepancy focuses on the worst case difference in expected value that two distributions assign to the same function in a reproducing kernel Hilbert space (RKHS). It is given by 
\begin{align}
\textsc{mmd}^2(\rmp, \rmq) &:= \sup_{\|f\|_{\mathcal{H}_k} \le 1} |\mathbb{E}_{\rmp}[f(X)] -  \mathbb{E}_{\rmq}[f(X)]|^2\\
   & =\E{X, X' \sim \rmp}{k(X, X')} - 2\E{X \sim \rmp, Y \sim \rmq}{k(X, Y)}  + \E{Y, Y' \sim \rmq}{k(Y, Y')}. 
\end{align} where $\mathcal{H}_k$ is an RKHS and $\|f\|_{\mathcal{H}_k}$ is the norm of a function in that RKHS. To estimate the MMD between the model and the data, we draw samples from the model. We again default to the same number of samples as data points, $m=n$. The MMD requires $O(n^2)$ time to compute; a linear-time approximation also exists \citep{Gretton2012-do}.

\item \textbf{Kernelized Stein discrepancy (KSD).} The kernelized Stein discrepancy focuses on differences in the (Stein) score function between two distributions, i.e. differences in the gradient of their log densities.
It is given by
\begin{align}
    \ksd(\rmp, \rmq) = \E{X, X' \sim \rmp}{\Delta_{\rmq, \rmp}(X)^T k(X, X') \Delta_{\rmq, \rmp}(X')}, 
\end{align}
where $\Delta_{\rmq, \rmp}(x) := \nabla_x \log \rmp(x) - \nabla_x \log \rmq(x)$. 
To estimate the KSD between the data and the model, we evaluate the empirical average of $\Delta_{\rmq, \rmp}(x)^T k(x, x') \Delta_{\rmq, \rmp}(x')$ over every pair of data points $(x, x')$.  For $n$ data points, this requires $O(n^2)$ time to compute \citep{Liu2016}, and there also exists a near-linear time approximation \citep{huggins2018random}. 
\end{itemize}
Both MMD and KSD rely on a kernel, which allows the user to control which aspects of a distribution to prioritize, but it also introduces the need for hyperparameter selection. Kernel hyperparameters are absent for the Wasserstein distance; however, one must instead choose the transportation cost.

Although the MMD and the KSD are sensitive to a rescaling of the kernel, the generalized Bayes factor is scale-invariant, since in \Cref{gBF} we normalize the posterior expected discrepancy by the prior expected discrepancy.

\begin{remark}[Choice of rate] \label{remark:set_r}
    For the MMD and KSD we have a consistent generalized Bayes factor so long as we choose a rate $0 < r < 1/2$, i.e. slower than the parametric rate (\Cref{thm:gbf}). Choosing a rate just below $1/2$ enables rapid convergence in the misspecified case, while still preserving consistency in the well-specified case.
For Wasserstein, we need to choose a slower rate when the dimension $\kappa$ of the data space is greater than four, namely $r \in (0, 2/(\kappa \lor 4))$ for $\X \subseteq \mathbb{R}^\kappa$ (\Cref{thm-Wp} in \Cref{sect-Wasserstein}).
\end{remark}

\paragraph{Nonparametric posterior.} Next we consider replacing the nonparametric term $\Pi_\pert\left(  \psi \left(\p\right)  \mid x_{1:n} \right)$ in the NPP posterior.
This term can be challenging to compute. The robustness and efficiency of the NPP model rely on the nonparametric perturbation $\pi_\pert(\p \mid \p_\theta)$ having support over all distributions on $\cX$,
so posterior computation requires integration over a very large space. On the other hand, since the mixing weight $\hat \eta_n$ does not depend on the nonparametric perturbation, the nonparametric component no longer needs to serve as a fully flexible density estimator; rather, it only needs to estimate the target consistently. 

We propose to replace the nonparametric perturbation with a semiparametric model that can consistently estimate $\psi(\p_0)$. 
That is, we consider a model $\p \sim \hat \pi_\pert (\p), x_{1:n} \sim \p$ such that $\hat{\Pi}_\pert(\psi(\p) \mid x_{1:n}) \to \delta_{\psi(\p_0)}$.
We replace the term $\Pi_\pert\left(  \psi \left(\p\right)  \mid x_{1:n} \right)$ in the NPP posterior with $\hat{\Pi}_\pert(\psi(\p) \mid x_{1:n})$.

To see how this approach can simplify computation, consider data consisting of a treatment and a response, $x_i = (y_i, a_i)$, and a target functional that is the expected response to a given treatment, $\psi(\p) = \mathbb{E}[Y \mid a_\star] = \int y\, \p(y \mid a_\star) \di y$. 
An NPP model would require us to place a prior over all distributions on $\mathcal{A} \times \mathcal{Y}$. 
However, to estimate the target functional consistently, it is enough to use a semiparametric model $y_i \sim \mathrm{Normal}(f(a_i), 1), f \sim \pi_f(f)$, and place a nonparametric prior on functions $f: \mathcal{A} \to \mathcal{Y}$. For example, $\pi_f$ could be a Gaussian process prior or a Bayesian additive regression tree \citep{Rasmussen2006,Chipman2010}.
These semiparametric models cannot describe all distributions over $\mathcal{A} \times \mathcal{Y}$, e.g. they cannot describe a distribution where the conditional $\p(y \mid a_\star)$ is not Gaussian, but they can still estimate any conditional expectation $\mathbb{E}[Y \mid a_\star]$. Moreover, the posterior is convenient to calculate, as we can rely on existing software for Gaussian processes and Bayesian additive regression trees.
Finally, compared to a fully nonparametric model, semiparametric models often require less data to produce accurate inference of the target estimand. 

%\blue{We note that $\hat{\Pi}_\pert(\psi(\p) \mid x_{1:n})$ is not the same as $\Pi_\pert\left(  \psi \left(\p\right)  \mid x_{1:n} \right)$, thus the finite-sample performance of the two posteriors could be different. However, in many applications of functional inference, this discrepancy is an acceptable cost: the primary goal is to obtain a consistent sequence of point estimates for the target estimand from $\Pi_\pert(\psi(\p) \mid x_{1:n})$, rather than to rely on its finite-sample uncertainty quantification.  }

\paragraph{Summary.} In summary, our generalized Bayes NPP posterior is
\begin{equation}
    \hat \Pi\left(\psi \left(\p \right) \mid x_{1:n}\right) := \hat \eta_n \Pi_{\param}\left( \psi \left(\p_\theta\right)  \mid x_{1:n} \right)  + (1 -  \hat \eta_n) \hat \Pi_\pert \left(  \psi \left(\p\right)  \mid x_{1:n} \right) ,
\end{equation}
where $\hat \eta_n$ comes from the generalized Bayes factor and $\hat \Pi_\pert$ comes from a semiparametric model.
We refer to this generalized posteriors as the \textit{generalized NPP (gNPP)} posterior.
In \Cref{sect-theory} we prove formally that the gNPP posterior preserves the robustness and efficiency of the NPP posterior: for any $\p_0$, $\hat \Pi\left(\psi \left(\p \right) \mid x_{1:n}\right) \to \delta_{\psi(\p_0)}$, and for $\p_0 \in \mathcal{M}_\param$, $\hat \Pi\left(\psi \left(\p \right) \mid x_{1:n}\right)$ converges at a rate $1/\sqrt{n}$.

The key to the gNPP is replacing the Bayes factor between parametric and nonparametric models with a generalized Bayes factor based on a divergence.
In doing so, we avoid the need to compute the marginal likelihood of the parametric model or of the nonparametric perturbation.
We also reduce the requirements and constraints on the perturbation model we can use: it no longer needs to be fully nonparametric, and the parametric model does not need to be nested inside it.
Instead, we can use a more computationally tractable semiparametric model, as it consistently estimates the target.

%Implementing the gNPP approximation is straightforward, since the only thing we need from the parametric model is a way to approximate its standard posterior and a way to calculate or draw samples from its predictive distribution. Moreover, we can rely on off-the-shelf tools for semiparametric Bayesian modeling, to approximate the posterior $\hat \Pi_\pert \left(  \psi \left(\p\right)  \mid x_{1:n} \right)$.

\section{Synthetic Data Illustration} \label{sec-synthetic}

\begin{figure}[!tp]
    \centering
    \begin{subfigure}[b]{0.45\columnwidth}
    \centering
    \caption{} \label{fig:bf_pt}
    \includegraphics[width=\columnwidth]{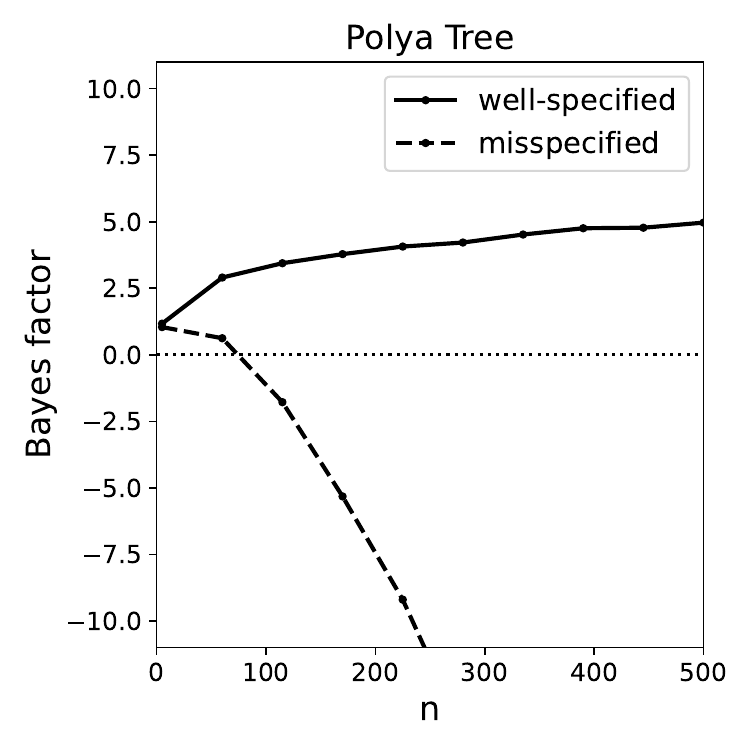}
    \end{subfigure}
    \begin{subfigure}[b]{0.45\columnwidth}
    \centering
    \caption{} \label{fig:bf_mmd}
    \includegraphics[width=\columnwidth]{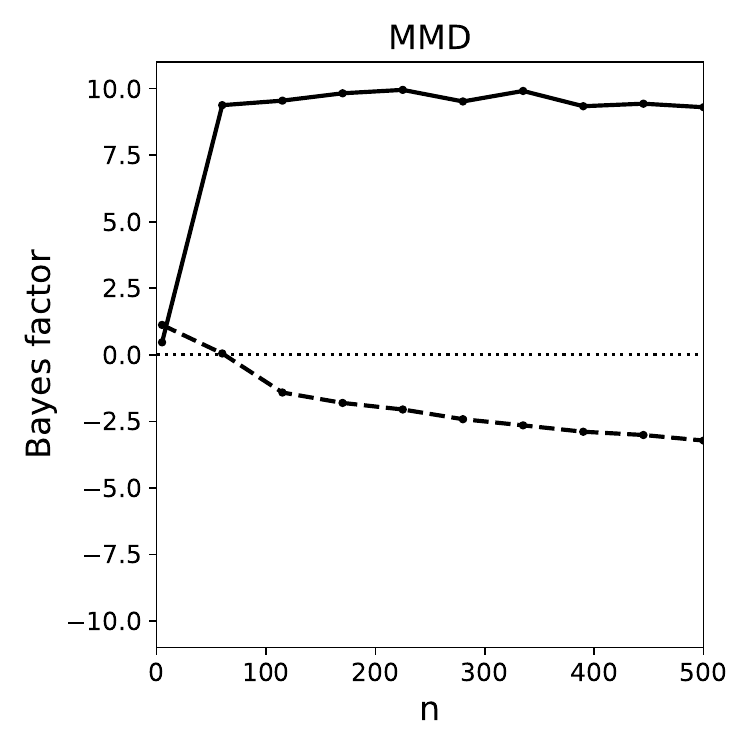}
    \end{subfigure}\\
    \begin{subfigure}[b]{0.45\columnwidth}
    \centering
    \caption{} \label{fig:bf_ksd}
    \includegraphics[width=\columnwidth]{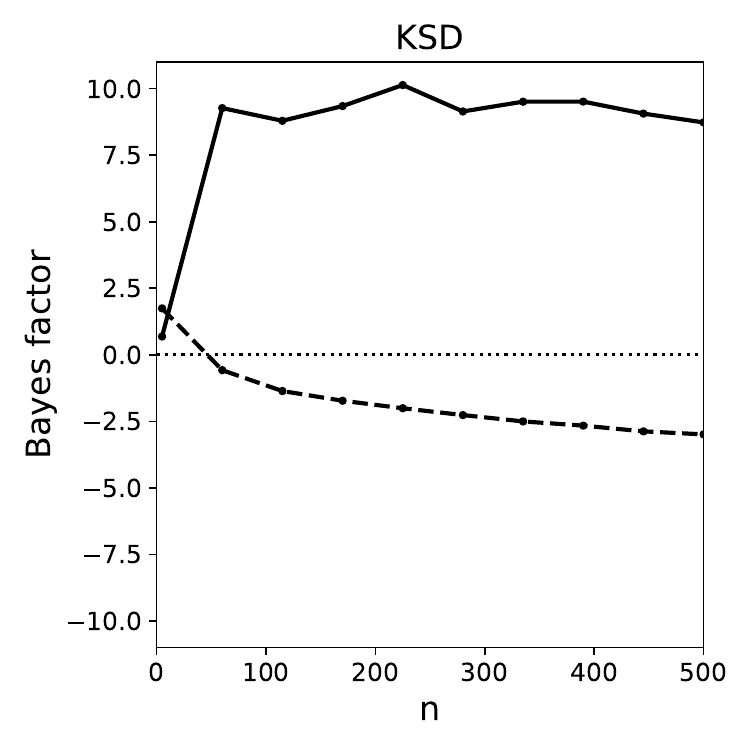}
    \end{subfigure}
    \begin{subfigure}[b]{0.45\columnwidth}
    \centering
    \caption{} \label{fig:bf_wass}
    \includegraphics[width=\columnwidth]{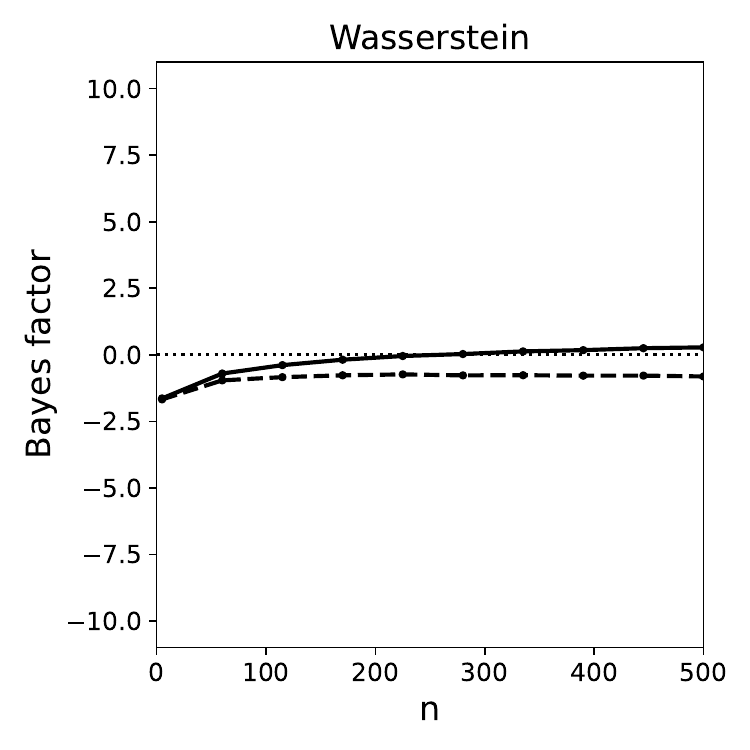}
    \end{subfigure}
    \caption{\textbf{Synthetic data (generalized) Bayes factors.} The log Bayes factor and log generalized Bayes factor comparing the parametric model to a nonparametric alternative, where positive values indicate the parametric model is favored. (a) NPP model with a Polya Tree. (b,c,d) gNPP models with the MMD, KSD and Wasserstein.} \label{fig:bf}
\end{figure}

\begin{figure}[!tp]
    \centering
    \captionsetup{position=top}
    \captionsetup[subfigure]{position=top, skip=-2pt}

    \setlength{\abovecaptionskip}{-3pt}
    \setlength{\belowcaptionskip}{-4pt}
    \begin{subfigure}[t]{0.42\columnwidth}
        \centering
        \caption{}\label{fig:pt_spec_kl}
        \includegraphics[width=\linewidth]{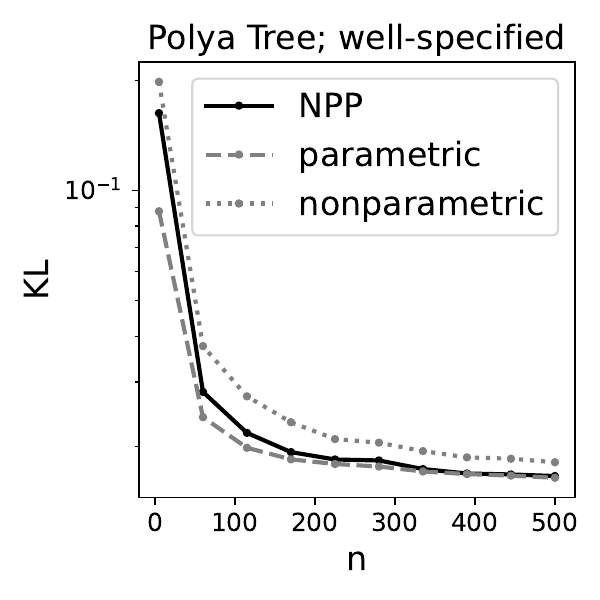}
    \end{subfigure}\hfill
    \begin{subfigure}[t]{0.42\columnwidth}
        \centering
        \caption{}\label{fig:pt_miss_kl}
        \includegraphics[width=\linewidth]{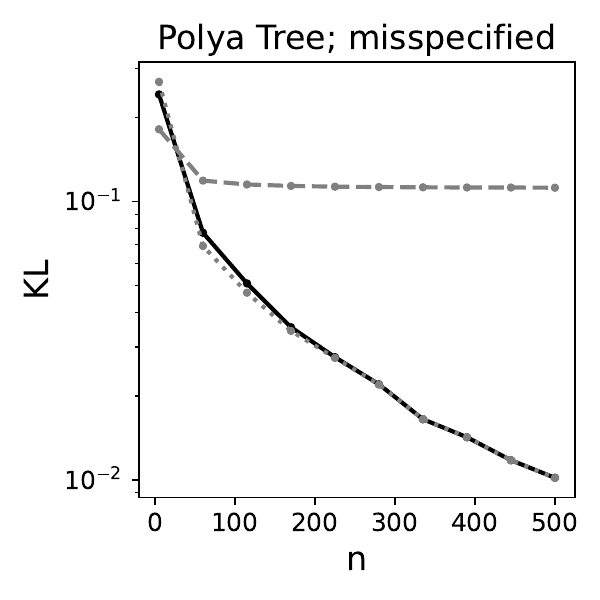}
    \end{subfigure}

    \vspace{-5pt}

    \begin{subfigure}[t]{0.42\columnwidth}
        \centering
        \caption{}\label{fig:mmd_spec_error}
        \includegraphics[width=\linewidth]{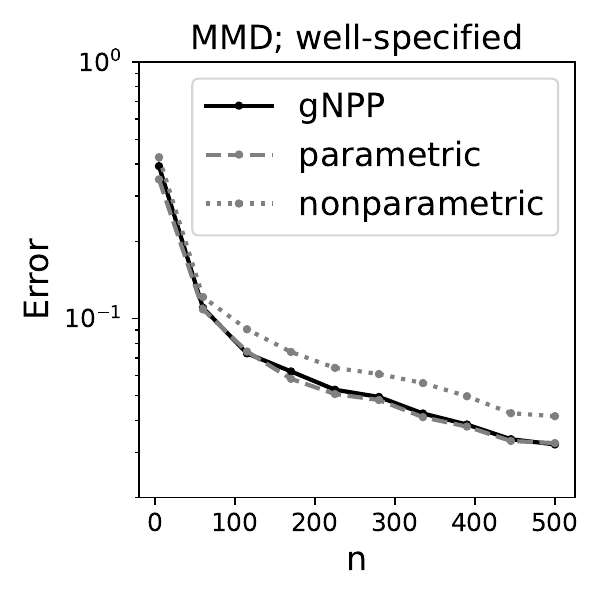}
    \end{subfigure}\hfill
    \begin{subfigure}[t]{0.42\columnwidth}
        \centering
        \caption{}\label{fig:mmd_miss_error}
        \includegraphics[width=\linewidth]{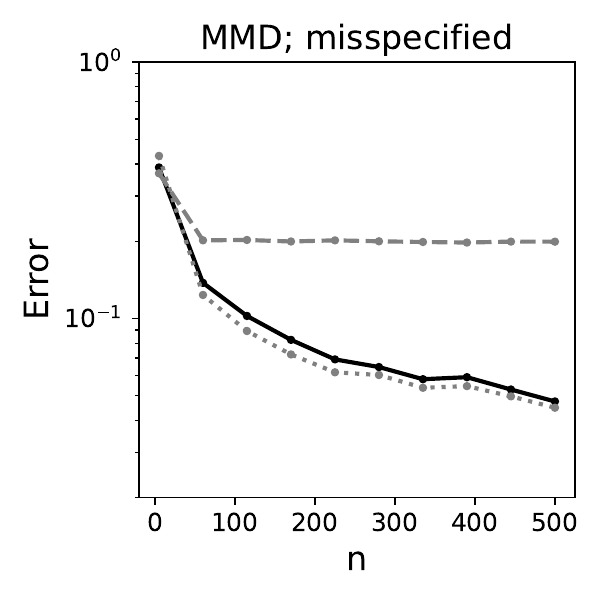}
    \end{subfigure}

    \vspace{-5pt}

    \begin{subfigure}[t]{0.42\columnwidth}
        \centering
        \caption{}\label{fig:mmd_spec_coverage}
        \includegraphics[width=\linewidth]{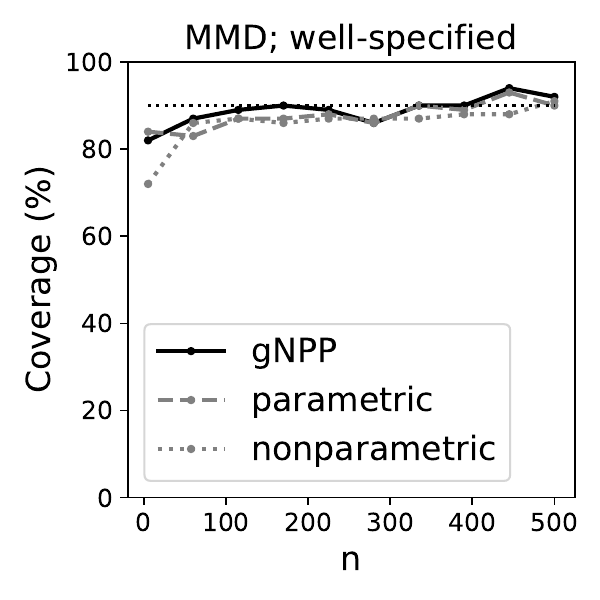}
    \end{subfigure}\hfill
    \begin{subfigure}[t]{0.42\columnwidth}
        \centering
        \caption{}\label{fig:mmd_miss_coverage}
        \includegraphics[width=\linewidth]{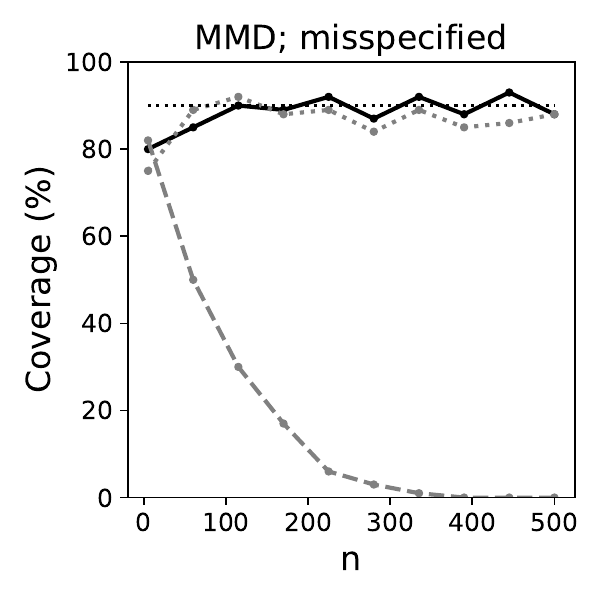}
    \end{subfigure}
            \vspace{-12pt}
    \caption{\textbf{Synthetic data results.} (a,b) KL divergence between the true data density and the posterior predictive of each model. (c,d) Absolute difference between the posterior mean estimate of the median and the true median, for each model, using the MMD gNPP. (e,f) Calibration of the MMD gNPP. We plot how often, across independent simulations, the posterior credible interval covers the true median. The nominal coverage is 90\% (dashed).}
\end{figure}

We illustrate the behavior of NPP models and the gNPP approximation in a synthetic data setting. We follow the basic setup of \citet{Lyddon2018}.
Consider a Gaussian parametric model,
\begin{equation}
\theta \sim \mathcal{N}(0, 1), \quad x_{1:n} \overset{iid}{\sim} \mathcal{N}(\theta, 1).
\end{equation}
We study inference in the well-specified setting where the true distribution $\p_0$ of $x_{1:n}$ is $\mathcal{N}(0, 1)$, and in the misspecified setting, where $\p_0$ is a \textit{skew} normal with the same mean and variance but skew parameter 10.
Details on the simulations and models are in \Cref{sec:synth_apx}, and code for these experiments is at \code.

\subsection{Polya Tree NPP}

We first study an NPP model that perturbs the parametric model with a Polya tree. The advantage of Polya trees is that they admit closed form marginal likelihoods,
making an accurate approximation of the Bayes factor in the NPP posterior possible. This provides a careful check of our theoretical results.
The Polya tree NPP model is,
\begin{equation} \label{pt_npp}
\begin{aligned}
    \theta &\sim \mathcal{N}(0, 1) \quad b \sim \text{Bernoulli}(\eta = 0.1),\\
    &\begin{cases}
\mathrm{p} = \mathcal{N}(\theta, 1) & \text{if } b = 1 \\
\p \sim \mathrm{PolyaTree}(\mathcal{N}(\theta, 1), h) \quad h \sim \mathrm{Exponential}(1) & \text{if } b = 0
    \end{cases}, \\
x_{1:n} &\overset{iid}{\sim} \p.
\end{aligned}
\end{equation}
Here, we parameterize the Polya tree as in \citet{Berger2001}; it is a distribution over distributions, with mean $\mathcal{N}(\theta, 1)$ and scale parameter $h$.
We draw the scale parameter $h$ from an exponential prior.
We compute $\Pi_{\pert}(x_{1:n})$ using the procedure in \citet{Berger2001}.
% The marginal distribution over $\p$, given $\theta$ and $h$, can be computed in closed form, and we integrate over $\theta$ and $h$ using the importance sampling procedure of \citet{Berger2001} to obtain the Bayes factor (\Cref{Bayes-factor}).

We first examine convergence of the posterior distribution to the truth, measuring the KL divergence between the true distribution and the posterior predictive of the NPP model, averaged over 100 independent datasets.
We compare to the performance of a parametric model (\Cref{pt_npp} given $b = 1$) and to the performance of a nonparametric model (\Cref{pt_npp} given $b = 0$).

\Cref{fig:bf_pt} shows that the NPP prefers the parametric model when the latter is well-specified, while it prefers the nonparametric model in the misspecified case. 
With small amounts of data ($n=5$), the Bayes factor prefers the parametric model even if it is misspecified, and even though the prior $\eta=0.1$ prefers the nonparametric model. This is an example of the Bartlett-Lindley effect: the parametric model is preferred even when it is misspecified since it provides a better approximation given the amount of available data.

The NPP model is efficient: in the well-specified case, it matches the faster convergence of the parametric model (\Cref{fig:pt_spec_kl}).
The NPP model is also robust: in the misspecified case, it converges to the true distribution, even though the parametric model does not (\Cref{fig:pt_miss_kl}).
Note that at very small amounts of data ($n = 5$), the parametric model provides better estimates than the nonparametric model despite its misspecification, as it offers a useful approximation of the underlying distribution.
The NPP inherits some of these benefits, outperforming the nonparametric model at $n = 5$.

\subsection{Generalized Bayes Approximation to the NPP} 

We next study the generalized Bayes approximation to the posterior (gNPP). To design our approximation, we must first choose a functional of interest.
Following \citet{Lyddon2018}, we focus on the median, $\psi(\p) = \argmin_\alpha \mathbb{E}_{\p}[|X - \alpha|]$.
Note this is a nonlinear functional of $\p$.
The parametric model assumes the mean and median of the distribution are both $\theta$, but in the misspecified case, when the true data distribution $\p_0$ is skew normal, the median is about $-0.2$.
For the nonparametric component, we use a Bayesian bootstrap, a Dirichlet process with concentration 0~\citep{Rubin1981}.
This offers consistent estimation of the target functional, and it is straightforward to draw samples from the posterior.
For the generalized Bayes factor, we consider the 2-Wasserstein divergence, the MMD with an inverse multiquadric (IMQ) kernel, and the KSD with the same kernel;
the IMQ, $k(x, x') = (c^2 + \|x - x'\|^2)^{-1/2}$, is a characteristic kernel that is well-suited for KSD-based inference \citep{Gorham2017-sd}.
We compare the gNPP to two baselines: (a) the parametric model alone, and (b) the Bayesian bootstrap, a nonparametric model that places no prior weight on the parametric family. 
%We compare to the parametric model and to a nonparametric model, the Bayesian bootstrap on its own.

We quantify the performance of models' point estimates in terms of the absolute difference between the posterior mean of the median and the true median, $|\mathbb{E}[\psi(\p) \mid x_{1:n}] - \psi(\p_0)|$.
The behavior of the gNPP follows that of the NPP.
In the well-specified case, the gNPP matches the fast convergence of the parametric model (\Cref{fig:mmd_spec_error,fig:point_est_synth}).
In the misspecified case, it converges to the true median (\Cref{fig:mmd_miss_error,fig:point_est_synth}).
%In the misspecified case, it also slightly outperforms the nonparametric model at small amounts of data, when using the MMD and KSD.
%The gNPP achieves adaptation thanks to the generalized Bayes factor (\Cref{fig:bf}), which matches the asymptotic behavior of the Bayes factor.
At small $n$ we see a Bartlett-Lindley effect in the MMD and KSD generalized Bayes factors, as they prefer the misspecified but simple parametric model at $n=5$, despite the prior $\eta=0.1$ (\Cref{fig:bf_mmd,fig:bf_ksd}). This explains their performance improvement over the nonparametric model at low $n$. 
Note the effect is not theoretically guaranteed to always occur, and in this case is absent from the Wasserstein gNPP (\Cref{fig:bf_wass}).

We also examined uncertainty quantification. We checked whether posterior credible sets are calibrated and achieve frequentist coverage. We computed how often, across independent simulations, the true median fell in the posterior 95\% credible interval; with correct calibration, it should fall in the interval 95\% of time.  
In the well-specified case, the gNPP is well calibrated, as is the parametric and nonparametric model (\Cref{fig:mmd_spec_coverage,fig:coverage_synth}).
In the misspecified case, the parametric model is miscalibrated, but the gNPP inherits the good calibration of the nonparametric model (\Cref{fig:mmd_miss_coverage,fig:coverage_synth}). 

A key hyperparameter in the gNPP is the rate $r$.
We set $r=0.49$ for the MMD and KSD versions, based on the intuition that $r$ should be as large as possible to enable efficient convergence in the misspecified case, while still below $1/2$, the parametric rate, to preserve convergence in the well-specified case (\Cref{remark:set_r}).
While this default worked well for the MMD and KSD versions, in the Wasserstein version it led to a gBF that failed to favor the parametric model in the well specified case even with $n = 500$ datapoints, and even though the gBF slowly trended towards larger value (\Cref{fig:wass_rate_comparison}).
Instead, setting $r=0.1$ led to improved performance by effectively discriminating between the well-specified and misspecified cases at lower values of $n$. %(\Cref{fig:bf_wass}).
So although $r = 0.49$ may be a reasonable default, problem-specific tuning can improve performance.

\section{Theory}\label{sect-theory}

We establish the frequentist properties of NPP models and the gNPP approximation. In particular, we prove they are 
(a) robust, in the sense that the posterior converges to the true data distribution, and
(b) efficient, in the sense that the posterior converges at a parametric rate when the underlying parametric model is well-specified.
We assume a population distribution $\p_0$ over $\X$ that produces i.i.d. observations $x_{1:n}$. In this section we assume, for simplicity of exposition, that $\X$ is an open subset of $\R^\kappa$, where the Lebesgue densities are well-defined. But note that this is not a requirement for implementing the NPP or the gNPP, with the Wasserstein or MMD, in practice. Finally, we assume a unique parameter $\theta_0$ in an open set $\Theta \subseteq \R^d$ that minimizes the KL divergence to the population distribution, $\KL{\p_0}{\p_\theta}$.
Detailed statements and proofs of each result are in \Cref{sect-proof-theory}.

\subsection{NPP models} \label{sect-theory-npp}
\paragraph{Model selection consistency.} We first analyze the mixing weight $\eta_n$ in the NPP. \Cref{thm:lambda-consistency} establishes general conditions for its model selection consistency: $\eta_n$ asymptotically selects the parametric model when that model is correct, and the nonparametric model otherwise. 
The assumptions (\Cref{assumption:prior,assumption:entropy,assumption:rate-diff} in \Cref{sect-BNP}) follow the standard ``prior mass and testing’’ framework for establishing posterior contraction rates in Bayesian nonparametrics \citep{Ghosal2001}. Informally, the entropy condition (\Cref{assumption:entropy}) characterizes the complexity of local Hellinger neighborhoods in the parametric and nonparametric model classes and thereby defines the corresponding contraction rates. We further need the NPP prior to put sufficient mass in a reverse KL ball around $\p_0$ (\Cref{assumption:prior}), and, when the parametric model is well-specified, the nonparametric prior should put much less mass around the truth than the parametric model (\Cref{assumption:rate-diff}).  The result then follows directly from \citet[Corollary 3.1]{Ghosal2008-bf}.
%The subsequent examples show that using Dirichlet process normal mixture as the nonparametric perturbation ensures model selection consistency, while a standard Dirichlet process fails. Finally, \Cref{thm-NPP} proves that NPP models achieve efficiency and robustness.
\begin{proposition}[$\eta_n$ is consistent for model selection] \label{thm:lambda-consistency}
%The mapping $\theta \mapsto \KL{\p_0}{\p_\theta}$ is injective and continuous
Assume the marginal density $\p_{\param}(x_{1:n}) := \int_\Theta \p_\theta(x_{1:n}) d \Pi_{\param}(\theta)$ is well-defined. Under \Cref{assumption:prior,assumption:entropy,assumption:rate-diff} given in \Cref{sect-BNP}, 
\begin{enumerate}
\item $\eta_n \to 1$ a.s. $[\P_0^\infty]$, if $\p_0 \in \cM_{\param}$. 
\item  $\eta_n \to 0$ a.s. $[\P_0^\infty]$, if $\p_0 \notin \cM_{\param}$. 
\end{enumerate} 
\end{proposition}

We check if the assumptions of \Cref{thm:lambda-consistency} are satisfied for non-parametric perturbations based on a Dirichlet process and on a Dirichlet process mixture model. 
\begin{example}[Dirichlet process perturbations are not consistent] 
\label{examle:DP}
Consider an NPP model (\Cref{npp}) with a Dirichlet process perturbation $\p \sim \textsc{dp}(\p_\theta, \alpha)$. For this choice of perturbation, the mixing weight fails to satisfy model selection consistency.
The technical assumption that fails is \Cref{assumption:prior} which requires the DP prior to put sufficient mass around $\{\p: \KL{\p_0}{\p} < \epsilon^2 \}$ for small $\epsilon$. But, since the DP prior is only supported on discrete measures, it puts zero mass on this reverse KL ball for any non-discrete $\p_0$.  The full result is provided in \Cref{prop:BF-DP} in \Cref{sect-proof-theory-npp} with a proof. 
%The technical assumption that causes this failure of consistency is \Cref{assumption:rate-diff} which works for a typically nonparametric contraction rate but breaks at the $n^{-1/2}$-contraction rate of the Dirichlet process posterior \cite[Example 8.5]{Ghosal2017}. The intuitive reason is that, despite the inherent misspecification for continuous $\p_\theta$, the Dirichlet process behaves like a parametric model with a contraction rate of $n^{-1/2}$. Thus, there is insufficient power to distinguish between the Dirichlet process model and a parametric model by considering the marginal probability alone.
\end{example}
\noindent One implication of this result is that the robust Bayes approach studied by \citet{Lyddon2018}, in which a parametric model is perturbed by a Dirichlet process, is robust but not necessarily efficient.

\begin{example}[Dirichlet process normal mixture perturbations are consistent]\label{ex:BF-DPNM}
Consider a nonparametric perturbation based on a Dirichlet process normal mixture, as described in \Cref{npp,eqn:dpmm_npp}. Unlike the Dirichlet process, this perturbation describes continuous densities over $\cX$. 
%In this case, the mixture weight $\eta_n$ takes the form: 
%\begin{equation}
%\eta_n = \frac{\eta \int \prod_{i = 1}^n \p_\theta(x_i) \, d\Pi_{\param}(\theta)}{(1 - \eta) \int \int \prod_{i = 1}^n \left(\int \phi\left(x_i - z; \tau I_d\right) \, d G(z)\right) d \Pi_{\pert} (G \mid \theta; \alpha) \, d \Pi_{\param}(\theta)}.
%\end{equation}
When $\p_0$ is smooth, \Cref{prop:BF-DPNM} implies that the Dirichlet process normal mixtures satisfy the assumptions of \Cref{thm:lambda-consistency}, implying $\eta_n$ is model selection consistent. 
\end{example}

\paragraph{Robustness and Efficiency. }  We now establish robustness and efficiency for the NPP posterior. Recall the target of inference is defined via a functional $\psi: \cP(\X) \mapsto \R^s, s < \infty$.
To establish efficiency, we prove a Bernstein-von Mises (BvM) theorem: when $\p_0 \in \mathcal{M}_\param$, the posterior over the functional $\psi(\p)$ is asymptotically normal, with a standard deviation proportional to $1/\sqrt{n}$. This result is closely related to semiparametric BvM theorems in the literature, which characterize the asymptotic normality of low-dimensional functionals in nonparametric Bayesian models \citep{Bickel2012, Rivoirard2012, Castillo2015}. To establish robustness, we show that even when $\p_0 \notin \mathcal{M}_\param$, the posterior concentrates at $\psi(\p_0)$.

The NPP inherits its asymptotic normality from the underlying parametric model. The standard Bernstein-von Mises theorem says the parametric posterior $\Pi_{\param}(d \theta \mid x_{1:n})$ is asymptotically normal, so under smoothness conditions, the posterior over $\psi(\p_\theta)$ will be asymptotically normal as well. 
\begin{assumption}[The parametric model is asymptotically normal] \label{assumption-AN}
There exists $\theta_0 \in \Theta$ and a positive definite matrix $V_{\theta_0}$ such that $n \to \infty$, 
\begin{enumerate}
    \item[(a)]  $\sqrt{n}(\hat \theta_{\mle} -\theta_0) \wto \cN \left(0, V_{\theta_0}^{-1}\right)$, where $\wto$ denotes weak convergence and $\hat \theta_{\mle}$ is the maximum likelihood estimate of $\theta$, and
    \item[(b)]  $d_{TV}\left(\sqrt{n} \left( \theta - \hat \theta_{\mle} \right), \cN \left(0, V_{\theta_0}^{-1}\right) \right) \overset{\P_0}{\to} 0$ for $\theta \sim  \Pi_{\param}\left(\cdot \mid x_{1:n} \right)$, with convergence in first and second moments in \( [\P_0] \)-probability. 
\end{enumerate}
\end{assumption}
\noindent Conditions (a) and (b) assume the parametric MLE and posterior are asymptotically normal at the $\sqrt{n}$-rate. Condition (b) is the Bernstein-von Mises theorem for regular parametric models, and guarantees the frequentist coverage of posterior credible regions will asymptotically match its nominal level. The convergence in first two moments follows, for example, from convergence of the posterior to a normal distribution in the Wasserstein metric. To streamline the presentation, we write $\dot \chi_\theta := \nabla \chi_\theta$ and $\ddot \chi_\theta := \nabla^2 \chi_\theta$ for a function $\chi_\theta$ of $\theta$.

\begin{assumption}[The target functional is smooth] \label{assumption-chi}
The function $\chi_\theta := \psi(\p_\theta) - \psi(\p_{\hat \theta_{\mle}})$ is twice differentiable, where $ \dot \chi_{\theta_0}^\top V_{\theta_0}^{-1} \dot \chi_{\theta_0}$ is positive definite, $\ddot \chi_\theta$ is continuous at $\theta_0$ and $\|\ddot \chi_{\theta_0}\|_2 < \infty$ in \( [\P_0] \)-probability.
\end{assumption}

We now establish our main result. To describe convergence of the NPP posterior to a normal distribution, we use the bounded Lipschitz distance $d_{BL}$, which metrizes the topology of weak convergence (\Cref{def:BL-dist} and \Cref{thm:BL-weak-top}). 
\begin{theorem}[NPP models are efficient and robust] \label{thm-NPP}
Let \Cref{thm:lambda-consistency} and \Cref{assumption-AN,assumption-chi} hold. Let $\Pi\left(\tilde \psi_n(\p) \mid x_{1:n}\right)$ denote the pushforward measure of the NPP posterior through $\tilde \psi_n(\p) := \sqrt{n}(\psi(\p) - \psi(\p_{\hat \theta_{\mle}}))$. If $\p_0 \in \cM_{\param}$, then
\begin{equation}
d_{BL} \left(\Pi\left(\tilde \psi_n(\p) \mid x_{1:n} \right),\cN \left(0, \dot \chi_{\theta_0}^\top V_{\theta_0}^{-1} \dot \chi_{\theta_0} \right) \right) \overset{\P_0}{\to} 0 \quad \text{(Efficiency)}.
\end{equation}
Further assume the nonparametric perturbation $\Pi_{\pert}(\psi(\p) \mid x_{1:n})$ is consistent at $\psi(\p_0)$. Then if $\p_0 \notin \cM_{\param}$, 
\begin{equation}
d_{BL} \left(\Pi\left(\psi(\p) \mid x_{1:n}\right), \delta_{\psi(\p_0)}\right) \overset{\P_0}{\to} 0 \quad \text{(Robustness)}.
\end{equation}
Hence the NPP posterior $\Pi\left(\psi(\p) \mid x_{1:n}\right)$ is also consistent at $\psi(\p_0)$. 
\end{theorem}
\noindent When the parametric model is correctly specified, \Cref{thm-NPP} tells us that the NPP posterior of the target estimand achieves the $\sqrt{n}$-rate and the optimal variance in the asymptotic minimax sense \citep{VdV2000}.  When the parametric model is misspecified, the NPP posterior is still robust as it converges to the true $\psi(\p_0)$.
Sufficient conditions for the consistency of $\Pi_{\pert}\left(\psi(\p) \mid x_{1:n}\right)$ are (a) consistency of the nonparametric posterior for all $\p_\theta$, e.g. Dirichlet process normal mixture models, and (b) continuity of the functional $\psi$ with respect to the weak topology.

The efficiency guarantee could be extended to the misspecified setting if the nonparametric posterior $\Pi_{\pert}(\psi(\rmp)\mid x_{1:n})$ achieves semiparametric efficiency for the target functional $\psi$. This could follow from a semiparametric Bernstein–von Mises theorem for the posterior of $\psi(\rmp)$ around a sequence of asymptotically efficient regular estimators $\hat\psi_n$ in the sense of the convolution theorem \citep[Theorem~25.20]{VdV2000}. Specifically, we require that, after centering at $\hat\psi_n$ and scaling by $\sqrt{n}$, the conditional distribution of $\psi(\rmp)$ under $\Pi_{\pert}(\cdot\mid x_{1:n})$ is asymptotically Gaussian with variance equal to the semiparametric efficiency bound. Given such a result, the efficiency statement for NPP would then follow by combining: (i) the semiparametric BvM for $\Pi_{\pert}(\psi(\rmp)\mid x_{1:n})$, and (ii) consistency of the (generalized) Bayes factor, which ensures that the posterior concentrates on the appropriate limiting model. An example of an efficient nonparametric posterior is the one-step corrected posterior of \cite{Yiu2023Semiparametric}, which achieves semiparametric efficiency under its stated assumptions.

%This is satisfied, for example, by the one-step corrected posterior \citep{Yiu2023Semiparametric}. 

%Sufficient conditions for the consistency of $\Pi_{\pert}\left(\psi(\p) \mid x_{1:n}\right)$ are (1) consistency of the nonparametric posterior for all hyperparameters $h > 0$ (e.g., log-spline models or Dirichlet process normal mixture models for smooth $\p_0$) and (2) the continuity of the functional $\psi$ with respect to the weak topology.

\subsection{gNPP approximation} \label{sect-theory-gnpp}

We establish the efficiency and robustness of the gNPP approximation, proving an analogue of \Cref{thm-NPP} for our generalized Bayes inference approach (\Cref{thm-gNPP}).
To do so, we first show that the generalized Bayes factor satisfies model selection consistency (\Cref{contraction-div}). 
Our results apply to a broad class of divergences, but we verify the conditions and provide explicit contraction rates for the Wasserstein, MMD and KSD (\Cref{thm-Wp,thm-mmd,thm-ksd} in \Cref{apx:empirical_divergences}).
The contraction rates justify our guidance for setting the rate hyperparameter~$r$.

\paragraph{Model selection consistency.} 
%To develop a general theory for gNPPs, we present our results using an abstract divergence, $\diverge$. We then apply these results to the cases of Wasserstein, MMD, and KSD under divergence-specific assumptions.
We first analyze the model selection consistency of the mixing weight $\hat \eta_n$ in the gNPP. Recall $\hat{\eta}_n := 1/(1 + \gBF_n^{-1})$, where $\gBF_n$ depends on a posterior expected empirical divergence $\E{\param}{\diverge_{m,n}(\p_\theta, \p_0) \mid x_{1:n}}$ (\Cref{gBF}). 
We are interested in the asymptotic behavior of this posterior divergence, showing it correctly detects model misspecification.

To begin, we study general divergences $\diverge$ that are semimetrics or depend on one, in the sense that $\diverge(\p, \q) = \rho^k(\p, \q)$ for $k \in \N$ and $\rho$ a semimetric.
The Wasserstein and MMD fit this form; the KSD is asymmetric and hence does not, so we later extend our arguments.
We also assume in the general theory that the estimated divergence $\diverge_{m,n}(\p, \q)$ takes the form of a \textit{plug-in} estimator $\diverge_{m,n}(\p, \q) = \diverge\left( \p^m, \q^n \right)$,
where $\p^m$ is the $m$-sample empirical distribution of $\p$, and $\q^n$ is the $n$-sample empirical distribution of $\q$. 
The empirical $p$-Wasserstein distance is directly given by this plug-in estimator, while the plug-in form of the MMD is its V-statistic \citep{Gretton2012-do}. We later extend our arguments to the MMD U-statistic, which matches the plug-in form with a vanishing error, $O(n^{-1} + m^{-1})$ \citep{Gretton2012-do,Serfling2009}.
%The empirical $p$-Wasserstein distance is directly given by this plug-in estimator, while the U-statistic approximations for the MMD and KSD match the plug-in form (the V-statistic) with errors of $O(n^{-1} + m^{-1})$ and $O(n^{-1})$, respectively.

%We may also use one sample of size $n$ to approximate $\diverge$, then we denote $\diverge_n$ and $\E{\param}{\diverge_{n}(\p_\theta, \p_0) \mid x_{1:n}}$.  
We make regularity assumptions on the divergence.
\begin{assumption}[The divergence is well-behaved] \label{assumption-semimetric}
Assume $\diverge(\p, \q) = \rho^k(\p, \q)$ where $k \in \N$ and $\rho$ is a semimetric, and that
\begin{enumerate}
\item[(a)] $\rho$ is continuous in the weak topology and $\sup_{\theta \in \Theta}\rho(\p_\theta,\p_0) < \infty$.
\item[(b)] There exists an $M_n n^{-1/2}$ neighborhood of $\theta_0$ where $M_n \to \infty$ such that the mapping $\theta \mapsto  \rho(\p_\theta, \p_{\theta_0})$ is twice differentiable and $\left\| \nabla_\theta^2 \rho(\p_\theta, \p_{\theta_0}) \right\|_2$ is uniformly bounded.
\end{enumerate}
\end{assumption}
%The Wasserstein distance and MMD satisfy the semimetric assumption while KSD fails due to asymmetry in the arguments, thus we will provide a slightly different set for KSD. 
\noindent \Cref{assumption-semimetric}(a) requires the divergence to be uniformly bounded. For example, with the Wasserstein distance, the condition is satisfied if $\X$ is compact. For MMD, the condition is satisfied if the kernel is uniformly bounded.
\Cref{assumption-semimetric}(b) requires local Lipschitz smoothness of the function $\theta \mapsto \rho(\p_\theta, \p_{\theta_0})$ within an asymptotically vanishing ball around $\theta_0$, where the radius is at least $n^{-\alpha/2}$, with $\alpha < 1$.  This condition is typically satisfied for commonly used discrepancies under standard smoothness assumptions on the parametric family, such as those used to establish central limit theorems for minimum $\rho$-discrepancy estimators \citep[e.g.][]{Barp2019MMD}. The precise meaning of a “smooth parametric model” depends on the choice of $\rho$.  For example, if $\rho$ is the MMD induced by a translation-invariant kernel $k$, a natural notion of smoothness is to parametrize $\P_\theta$ as a pushforward $\P_\theta=T_{\theta,\#}U$ for a reference measure $U$. In this setting, by following the same argument of \citet[Theorem 2.]{Briol2019}, it suffices for \Cref{assumption-semimetric}(b) to hold if (a) $k$ has bounded mixed derivatives up to order $2$; (b) there exists a neighborhood $O_n$ of $\theta_0$ of radius $M_n n^{-1/2}$ such that $T_\theta$ is twice continuously differentiable in $\theta$ for all $\theta\in O_n$; and (c) the derivatives of $T_\theta$ satisfy suitable integrability/dominance conditions, e.g., $\int \sup_{\theta\in O_n}\|\nabla_\theta^i T_\theta(u)\|_2\,d U(u) < \infty$ for $i\in\{1,2\}$. Similar conditions appear for the KSD in \cite[Theorem 4 and 5]{Barp2019MMD} and for the Wasserstein in \cite[Theorem 2.3.]{bernton2019parameter}. 
% if $\rho(\rmp_\theta, \rmp_{\theta_0}) = \KL{\rmp_{\theta_0}}{\rmp_\theta}$, then a sufficient condition for Assumption~3(b) is that $\theta\mapsto \log \rmp_\theta(x)$ is twice continuously differentiable on $\{\|\theta-\theta_0\|_2\le M_n n^{-1/2}\}$ for $\rmp_{\theta_0}$-a.e.\ $x$, and that the Hessian is uniformly bounded in the neighborhood,  $\sup_{\|\theta-\theta_0\|_2\le M_n n^{-1/2}}\;\EE{\rmp_{\theta_0}}{\|\nabla_\theta^2 \log \rmp_\theta(X)\|_2}<\infty$. If

We also assume that the empirical divergence converges to the true divergence, at least when the model likelihood is in a neighborhood of the truth.
\begin{assumption}[The empirical divergence converges] \label{assumption-rate}
There exists a sequence $r_{m,n} \to 0$ such that
\begin{enumerate}
\item[(a)] For any sequence $\theta_n \to \theta_0$, as $m, n \to \infty$ with $m/(m + n) \to c \in (0,1)$, we have $\E{}{\diverge_{m,n}\left(\p_{\theta_n}, \p_0 \right)} = \diverge\left(\p_{\theta_0}, \p_0 \right) + O(r_{m,n})$. The expectation is over $m$ samples from $\p_{\theta_n}$ and $n$ samples from $\p_0$.
\item[(b)] There exists a function $\cV(\theta)$ that is finite at $\theta_0$ and continuous at $\theta_0$ such that for all $(m,n)$ large enough, $\var \left(\diverge_{m,n} \left(\p_\theta, \p_0 \right) \right) \leq r_{m,n}^2 \cV(\theta)$, where $\var$ denotes the variance.
\end{enumerate}
\end{assumption}
\noindent \Cref{assumption-rate} requires an asymptotic bound on the expectation of $\diverge_{m,n}\left(\p_\theta, \p_0\right)$ and a finite-sample variance bound around $\theta_0$, both converging at a rate $r_{m,n}$. These assumptions are local, applying only to a neighborhood of $\theta_0$. They are also weaker than assuming uniform convergence of the mean or the variance of $\diverge_{m,n}\left(\p_\theta, \p_0\right)$ in a neighborhood around $\theta_0$.

When we specialize \Cref{assumption-rate} to specific divergences, we can verify the assumption more closely. 
For example, for the MMD, the plug-in V-statistic matches the U-statistic up to an ignorable error, and for the U-statistic, \Cref{assumption-rate}(a) is satisfied, since the statistic is unbiased, and \Cref{assumption-rate}(b) follows from the variance formula of two-sample U-statistics and kernel regularity conditions.

We now study the posterior expected value of the empirical divergence, under the parametric model. To make the analysis more tractable theoretically, we assume the samples used to approximate $\p_0$ in $\diverge_{m,n}(\p_\theta, \p_0)$ are independent of those used to compute the posterior. This can be achieved by splitting data and cross-fitting~\citep{Chernozhukov2018}. However, we expect the model selection consistency of $\hat \eta_n$ (\Cref{thm:gbf}) to remain valid even without sample splitting, and in the empirical studies we do not apply it.
\begin{theorem}[The posterior expected empirical divergence converges at a rate $r_{m,n} \lor n^{-1}$]
\label{contraction-div}
Let \Cref{assumption-AN,assumption-semimetric,assumption-rate} be satisfied. As $m,n \to \infty$ with $n/(n+m)\to c \in (0, 1)$, $   \E{\param}{\diverge_{m,n}(\p_\theta, \p_0) \mid x_{1:n}} $ converges in $[\P_0^\infty]$-probability to $\diverge(\p_{\theta_0}, \p_0)$ at the rate of $r_{m,n} \lor n^{-1} := \max(r_{m,n}, n^{-1})$: 
\begin{equation}
   \E{\param}{\diverge_{m,n}(\p_\theta, \p_0) \mid x_{1:n}} = \diverge(\p_{\theta_0}, \p_0) + O_{\P_0}(r_{m,n} \lor n^{-1}).
\end{equation}
\end{theorem}
\noindent So the posterior empirical divergence converges to the true minimal divergence between the model class $\mathcal{M}_\param$ and $\p_0$, at a rate that is the slower of (a) $r_{m,n}$, the convergence rate of $\diverge_{m,n}$, and (b) $n^{-1}$. 

\Cref{contraction-div} applies to general divergences. For the Wasserstein, MMD, and KSD in particular, we derive the following contraction rates (detailed theorems and conditions are in \Cref{apx:empirical_divergences}).
\begin{enumerate}
\item \textbf{$p$-Wasserstein Distance.} The empirical Wasserstein divergence $\E{\param}{\W_p^p(\p_\theta^m, \p_0^n) \mid x_{1:n}}$ converges to $\W_p^p \left(\p_{\theta_0}, \p_0\right)$ at a rate of $O_{\P_0}\left(n^{-2/(\kappa \lor 4)} + m^{-2/(\kappa \lor 4)} \right)$, where $\kappa$ is the dimension of $\X$ (\Cref{sect-Wasserstein}).

\item \textbf{MMD.} The empirical MMD $\E{\param}{\mmd_U^2 \left(\p_\theta^m, \p_0^n \right) \mid x_{1:n}}$, based on the U-statistic for the MMD, converges to $\mmd^2 \left(\p_{\theta_0}, \p_0\right)$ at a rate of $O_{\P_0}\left(n^{-1/2} + m^{-1/2}\right)$ (\Cref{sect-mmd}).

\item \textbf{KSD.} The empirical KSD $\E{\param}{\ksd_U(\p_0^n, \p_\theta) \mid x_{1:n}}$, based on the U-statistic for the KSD, converges to $\ksd(\p_0, \p_{\theta_0})$ at a rate of $O_{\P_0}(n^{-1/2})$ (\Cref{sect-KSD}).
\end{enumerate}
For simplicity, in our implementation, we set $m = n$ for the two-sample empirical divergences (Wasserstein and MMD), so the rate $r_{m,n}$ simplifies to $r_n := r_{n,n}$.  
In this case, for the Wasserstein, we can take $r_n = n^{-2/(\kappa \lor 4)}$, and for the MMD and KSD, we can take $r_n = n^{-1/2}$. 

We have shown that the posterior expected empirical divergence converges to the true divergence at a known rate, $r_{n}$.
We can now set the rate hyperparameter $r$ in the generalized Bayes factor to be slower than this known rate, i.e. $r \in \left(0,\frac{2}{\kappa \lor 4} \right)$ for the Wasserstein and $r \in (0, 1/2)$ for MMD and KSD. Then, the generalized Bayes factor is model selection consistent.
\begin{theorem}[$\hat \eta_n$ is consistent for model selection] \label{thm:gbf}
Let $\E{\param}{\diverge_n(\p_\theta, \p_0)}$ be bounded in $[\P_0^\infty]$-probability, $\eta > 0$ and assume $\E{\param}{\diverge_n(\p_\theta, \p_0) \mid x_{1:n}} = \diverge(\p_{\theta_0}, \p_0) + r_n$, where $\diverge(\cdot, \cdot)$ is a statistical divergence. Choose $r > 0$ such that $r_n (n+1)^r = o(1)$. Then, as $n \to \infty$:
\begin{enumerate}
\item $\hat \eta_n \to 1$ a.s. $[\P_0^\infty]$, if $\p_0 \in \cM_{\param}$. 
\item  $\hat \eta_n \to 0$ a.s. $[\P_0^\infty]$, if $\p_0 \notin \cM_{\param}$. 
\end{enumerate}
\end{theorem}
\noindent Note $\E{\param}{\diverge_n(\p_\theta, \p_0)}$ is bounded in probability when, for example, $\diverge_n(\p_\theta, \p_0)$ is uniformly bounded or satisfies a uniform law of large numbers. \Cref{remark-mmd-synthetic} in \Cref{sect-mmd} verifies the conditions of \Cref{thm:gbf} for the model used in the synthetic study of \Cref{sec-synthetic}, with the generalized Bayes factor constructed with the MMD distance. 

\paragraph{Robustness and Efficiency.} The gNPP approximation uses the generalized Bayes factor to adaptively trade off between parametric and nonparametric models. We have shown the generalized Bayes factor is model selection consistent. As a result, the gNPP approximation converges efficiently when the parametric model is well-specified, but still converges to the truth when it is misspecified.
\begin{theorem}[gNPP approximations are efficient and robust]\label{thm-gNPP}
Let \Cref{assumption-AN,assumption-chi,assumption-semimetric,assumption-rate} hold for $m = n$ and $r_n := r_{n,n}$. Assume also that $\E{\param}{\diverge_n(\p_\theta, \p_0)}$ is bounded in $[\P_0]$-probability. Let $r < 1$ be a positive constant such that $r_n (n+1)^r = o(1)$. Let $\hat \Pi\left(\tilde \psi_n(\p) \mid x_{1:n}\right)$ denote the pushforward measure of $\tilde \psi_n(\p) := \sqrt{n}(\psi(\p) - \psi(\p_{\hat \theta_{\mle}}))$ onto the gNPP posterior. If $\p_0 \in \cM_{\param}$, then
\begin{equation}
d_{BL} \left(\hat \Pi\left(\tilde \psi_n(\p) \mid x_{1:n} \right),\cN \left(0, \dot \chi_{\theta_0}^\top V_{\theta_0}^{-1} \dot \chi_{\theta_0} \right) \right) \overset{\P_0}{\to} 0 \quad \text{(Efficiency)}.
\end{equation}
Furthermore, assume that the nonparametric posterior $\hat \Pi_{\pert}(\psi(\p) \mid x_{1:n})$ is consistent at $\psi(\p_0)$. Then if $\p_0 \notin \cM_{\param}$, 
\begin{equation}
d_{BL} \left(\hat \Pi\left(\psi(\p) \mid x_{1:n}\right), \delta_{\psi(\p_0)}\right) \overset{\P_0}{\to} 0 \quad \text{(Robustness)}, 
\end{equation}
hence the gNPP posterior $\Pi\left(\psi(\p) \mid x_{1:n}\right)$ is also consistent at $\psi(\p_0)$. 
\end{theorem}
\noindent We can replace \Cref{assumption-semimetric,assumption-rate} with assumptions specific to the Wasserstein, MMD and KSD, as detailed in \Cref{apx:empirical_divergences}. 
% TODO: could maybe cut the following sentence for space, if necessary
%For Wasserstein, we use \Cref{assumption-semimetric,assumption-wass} to prove the posterior expected divergence converges at a given rate (\Cref{thm-Wp}), implying \Cref{thm-gNPP}. For the MMD, we use \Cref{assumption-semimetric,assumption-mmd,assumption-mmd2} (\Cref{thm-mmd}). For the KSD, we use \Cref{assumption-KSD,assumption-KSD2,assumption-KSD3} (\Cref{thm-ksd}).

\section{Application: Estimating the Effects of Gene Expression}

A cell's behavior depends crucially on its genes' \textit{expression levels}, or the number of RNA transcripts of each gene in the cell.
An important question in cell biology is how the expression level of one gene affects another.
Answering such questions can help scientists understand cells' behavior, and may help develop drugs which modify gene expression levels or activity.

We develop a robust method to estimate the causal effect of one gene's expression level on another's.
To do so, we leverage observational data about the expression level of genes in individual cells, measured with single cell RNA sequencing (scRNAseq)~\citep{kolodziejczyk2015technology}.
We focus on analyzing primary patient data, which is often hard to obtain and hence limited in scale.
Bayesian parametric models offer data-efficient inference, but robustness is crucial given the importance of the estimates to human health.

We specify the causal effect as a functional of the data distribution.
We start from a parametric Bayesian model that incorporates prior information from an auxiliary dataset, a \textit{single cell atlas}.
We robustify this parametric model by creating a gNPP approximation to the NPP posterior, with a BART-based nonparametric perturbation and an MMD-based generalized Bayes factor.

We apply the gNPP model to conduct an exploratory analysis of data from an ovarian cancer patient. We estimate the causal effects of genes that are potential therapeutic targets. Our results reveal that interventions on some genes may have unexpected effects in this patient.

Additional details are in \Cref{sec:empirical_apx}. 
Code and data can be found at \code.

\subsection{Setup}

\paragraph{Estimand.} 
We are interested in the effect of the expression level of a treatment gene on the expression level of an outcome gene.
The challenge for causal inference is that there may be confounders, factors which affect the expression of both the treatment and the outcome genes.
An important source of confounding is \textit{transcription factors}, proteins which directly modify the expression level of many other genes. To correct for this confounding, we adjust for the expression level of the transcription factor genes in each cell.

\begin{figure}[t]
\centering
\begin{subfigure}{0.49\columnwidth}
    \centering
    \includegraphics[width=\columnwidth]{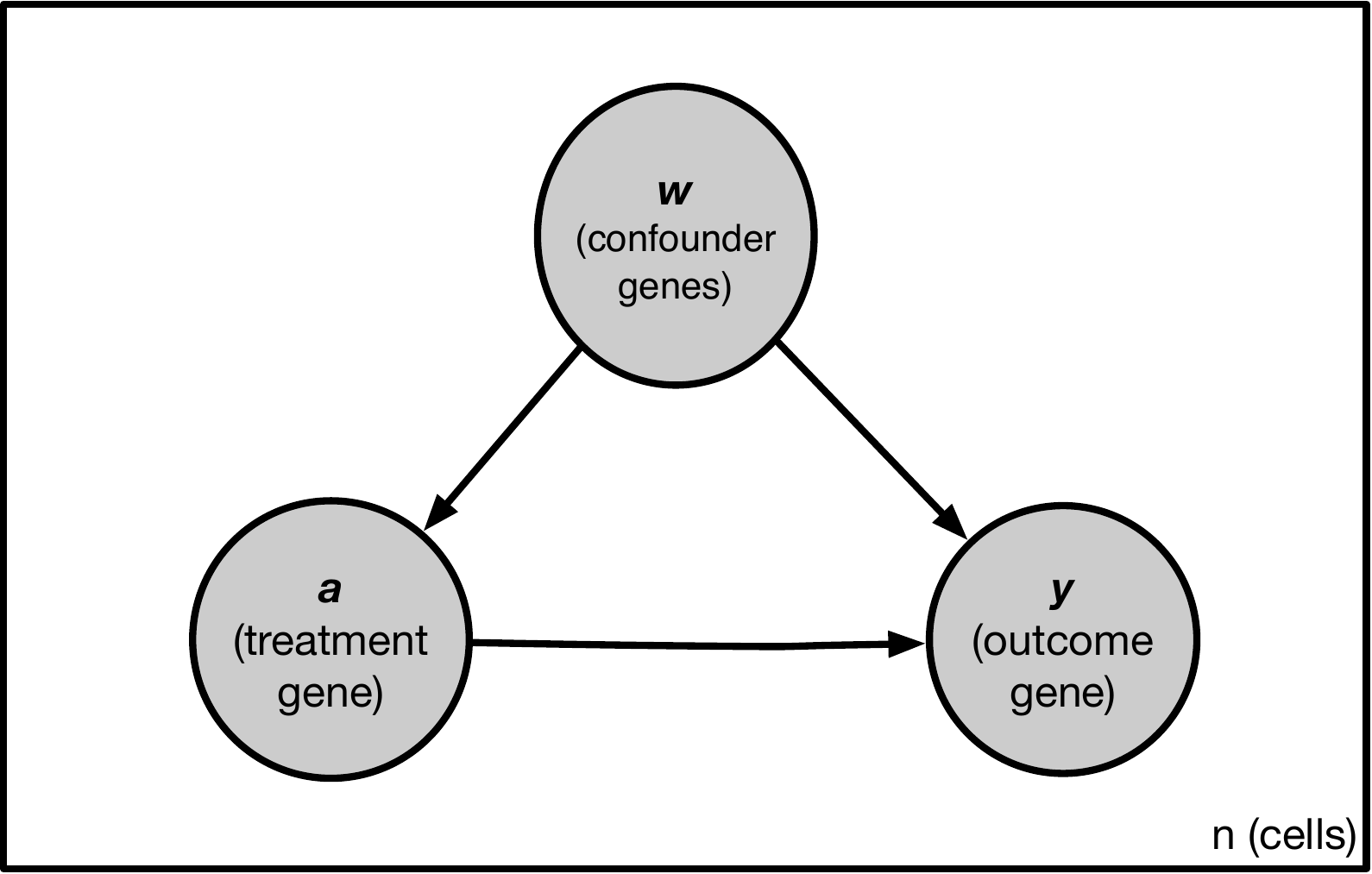}
    \caption{}
    \label{fig:gene-causal-graph1}
\end{subfigure}
\hfill
\begin{subfigure}{0.49\columnwidth}
    \centering
    \includegraphics[width=\columnwidth]{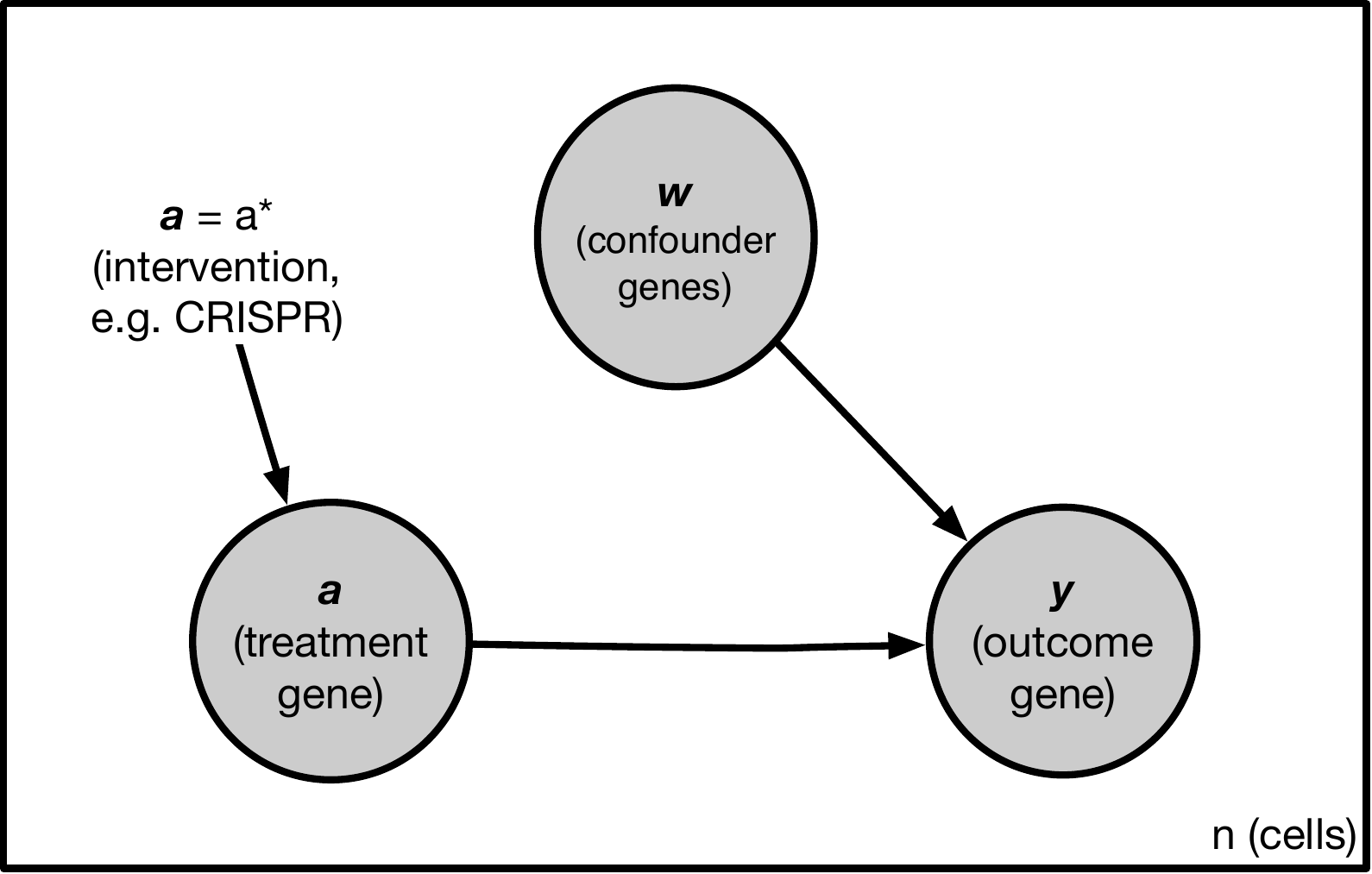}
    \caption{}
    \label{fig:gene-causal-graph2}
\end{subfigure}
\caption{\textbf{Causal model.} To analyze the effects of gene expression, we assume this causal graphical model. (a) The initial model where all variables are observed. (b) The model after intervening on the treatment variable $A$. The goal is to estimate the effect of the treatment gene on the outcome gene, as highlighted in blue.} \label{fig:gene-graph}
\end{figure}
% TODO: variable names should be lower case italic

Formally, we consider the causal graph in \Cref{fig:gene-graph}, where $a$ is the treatment gene expression level, $y$ is the outcome gene expression level, and $w$ is a vector of expression levels for confounding genes, i.e. transcription factors.
We are interested in the effect of an intervention which sets the expression level of $a$ to $a_\star$.
From \Cref{fig:gene-graph}, the distribution of the outcome after the intervention is given by the backdoor adjustment, $\p(y \mid \rmdo(a_\star)) = \int \p(y \mid a_\star, w) \p(w) \di w$ \citep{Pearl2009}.

We compare the average outcome when the treatment gene is highly expressed versus when it is unexpressed. To define ``highly expressed'' we take the 98th quantile of the empirical distribution of treatment expression, denoted $q_{98}(a)$.
Our target functional $\psi(\p)$ is the \textit{average treatment effect},
\begin{equation}
\begin{aligned}
        \psi(\p) = \ATE\left(\p\right) &:= \E{\p}{Y \s \rmdo(a_\star = q_{98}(a))} - \E{\p}{Y \s \rmdo(a_\star = 0)} \\
    &= \int \int y\, \p(y \mid q_{98}(a) , w) \p(w) \di y \di w - \int \int y\, \p(y \mid 0 , w) \p(w) \di y \di w,
\end{aligned}
\end{equation}
where $\p(a, y, w)$ is the joint distribution over treatment, outcome and confounders.

\paragraph{Parametric model} We first introduce a parametric Bayesian model to estimate the causal effect $\psi(\p_0)$. 
This parametric model incorporates information from a single cell atlas dataset, a large collection of scRNAseq data from donors without cancer.

Rather than using the full vector of confounding gene expression $w$, the model uses a low-dimensional representation $z(w)$ derived from the atlas.
Specifically, we set $z(\cdot)$ to be the projection of $w$ onto the first ten principal components of the atlas.  
Such low-dimensional representations of gene expression -- whether derived from principal component analysis or other approaches -- have been found to provide an effective summary of \textit{cell type} and \textit{cell state}~\citep{lopez2018deep}.
By using representations learned from a large atlas, biologists aim to better estimate the cell type and state underlying the noisy expression levels in a small dataset~\citep{hao2021integrated}.
We apply this idea to obtain a low-dimensional representation of confounding.

Our parametric Bayesian model depends linearly on the treatment and on the low-dimensional representation of the confounders,
\begin{equation}
    \p_\theta(y \mid a, w) = \mathcal{N}(c + \tau a + \gamma^\top z(w), \sigma^2),
\end{equation}
where $\theta := [c, \tau, \gamma]$ and $\mathcal{N}(\mu, \sigma^2)$ denotes a normal distribution with mean $\mu$ and variance $\sigma^2$.
The parameter $c$ is an intercept, $\tau$ is the coefficient on the treatment and $\gamma$ is a vector of coefficients on the confounder representation.
We use an improper prior $\pi(\theta) \propto 1$ and specify $\sigma \sim \text{HalfNormal}(0, 1)$. To compute the empirical divergence of the prior, we approximate the improper prior by a uniform distribution over the hypercube with slice $[-100, 100]$. We place a Dirichlet process prior on $\p(a, w)$ with concentration approaching zero, such that the posterior over the ATE is obtained by the Bayesian bootstrap.

We expect the parametric model to learn efficiently even from limited numbers of cells, since the cell representation $z(w)$ is low-dimensional and hence so is the model. 
However, the parametric model may be misspecified, for instance if the ovarian cancer patient has a cells in a pathological state that is not present in the healthy atlas.
In this case, the representation $z(w)$ may project out important variation in $w$, ignoring a source of confounding and leading to unreliable inferences.

\paragraph{gNPP model} 
We introduce a gNPP model for robust inference. For the nonparametric model, we use a Bayesian additive regression tree (BART) with a fixed propensity score correction, an established method for nonparametric Bayesian causal inference~\citep{Hill2011,Hahn2020}.
Since single cell RNA expression data is highly non-normal, we adjust the BART model with a Yeo-Johnson transformation~\citep{Yeo2000}.
We place a Dirichlet process prior on $\p(a, w)$ with concentration zero, such that the posterior over the ATE is obtained by a Bayesian bootstrap.
The model does not incorporate prior information from the atlas.
For the generalized Bayes factor, we use the MMD with rate $r = 0.49$.

\paragraph{Data and preprocessing}
We study an scRNAseq dataset consisting of 544 T cells from a patient with ovarian cancer~\citep{Vazquez2022}. 
The atlas contains 261{,}611 T cells collected across 17 sites/tissues from 12 organ donors without cancer~\citep{dominguez2022cross}.
We log transform and standardize the expression data following standard practice in the field~\citep{heumos2023best}.
As potential confounders, we adjust for 157 transcription factors with highly variable expression.
To find highly variable genes, we use the method of \citet{stuart2019comprehensive} as implemented in scanpy~\citep{wolf2018scanpy}. Briefly, the read counts are z-score normalized (using a regularized standard deviation), then genes are ranked by their empirical variance.

\subsection{Results}

A route to treating cancer is to intervene on a patient's immune system, such that it can better attack the cancer.
One potential strategy is to modify regulatory T cells, which suppress immune attack, to be more like cytotoxic T cells, which conduct immune attack.
FOXP3 is a transcription factor that makes T cells regulatory, so it may be a good drug target for achieving this goal~\citep{Revenko2022-ak}.
Will decreasing the expression of FOXP3 make this ovarian cancer patient's T cells more cytotoxic?
To address this question, we estimate the causal effect of FOXP3 on the expression of key genes that T cells use to clear tumors.
We also consider interventions on another gene in \Cref{apx:tcf7_sell}.

\begin{figure}[h]
\centering
\begin{subfigure}{0.49\columnwidth}
    \centering
    \includegraphics[width=0.85\columnwidth]{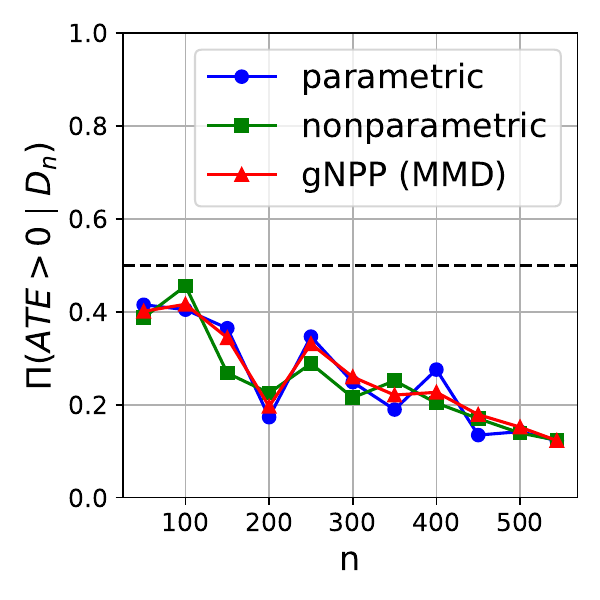}
    \caption{}
    %\caption*{$\hat \Pi\left(\ATE\left(\p\right) < 0\mid \cD_n \right)$ for \textit{FOXP3} on GZMH across varying $n$, for parametric, nonparametric, and gNPP models. }
    \label{fig:ATE_FOXP3_GZMH}
\end{subfigure}
\hfill
\begin{subfigure}{0.49\columnwidth}
    \centering
    \includegraphics[width=0.85\columnwidth]{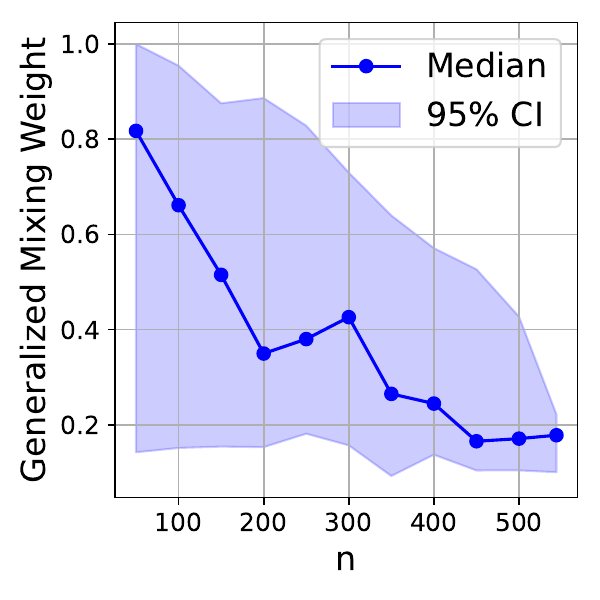}
    \caption{}
    %\caption*{Generalized mixing weights for \textit{FOXP3} on GZMH experiment.}
    \label{fig:gbf_FOXP3_GZMH}
\end{subfigure}
\caption{\textbf{Effect of FOXP3 on GZMH.} a. Posterior probability of the ATE being positive under the parametric, nonparametric, and gNPP models. $n$ denotes the size of the (subsampled) dataset. Values are the estimated median from 10 independent data subsamples and model samples. b. Generalized mixing weights, $\hat{\eta}_n$. The estimated confidence interval (CI) is across independent data subsamples and model samples. }
\label{fig:summary_FOXP3_GZMH}
\end{figure}

\paragraph{GZMH} We first consider the effect of FOXP3 on granzyme H and K (GZMH and GZMK), proteases which digests tumor proteins.
For granzyme H (GZMH), under the parametric model, the posterior probability that the ATE is positive is low, 0.12 (\Cref{fig:ATE_FOXP3_GZMH}). 
This suggests the ATE is likely negative, so an intervention that decreases FOXP3 expression will increase GZMH expression.
In other words, the parametric model infers that intervening on FOXP3 will make these T cells more capable of attacking tumors.
However, there are reasons to question this conclusion. 
The parametric model's $R^2$ is low, at 0.12, suggesting it explains only a small portion of the variance in the data. Moreover, its residual histogram and QQ plot suggest strong deviations from normality (\Cref{fig:lm-diagnostic}).
So the parametric model may be poorly specified.

The gNPP draws inferences that are robust to misspecification.
We find a small generalized mixing weights of 0.18, placing most of the posterior weight on the nonparametric model (\Cref{fig:gbf_FOXP3_GZMH}). 
However, the posterior probability of a positive ATE under the gNPP model remains nearly unchanged, at 0.12.
So, the conclusion that decreasing FOXP3 expression will increase GZMH expression is robust to possible model misspecification.

\begin{figure}[h]
\centering
\begin{subfigure}{0.49\columnwidth}
    \centering
    \includegraphics[width=0.85\columnwidth]{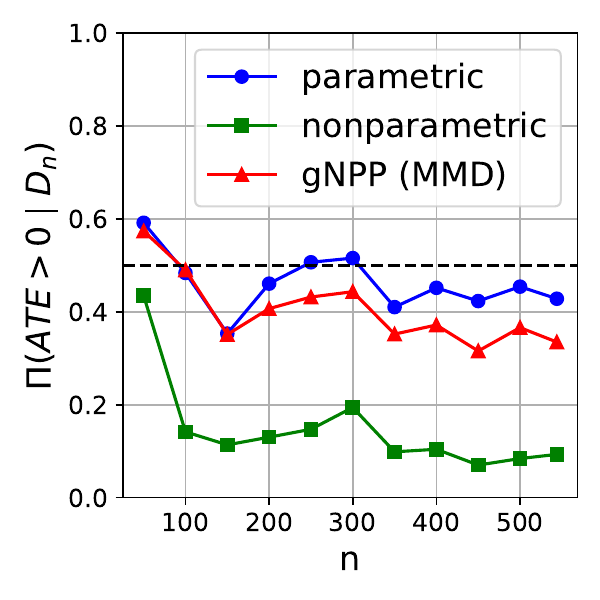}
    \caption{}
    \label{fig:ate_FOXP3_GZMK}
\end{subfigure}
\hfill
\begin{subfigure}{0.49\columnwidth}
    \centering
    \includegraphics[width=0.85\columnwidth]{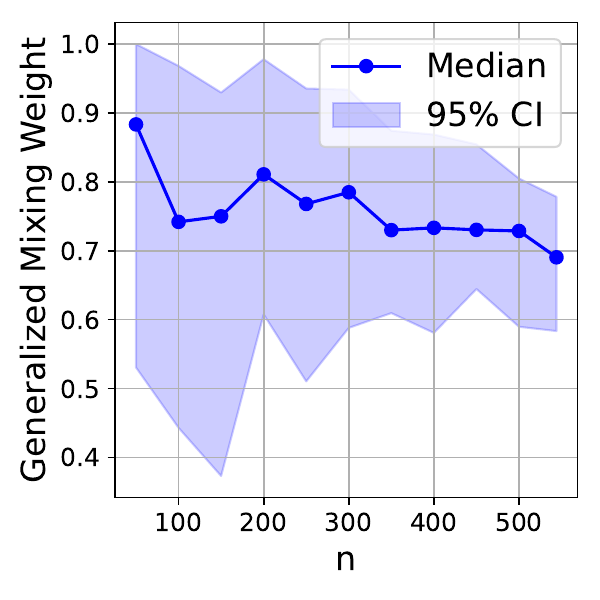}
    \caption{}
    \label{fig:gbf_FOXP3_GZMK}
\end{subfigure}
\caption{\textbf{Effect of FOXP3 on GZMK.} a. Posterior probability of the ATE being positive under the parametric, nonparametric, and gNPP models. $n$ denotes the size of the (subsampled) dataset. b. Generalized mixing weights, $\hat{\eta}_n$. CI: confidence interval across independent data subsamples and model samples.}
\label{fig:summary_FOXP3_GZMK}
\end{figure}

\paragraph{GZMK} We next consider the effect of FOXP3 on granzyme K (GZMK). 
The posterior probability that the ATE is positive, under the parametric model, is 0.43 (\Cref{fig:ate_FOXP3_GZMK}). The fact that this value is close to 0.5 suggests there is substantial uncertainty in the effect of FOXP3 on GZMK, and there may be no effect at all.
The gNPP, however, revises this estimate downward to 0.33, providing more confidence that decreasing FOXP3 expression will increase GZMK expression.
In this case, the gNPP still places considerable weight on the parametric model, with a generalized Bayes factor of 0.69 (\Cref{fig:gbf_FOXP3_GZMK}).

We subsampled the data down to smaller numbers of cells and reran the analysis, averaging over ten independent subsamples.
We found that with smaller amounts of data the gNPP model relies even more heavily on the parametric model, producing nearly the same posterior probability of a positive ATE (\Cref{fig:ate_FOXP3_GZMK}). The gNPP's estimates only begin to diverge from the parametric model at around $n = 200$ cells.

\begin{figure}[h]
\centering
\begin{subfigure}{0.49\columnwidth}
    \centering
    \includegraphics[width=0.85\columnwidth]{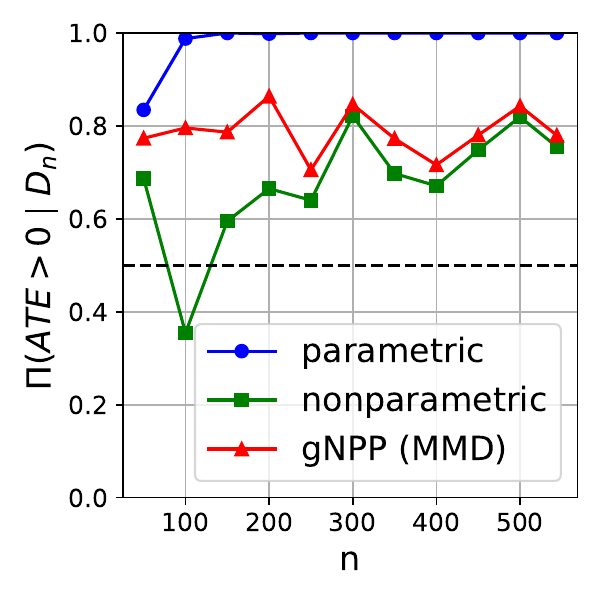}
    \caption{}
    \label{fig:ate_FOXP3_NKG7}
\end{subfigure}
\hfill
\begin{subfigure}{0.49\columnwidth}
    \centering
    \includegraphics[width=0.85\columnwidth]{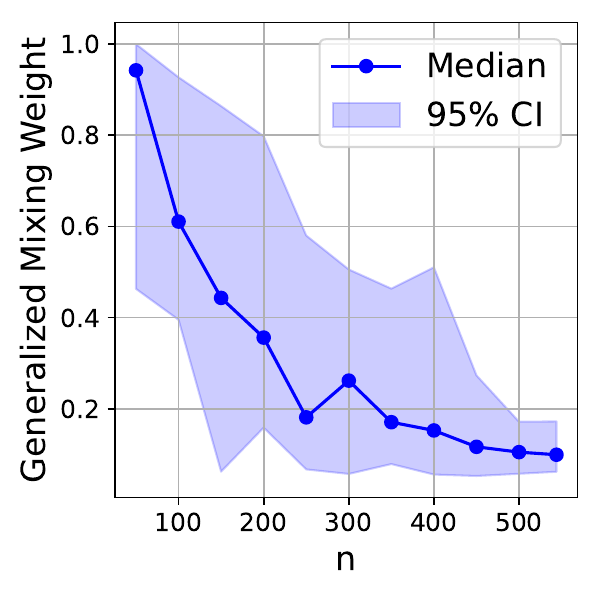}
    \caption{}
        \label{fig:gbf_FOXP3_NKG7}
\end{subfigure}
\caption{\textbf{Effect of FOXP3 on NKG7.} a. Posterior probability of the ATE being positive under the parametric, nonparametric, and gNPP models. $n$ denotes the size of the (subsampled) dataset. b. Generalized mixing weights, $\hat{\eta}_n$. CI: confidence interval across independent data subsamples and model samples.}
\label{fig:summary_FOXP3_NKG7}
\end{figure}

\paragraph{NKG7} We next consider the effect of FOXP3 on the granule protein NKG7, another key component of T cells' attack.
The parametric model suggests that decreasing FOXP3 expression will actually decrease NKG7, with the posterior probability of a positive ATE of 0.99 (\Cref{fig:ate_FOXP3_NKG7}).
Biologically this is unexpected, since regulatory T cells do not express much NKG7 in general \citep{Szabo2019}. 
Correspondingly, the gNPP increases our uncertainty in the finding, revising the posterior probability down to 0.77 (\Cref{fig:ate_FOXP3_NKG7}).
Here, the gNPP has strong confidence that the parametric model is wrong, with a generalized mixing weight of 0.1.

The gNPP adaptively smooths between the parametric and nonparametric posteriors, so its posterior probability of a positive ATE is quite stable here even if we sub-sample the data to below 100 datapoints (\Cref{fig:ate_FOXP3_NKG7}). The nonparametric model, by contrast, produces more variable estimates: at $n = 100$, it produces a posterior probability of 0.35, compared to the full dataset value of 0.76.

%The finding of positive effect is partly supported in \cite{Ng2020}, but more investigation is needed. 

\section{Discussion}
We showed how adding a nonparametric perturbation to a parametric Bayesian model can robustify the model, guarding against misspecification without sacrificing data efficiency.
We then developed a generalized Bayes posterior that achieves these same properties, but enjoys more scalable computation and implementation.
The basic technique is: (a) use a backup nonparametric or semiparametric model that is consistent for the target of inference and (b) mix the parametric posterior with the nonparametric posterior based on how well the parametric model describes the data.
Overall, gNPP models offer a practical approach to combining scientific domain knowledge with flexible machine learning models, by regularizing inferences from a nonparametric model towards a parametric model.

%The robustness and efficiency of gNPP models bear analogy to the properties of doubly robust estimators used for causal inference.
%Specifically, consider a doubly robust estimator that uses a parametric outcome model and a nonparametric propensity model. When the outcome model is well-specified, the estimator will converge at a parametric rate. When the outcome model is misspecified, the estimator will still converge to the true effect, but at a slower rate \citep{Bang2005,Antonelli2020}.
%A gNPP model with a parametric outcome model perturbed by a nonparametric outcome model offers the same guarantees, even without a propensity model.
%But the gNPP framework is not restricted to causal functionals. 
%Moreover, the two approaches are not mutually exclusive: when NPP models are applied to causal functionals, doubly robust parametri doubly robust approach can be used to estimate the NPP model's posterior, e.g. \cite{Yiu2023Semiparametric}.

% Limitations
The proposed gNPP posteriors have several important assumptions and limitations, which future work may overcome. 
First, they assume the data is independent and identically distributed (i.i.d.). 
In practice, data may be correlated, and the distribution may shift across time or space. 
An important future direction is to extend the generalized Bayes factor to handle non-i.i.d. data.
Second, gNPP posteriors have several tuning parameters, including the rate $r$, the divergence $\textsc{D}$, and the divergence's hyperparameters, such as the choice of kernel for the MMD and the transportation cost for the Wasserstein.
We have offered some basic guidance for choosing these parameters, but future work could go further to advance additional sensitivity analyses and develop techniques to tune parameters automatically and adaptively in response to data or the target estimand \citep{Schrab2023-yi}. Third, our theoretical analysis of the gNPP has focused on its asymptotic behavior. Future theoretical and empirical work could investigate the non-asymptotic regime in depth, perhaps leading to improved designs for the generalized Bayes factor.
% to assess different choices of $r$, we included a sensitivity analysis of the generalized Bayes factor only for the Wasserstein distance (\Cref{fig:wass_rate_comparison} in the Appendix). A similar empirical sensitivity analysis that varies $r$ over $(0,1/2)$ could also be carried out for MMD and KSD to examine how this affects the generalized Bayesian factors, potentially under different kernels, and would provide further practical insight. Finally, 

\vskip 0.2in
\bibliography{refs_new}

\newpage

\appendix
\section{Scaling of the Generalized Bayes Factor} \label{sec-gbf-scaling}
In \Cref{sect-method}, we define the generalized Bayes factor using the transformation $\Xi$. Here, we show that the transformed variable aligns with the typical scaling behavior of the Bayes factor \citep{Hong2005-kf,Dawid2011-kb}. First note,
\begin{align*}
    - \log  \gBF_n = \frac{\E{}{\diverge_n\left(\rmp_\theta, \rmp_0 \right) \mid x_{1:n}}}{\E{}{\diverge_n\left(\rmp_\theta, \rmp_0 \right)}} \left(n + 1\right)^r - 1 + \log \left(\frac{\E{}{\diverge_n\left(\rmp_\theta, \rmp_0 \right) \mid x_{1:n}}}{\E{}{\diverge_n\left(\rmp_\theta, \rmp_0 \right)}}(n + 1)^{r} \right).  
\end{align*}
Suppose the assumptions of \Cref{thm-gNPP} hold. Then the empirical divergence satisfies
\begin{equation*}
    \E{}{\diverge_n\left(\rmp_\theta, \rmp_0 \right) \mid x_{1:n}} = \diverge(\rmp_{\theta_0}, \rmp_0) + r_n, 
\end{equation*}
for some rate $r_n \to 0$ such that $r_n (n+1)^r = o(1)$. In practice, for the divergences we study, $r_n = n^{-\tilde r}$, and we choose $0 < r < \tilde r$. 
This yields the following scaling behavior: if $\p_0 \in \cM_{\param}$, then $\log  \gBF_n  = O((\tilde r - r)\log n)$; if $\p_0 \notin \cM_{\param}$, then $- \log  \gBF_n = O( n^r)$.

This aligns with the standard Bayes factor, comparing a high-dimensional model to a nested low-dimensional model, where, if $\p_0$ is in the low-dimensional model class, then $\log  \BF_n  = O(\log n)$; otherwise $- \log  \BF_n = O( n)$~\citep{Hong2005-kf,Dawid2011-kb}.

The scaling behavior of the generalized Bayes factor also reveals the tradeoffs involved in setting the hyperparameter $r$. Choosing smaller values of $r$ leads to faster convergence in the well-specified case, but slower convergence in the misspecified case.

\section{Details on Synthetic Experiments} \label{sec:synth_apx}

In the NPP, we estimate the KL divergence between $\p_0$ and the posterior predictive using Monte Carlo, with 1000 heldout samples from the true distribution. We use $m = n$ to estimate the divergence in the gNPP, for the MMD version and Wasserstein version. We use the IMQ kernel $k(x, y) = (c^2 + \|x - y\|_2^2)^{-0.5}$ with the bandwidth $c$ set to the median distance between datapoints~\citep{Gorham2017-sd}.

\begin{figure}[h]
\centering
\begin{subfigure}[b]{0.32\textwidth}
    \centering
    \caption{} \label{fig:wass_spec_error}
    \includegraphics[width=\textwidth]{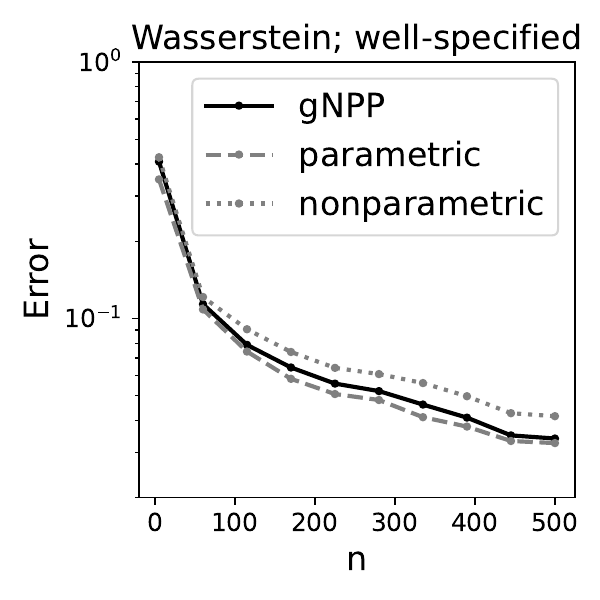}
\end{subfigure}
\begin{subfigure}[b]{0.32\textwidth}
    \centering
    \caption{} \label{fig:wass_miss_error}
    \includegraphics[width=\textwidth]{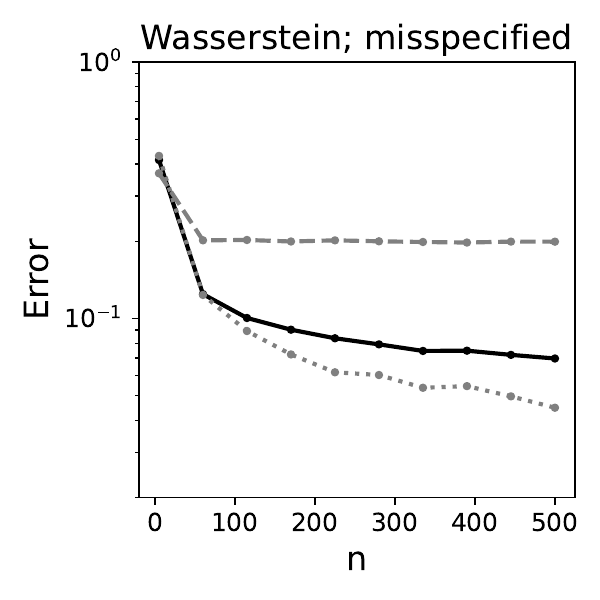}
\end{subfigure}\\
\begin{subfigure}[b]{0.32\textwidth}
    \centering
    \caption{} \label{fig:ksd_spec_error}
    \includegraphics[width=\textwidth]{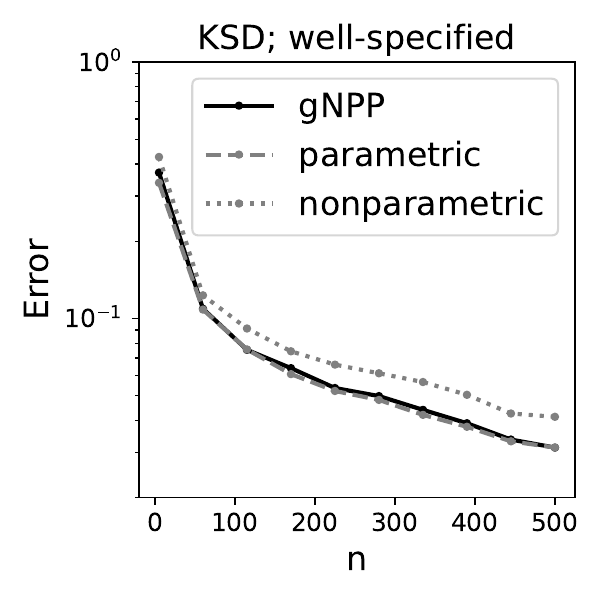}
\end{subfigure}
\begin{subfigure}[b]{0.32\textwidth}
    \centering
    \caption{} \label{fig:ksd_miss_error}
    \includegraphics[width=\textwidth]{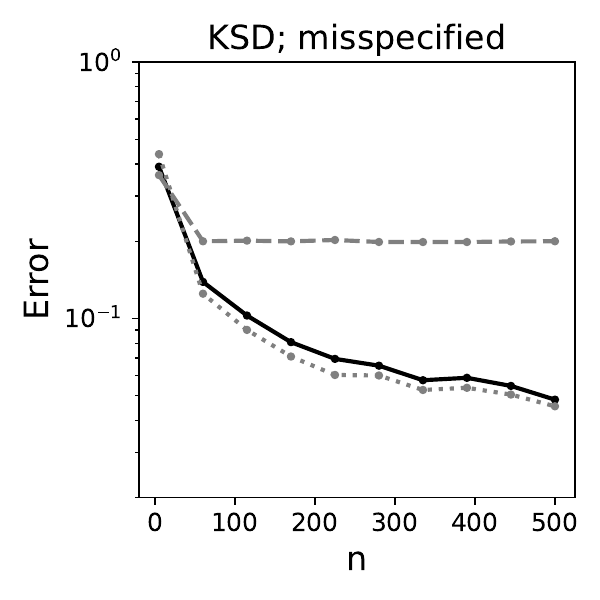}
\end{subfigure}
\caption{\textbf{Point estimation error of the gNPP.} Absolute difference between the posterior mean of the median and the true median, using the Wasserstein (a,b) and KSD (c, d) divergences, in the well-specified (a,c) and mis-specified cases (b,d).} \label{fig:point_est_synth}
\end{figure}

\begin{figure}[h!]
\centering
\begin{subfigure}[b]{0.32\textwidth}
\centering
    \caption{} \label{fig:ksd_spec_coverage}
    \includegraphics[width=\textwidth]{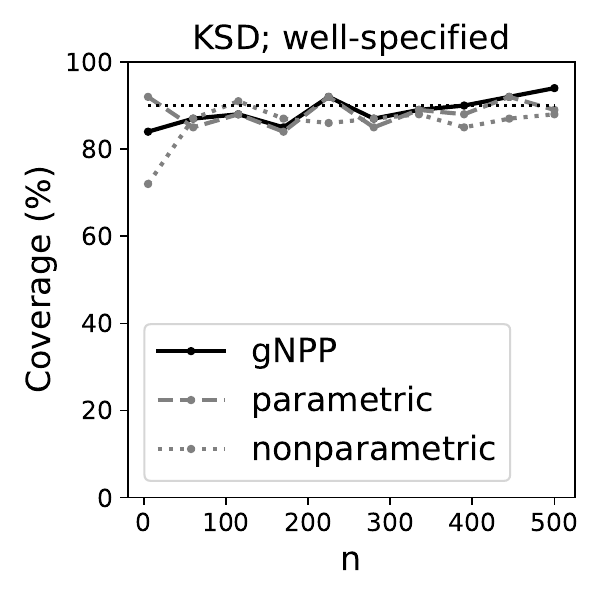}
\end{subfigure}
\begin{subfigure}[b]{0.32\textwidth}
    \centering
    \caption{} \label{fig:ksd_miss_coverage}
    \includegraphics[width=\textwidth]{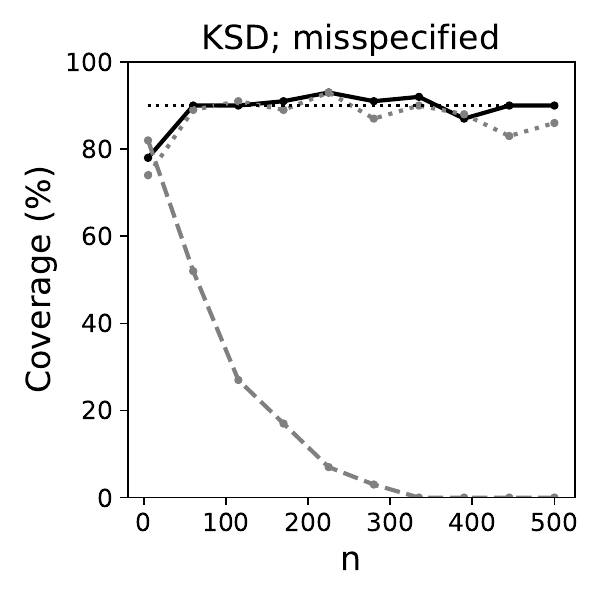}
\end{subfigure}\\
\begin{subfigure}[b]{0.32\textwidth}
\centering
    \caption{} \label{fig:wass_spec_coverage}
    \includegraphics[width=\textwidth]{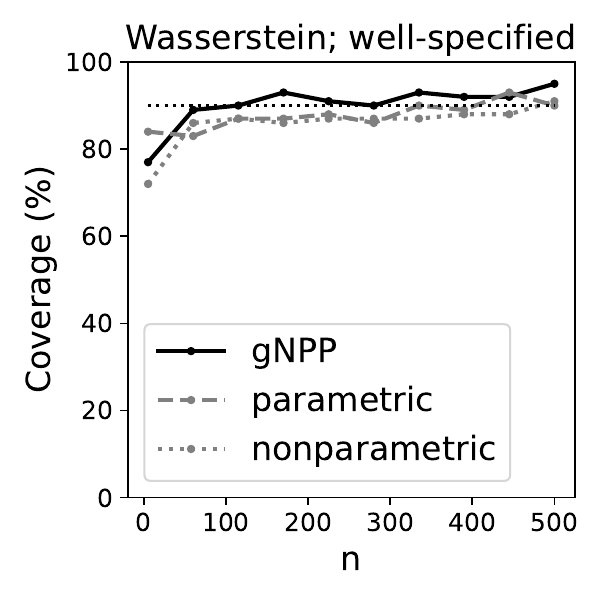}
\end{subfigure}
\begin{subfigure}[b]{0.32\textwidth}
    \centering
    \caption{} \label{fig:wass_miss_coverage}
    \includegraphics[width=\textwidth]{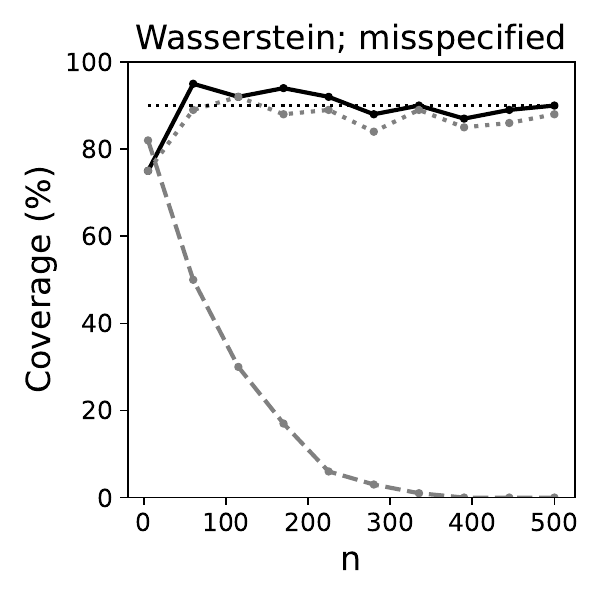}
\end{subfigure}
\caption{\textbf{Calibration of the gNPP.} We calculate how often the credible interval of the gNPP posterior includes the true median. } \label{fig:coverage_synth}
\end{figure}

\begin{figure}
    \centering
    \includegraphics[width=0.5\linewidth]{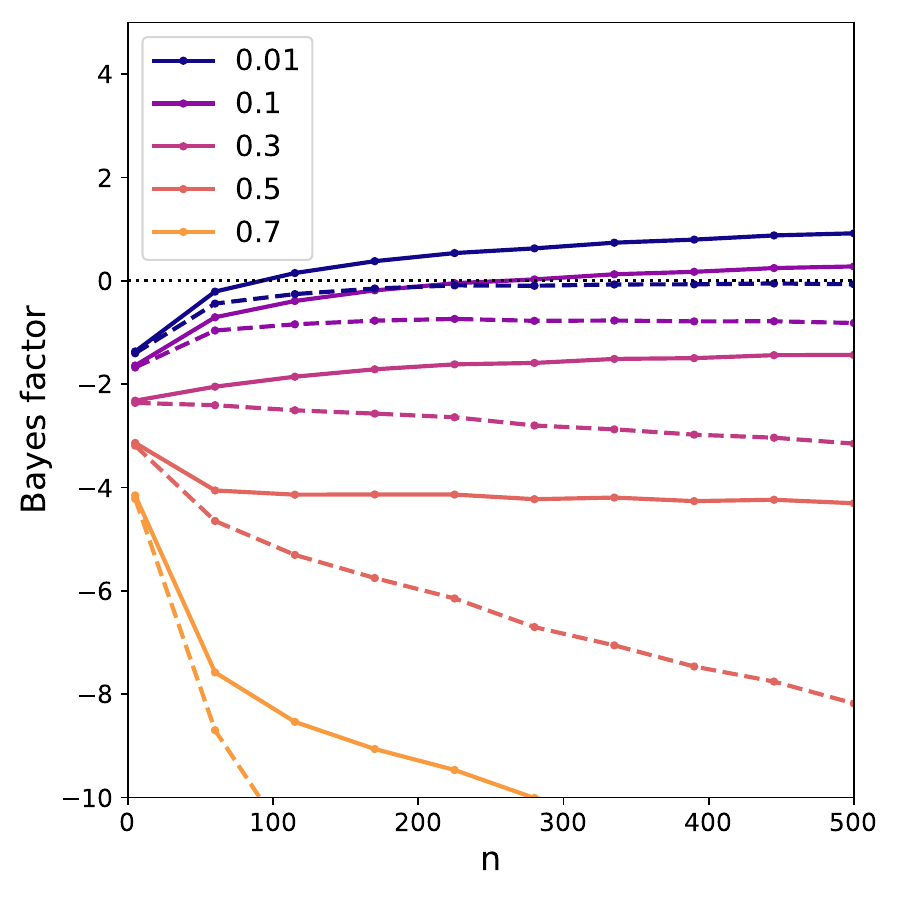}
    \caption{\textbf{Dependence of the generalized Bayes factor on the rate $r$}. We plot the log generalized Bayes factor with the Wasserstein distance, under different choices of rate hyperparameter $r$ (colors), in the well-specified case (solid) and misspecified case (dashed). Values above zero indicate the parametric model is favored.}
    \label{fig:wass_rate_comparison}
\end{figure}

\begin{figure}[h]
\centering
\begin{subfigure}[b]{0.32\textwidth}
    \centering
    \caption{}
    \includegraphics[width=\textwidth]{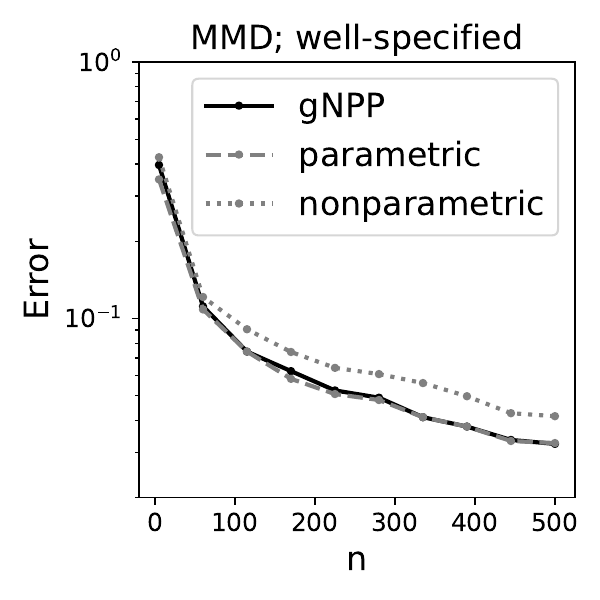}
\end{subfigure}
\begin{subfigure}[b]{0.32\textwidth}
    \centering
    \caption{}
    \includegraphics[width=\textwidth]{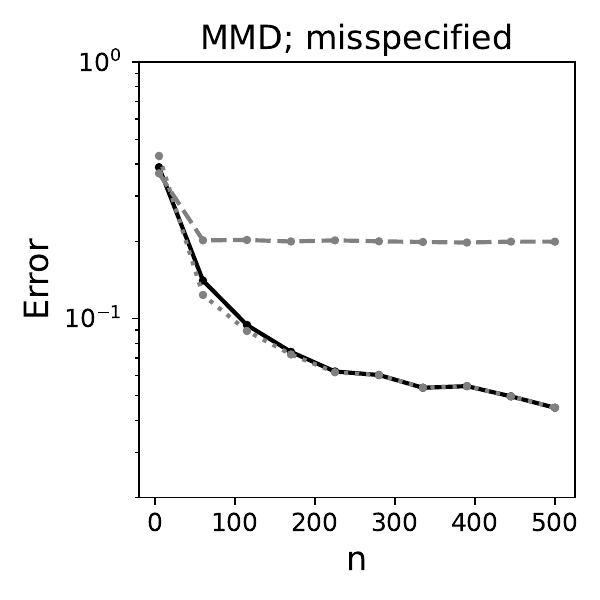}
\end{subfigure}
\caption{\textbf{Point estimation error of the gNPP with a binarized mixing weight.} Same as \Cref{fig:mmd_spec_error,fig:mmd_miss_error}, but using a binarized mixing weight in the gNPP.} \label{fig:point_est_binarized}
\end{figure}

As an additional comparison, we also examined the behavior of a modified version of the gNPP that uses a point estimate of the optimal model, $\mathbb{I}(\hat \eta_n > 0.5)$, in place of the standard mixing weight $\hat \eta_n$. We find similar behavior (\Cref{fig:point_est_binarized}).
\

\newpage
\section{Background}

We review key background for our theoretical results.
\subsection{Distances and Divergences}\label{subsect:divergence}
This section collect some definitions and results about statistical divergences.
\begin{definition}[KL divergence]
    The Kullback-Leibler (KL) divergence between two probability distributions $\rmp$ and $\rmq$ is defined as
\begin{equation*}
\KL{\rmp}{\rmq}  =\int \log \left(\frac{\rmp(t)}{\rmq(t)}\right) \rmp(t) dt . 
\end{equation*}
\end{definition}

\begin{definition}[Hellinger distance]
The Hellinger distance between two probability distributions $\rmp$ and $\rmq$ is defined as
\begin{equation*}
d_H(\rmp,\rmq) = \sqrt{\frac{1}{2} \int \left(\sqrt{\rmp(x)} - \sqrt{\rmq(x)}\right)^2} dx.
\end{equation*}
\end{definition}

\begin{definition}[Total variation distance]
The total variation distance between two probability distributions $\rmp$ and $\rmq$ is defined as
\begin{equation*}
d_{TV}(\rmp, \rmq) = \frac{1}{2} \int |\rmp(x) - \rmq(x)| dx.
\end{equation*}
\end{definition}
\begin{definition}[Bounded Lipschitz distance]
\label{def:BL-dist}
For a real-valued function $f: \R^d \mapsto \R$, the bounded Lipschitz norm is defined as 
\begin{equation}
    \|f\|_{BL} = \|f\|_\infty + \|f\|_L, 
\end{equation}
where $\|f\|_\infty = \sup_{x}|f(x)|$ and $\|f\|_L = \sup_{x \neq y} \frac{|f(x) - f(y)|}{\|x - y\|_2}$. 

For two distributions $\rmp, \rmq$ on $\R^d$, the bounded Lipschitz metric is defined as
\begin{equation}
d_{BL}(\rmp, \rmq) = \sup_{\|f\|_{BL} \leq 1} \left|\int f(x)(\rmp(x) - \rmq(x)) dx \right|.
\end{equation}
\end{definition}

\begin{theorem}\cite[Theorem 1.12.4]{VdV2023}
\label{thm:BL-weak-top}
For a sequence of distributions ${\rmp_n}$ and a distribution $\rmp_0$ on $\R^d$, the following are equivalent:
\begin{enumerate}
\item[(a)] $\rmp_n$ converges weakly to $\rmp_0$.
\item[(b)] $d_{BL}(\rmp_n, \rmp_0) \to 0$.
\end{enumerate}
\end{theorem}
\subsection{Bayesian Nonparametrics} \label{sect-BNP}

In this section we provide background on the frequentist analysis of Bayesian nonparametric models.

\paragraph{Notation.} Let $\cX \subseteq \R^d$. We denote the Hellinger distance, KL divergence, and $L_2$ norm between two densities $\rmp, \rmq \in \cP(\cX)$ as $d_H(\rmp, \rmq)$, $\KL{\rmp}{\rmq}$, and $\|\rmp - \rmq\|_2$, respectively, as reviewed in \Cref{subsect:divergence}. We also consider a generic metric $d$ on the space of probability densities, such as the Hellinger distance, $L_1$ distance, or $L_2$ distance for uniformly bounded densities.

The $\epsilon$-covering numbers and $\epsilon$-packing numbers of a metric space $\left(\cP, \rho \right)$ denoted by $N(\epsilon, \cP, \rho)$ and $D\left(\epsilon, \cP, \rho \right)$ are defined as the minimal number of balls of radius $\epsilon$ needed to cover $\cP$, and the maximal number of $\epsilon$-separated points, respectively. 

For any $\epsilon > 0$, we define the neighborhood of \(p_0\) as follows:
\begin{equation} \label{KL-neighborhood}
\begin{aligned}
    B_0(\rmp_0,  \epsilon, \cP) &= \left\{\rmp \in \cP: d_H\left( \rmp_0, \rmp \right)  < \epsilon^2 \right\}, \\
    B_k(\rmp_0, \epsilon, \cP) &= \left\{\rmp \in \cP:  \KL{\rmp_0}{\rmp} < \epsilon^2, \E{\rmp_0}{\left| \log(\rmp_0/\rmp) - \KL{\rmp_0}{\rmp}\right|^k } < \epsilon^k \right\}, \quad \text{for } k > 0. 
\end{aligned}
\end{equation}
The $B_0$ neighborhood defines an $\epsilon$-open ball around $\rmp_0$ under the Hellinger distance. 
The \(B_k\) neighborhoods (for \(k > 0\)) consider the KL divergence and higher-order moments of the log-likelihood ratio. These definitions help us reasons about the posterior contraction rates.  We keep the metric $d$ purposefully vague, but a common choice is the Hellinger divergence.

We define the nonparametric prior $\Pi_{\np}\left( \rmp\right)$ and $\cM_{\np}$ as the support of $\Pi_{\np}$. 

When we say the parametric model is regular, we mean that the model is finite-dimensional and satisfies sufficient conditions for the Bernstein-von Mises theorem. Sufficient conditions can be found in Ch10 of \cite{VdV2000}, Ch 8 of \cite{Ghosal2017} or Section 2 of \cite{Miller2021}.  
\paragraph{Prior mass and entropy assumptions. } 
Let $\epsilon_{n, \param}, \epsilon_{n, \np}$ be two sequences of positive numbers tending to zero, such that $\epsilon_{n, \param} < \epsilon_{n, \np}$.

\begin{assumption}[Model Entropy] \label{assumption:entropy}
There exists constant $C_{\param},C_{\np}$ such that
\begin{equation}
\begin{aligned}
&\sup_{\epsilon \geq \epsilon_{n, \param}} \log N\left( \frac{\epsilon}{3}, B_0\left(\rmp_0, 2\epsilon, \cM_{\param}  \right), d_H \right) \leq C_{\param} n \epsilon_{n, \param}^2,  \quad \text{and} \\
&\sup_{\epsilon \geq \epsilon_{n, \np}} \log N\left( \frac{\epsilon}{3}, B_0\left(\rmp_0, 2\epsilon, \cM_{n,\np}\right), d_H \right) \leq C_{\np} n \epsilon_{n, \np}^2.  
\end{aligned}
\end{equation}
where $\cM_{n,\np} \subseteq \cM_{\np}$  is a submodel that satisfies 
\begin{equation}
\begin{aligned}
  &\Pi_{\np}\left(\cM_{\np} \backslash \cM_{n,\np} \right)/\Pi_{\param}\left(B_2\left(\rmp_0, \epsilon_{n, \param}, \cM_{\param}\right)\right) \leq \exp\left(- 2 n \epsilon_{n, \np}^2\right). 
\end{aligned}
\end{equation}
\end{assumption}
\begin{assumption}[Prior Mass]\label{assumption:prior}
For the KL neighborhood in \Cref{KL-neighborhood}, the priors satisfy
\begin{equation}
\begin{aligned}
&\Pi_{\param}\left(B_2\left(\rmp_0, \epsilon_{n, \param}, \cM_{\param}\right)\right) \geq \exp\left(-n \epsilon_{n, \param}^2\right), \quad \text{if } \rmp_0 \in \cM_{\param}, \\
&\Pi_{\np}\left(B_2\left(\rmp_0, \epsilon_{n, \np}, \cM_{\np}\right)\right) \geq \exp\left(-n \epsilon_{n, \np}^2\right), \quad \text{if } \rmp_0 \notin \cM_{\param}.
\end{aligned}
\end{equation}
\end{assumption}

\Cref{assumption:prior,assumption:entropy} establish the standard prior mass and entropy conditions necessary for posterior contraction rates \citep{Ghosal2000-pf}. \Cref{assumption:entropy} defines the parametric and nonparametric rates. The second part of \Cref{assumption:prior} relaxes the entropy condition to a submodel supported on most of the prior mass, in view of \cite[Remark 10.4]{Ghosal2017}. \Cref{assumption:prior} requires the parametric prior to put sufficient mass around the truth when the parametric model is correctly specified. Similarly, the nonparametric prior is required to put sufficient mass around the truth when the parametric model is incorrectly specified. % Cases where the truth deviates from both the parametric or nonparametric models are excluded, as the posterior consistency fails.

It is useful to think of $\epsilon_{n, \param}, \epsilon_{n, \np}$ as the posterior contraction rates of the parametric and nonparametric models, respectively.  When the parametric model is regular, a choice of $\epsilon_{n, \param}$ is $n^{-1/2}d^{1/2} \log n$ by the Bernstein-von Mises theorem \citep{VdV2000}. The nonparametric contraction rate is slower by at least a logarithmic factor. For example, the posterior of a Dirichlet process mixture of normal priors contracts a rate of $n^{-1/2} (\log n)^{(d + 1 + 1/r_0)/2}$ where $d$ is the dimension and $r_0$ is some measure of smoothness of the true mixing distribution \citep{Shen2013}. 

%\Cref{assumption:uct} requires every neighborhood of $\rmp_0$ to be "tested away", i.e. $\rmp_0$ can a distinguished from other distributions. The notion of neighborhood depends on the topology under consideration, e.g. the weak topology. The testing assumption exists under mild conditions when the (e.g. Lemma 10.4 and 10.6 of \cite{VdV2000}), but it is highly non-trivial for Bayesian nonparametric models \citep{Ghosal2017}. 

\begin{assumption}[Rate Difference]
\label{assumption:rate-diff} 
We assume that for a sufficiently large $M > 0$,  
\begin{equation} \label{rate-diff1}
 \Pi_{\np}\left( B_0(\rmp_0,  M \epsilon_{n, \np}, \cM_{\np})   \right)  = o\left(\exp\left(- 3 n \epsilon_{n, \param}^2  \right)\right). 
\end{equation}
\end{assumption}
\Cref{assumption:rate-diff} establishes an upper bound for the nonparametric prior mass within a ball of radius $M \epsilon_{n, \np}$, which matches the prior mass lower bound in \Cref{assumption:prior}. As we approach the truth at a rate of $M \epsilon_{n, \np}$, the prior mass decreases at a rate exceeding \(\exp\left(-3n\epsilon_{n, \param}^2\right)\) within this neighborhood. 

%When the parametric model contracts at a rate of $\epsilon_{n, \param}^2 = \frac{d \log n}{n}$, this requirement implies $\Pi_{\np}\left( B_0(\mathbb{P}_0, M \epsilon_{n, \np},  \cM_{\np}) \right) = o\left(n^{-3d}\right)$. This is typically met in most nonparametric models, where the decay rate of the left side of \Cref{rate-diff1} is on the order of \(\exp\left(-M' n \epsilon_{n, \np}^2\right)\).

\section{Proof of \Cref{sect-theory}} \label{sect-proof-theory}

\subsection{Proofs of \Cref{sect-theory-npp}} \label{sect-proof-theory-npp}
This section contains two main results: one negative and one positive.
In \Cref{prop:BF-DP}, we show that for Dirichlet process perturbations, model selection consistency can fail (\Cref{examle:DP}). In contrast, \Cref{prop:BF-DPNM} shows that for Dirichlet process normal mixture perturbations, model selection consistency holds under mild conditions.

\begin{proposition} \label{prop:BF-DP}
\Cref{examle:DP} fails to satisfy model selection consistency as in \Cref{thm:lambda-consistency}. 
\end{proposition}

\begin{proof}[Proof of \Cref{prop:BF-DP}]
Consider the Polya urn construction of the Dirichlet process, 
\begin{equation}
x_1 \sim \rmp_\theta, \quad x_k \sim \frac{1}{1 + h} \rmp_\theta + \frac{h}{1 + h} \hat \rmp_{x_{1:k-1}}, \quad \text{for } k > 1, 
\end{equation}
where $\hat \rmp_{x_{1:k-1}}$ is the empirical distribution of $x_{1:k-1}$, and $h = 1/\alpha$~\citep{blackwell1973ferguson}.
For distinct $x_1, \cdots, x_n$, we have
\begin{equation}
\begin{aligned}
\eta_n 
&= \frac{\eta \int \rmp_\theta(x_{1:n}) \, d\Pi_{\param}(\theta)}{(1 - \eta) \int \tilde \rmp(x_{1:n}) \, d\Pi_{\pert}\left(\tilde \rmp \right)} \\
&= \frac{\eta}{1 - \eta}
\frac{\int \rmp_\theta(x_{1:n}) \, d\Pi_{\param}(\theta)}{\left( \frac{1}{1 + h}\right)^{n-1} \int \rmp_\theta(x_{1:n}) \, d\Pi_{\param}\left(\theta \right)}
= \frac{\eta}{1 - \eta} \left( \frac{1}{1 + h}\right)^{1 - n}. 
\end{aligned}
\end{equation}
The resulting $\eta_n$ does not depend on whether the parametric model is correct, and hence does not satisfy model selection consistency.
\end{proof}

We next establish that for Dirichlet process mixture model perturbations, model selection consistency holds if $\rmp_0$ is sufficiently smooth (\Cref{ex:BF-DPNM}). Loosely, a distribution is said to be \textit{supersmooth} when it is a Gaussian mixture over a thin-tailed mixing distribution, and \textit{$\beta$-smooth} when the density is thin-tailed and has sufficiently regular derivatives.

\begin{definition}\label{def:smoothness}
A density $\rmp_0$ on $\R^d$ is said to be 
\begin{itemize}
\item supersmooth if there exists $(F_0, \Sigma_0)$ such that $\rmp_0 = F_0 \ast \cN \left(0, \Sigma_0 \right)$ and $1 - F_0 \left([-z, z]^d\right) \lesssim \exp\left( - c_0 z^{a_1}\right)$ for every $z > 0$, with $c_0 > 0$ and $a_1 \geq 2$; 
\item $\beta$-smooth if the following holds: 
\begin{enumerate}
    \item The mixed partial derivative $D^k \rmp_0$ of order up to $k_+ \leq \lfloor \beta \rfloor$ satisfies
\begin{equation}
\begin{aligned}
\left|(D^k \rmp_0)(x + y) - D^k \rmp_0 \left(x \right)\right| 
&\leq L(x) \exp \left( c_1 \|y \|^2\right) \|y \|^{\beta - \lfloor \beta \rfloor}, \quad k_+ = \lfloor \beta \rfloor, \quad x , y \in \R^d,  \\
\rmP_0 \left( \frac{L + |D^k \rmp_0|}{\rmp_0}\right)^{(2 \beta + \epsilon)/\beta}
&< \infty, 
\end{aligned}
\end{equation}
for some function $L: \R^d \mapsto [0, \infty)$ and $c_1, \epsilon > 0$;
\item For every $\|x \| > a$, $\rmp_0(x) \leq c \exp \left( - b \|x\|^r \right)$, for some $a, b, c, r > 0$. 
\end{enumerate}
\end{itemize}
\end{definition}

\begin{proposition} \label{prop:BF-DPNM}
Assume that the true density $\rmp_0$ is either supersmooth or $\beta$-smooth. Suppose there exist positive constants $a_1,b_1,C_1$ such that with $\Pi_{\param}$-probability at least $1-\exp(-C_1 z^{a_1})$, the following hold:
\begin{enumerate}
\item $\rmp_\theta$ admits a continuous and strictly positive Lebesgue density on $\R^d$.
\item $\rmP_\theta\!\left([-z,z]^d\right)\ge 1-b_1\exp(-C_1 z^{a_1})$. 
\end{enumerate}

Let $\lambda_1(\Sigma)\ge \cdots \ge \lambda_d(\Sigma)$ denote the eigenvalues of $\Sigma$. Assume the prior $G$ on the scale matrix $\Sigma$ satisfies the following conditions:
\begin{enumerate}
\item There exist constants $a_2,b_2,C_2>0$ such that, for all $s>0$, $G\!\left(\lambda_d(\Sigma)\ge s\right)\le b_2 \exp\!\left(-C_2 s^{a_2}\right)$. 

\item There exist constants $a_3,b_3>0$ such that, for all $s>0$, $G\!\left(\lambda_1(\Sigma)\le s\right)\le b_3 s^{a_3}$. 

\item There exist constants $a_4,a_5,b_4,C_3,C_4>0$ such that, for all $0<s_1\le \cdots \le s_d$ and $0<t<1$,
\[
G\!\left(\bigcap_{j=1}^d \left\{ s_j < \lambda_j(\Sigma) < s_j(1+t)\right\}\right)
\ge b_4\, s_1^{a_4}\, t^{a_5}\, \exp\!\left(-C_3 s_d^{C_4/2}\right). 
\]
\end{enumerate}

Then \Cref{assumption:prior,assumption:entropy} hold for the Dirichlet process normal mixture perturbation (\Cref{npp,eqn:dpmm_npp}) and
\begin{enumerate}
\item $\epsilon_{n, \np} = n^{-1/2} (\log n)^{(d - 1 + 1/a_1)/2}$ if $\rmp_0 = F_0 \ast \cN \left(0, \sigma_0^2 I_d \right)$ is supersmooth with $\sigma_0^2 > 0$;
\item $\epsilon_{n, \np} = n^{-\beta/(2 \beta + d^\star)}(\log n)^{t_0}$ if $\rmp_0$ is $\beta$-smooth, where $d^\star = d \land 2$ and $t_0 = (\beta d^\star + \beta d^\star/r + d^\star + \beta)/(2 \beta + d^\star)$. 
\end{enumerate}
\end{proposition}

\noindent When the true density $\rmp_0$ is $\beta$-Hölder smooth, \cite{Shen2013} show that Dirichlet process normal mixtures converge to $\rmp_0$ at the rate $\epsilon_{n, \np} = n^{-\beta/(2 \beta + d^\star)}(\log n)^{t_0}$ for some constant $t_0 > 0$. Under the same rate, \Cref{prop:BF-DPNM} explicitly verifies \Cref{assumption:prior,assumption:entropy}. To satisfy \Cref{thm:lambda-consistency}, \Cref{assumption:rate-diff} must also hold; for this it suffices that $\epsilon_{n, \np}$ lower-bound the parametric posterior contraction rate \citep[Theorem 8.35]{Ghosal2017}. In particular, the parametric model will in general contract at the rate $\epsilon_{n, \param}^2 = d \log n/n$, in which case \Cref{assumption:rate-diff} is implied by the simplified condition $\Pi_{\np}\left( B_0(\rmP_0, M \epsilon_{n, \np},  \cM_{\np}) \right) = o\left(n^{-3d}\right)$. This is reasonable to expect since the decay rate on the left-hand side of \Cref{assumption:rate-diff} is typically of order \(\exp\left(-M' n \epsilon_{n, \np}^2\right)\). Unfortunately, \Cref{assumption:rate-diff} is difficult to verify directly for the Dirichlet process normal mixture, since it depends on the particular choice of priors on the mixing distribution and the scale matrix. For example, two Gaussian mixtures can be extremely close in Hellinger distance even when their mixing measures are far apart \citep{Soloff2025}.

\begin{proof}
Let $\Pi_{\pert}$ denote the prior on $\rmp$. We first verify \Cref{assumption:prior}. 

Let $E$ be the event
\begin{equation}
E
:= \left\{\theta:\;
1-\rmP_\theta\!\left([-z,z]^d\right)\le b_1 \exp\!\left(-C_1 z^{a_1}\right)\ \text{for all } z>0,\ 
\rmp_\theta(x)>0\ \text{and }\rmp_\theta \text{ is continuous for all } x\in\R^d
\right\}.
\end{equation}
By assumption, $\Pi_{\param}(E)>0$. Conditioning on $\theta\in E$, \citet[Proposition~9.14]{Ghosal2017} implies that there exist constants $A,C>0$ such that
\begin{equation*}
    \left(\textsc{dp}(\rmp_\theta,\alpha)\times G\right)\left(B_2 \left(\rmp_0, A\epsilon_{n,\np}, \cM_{\np}\right)\right)
\ge \exp\left(-C n\epsilon_{n,\np}^2\right). 
\end{equation*}
Integrating this lower bound over $E$ yields, for some other constant $C'>0$,
\begin{equation}
\Pi_{\pert}\!\left(B_2\!\left(\rmp_0, A\epsilon_{n,\np}, \cM_{\np}\right)\right)
\ge \Pi_{\param}(E)\,\exp\!\left(-C n\epsilon_{n,\np}^2\right)
\ge \exp\!\left(-C' n\epsilon_{n,\np}^2\right).
\end{equation}
This verifies \Cref{assumption:prior}. 

Now we verify \Cref{assumption:entropy} by following the argument in Section 9.4.4 of \cite{Ghosal2017}. Conditioning on the event $E$, we construct a sieve $\cM_{n,\np}$ consisting of densities of the form
$\rmp = F \ast \cN(0,\Sigma)$, where $F$ is a discrete mixing distribution
$F=\sum_{j=1}^\infty w_j\delta_{z_j}$ and
\begin{equation*}
\sum_{j=N+1}^\infty w_j < \epsilon^2,\qquad
z_1,\ldots,z_N\in[-a,a]^d,\qquad
\sigma^2 \le \lambda_d(\Sigma)\le \lambda_1(\Sigma)<\sigma^2(1+\epsilon^2)^n, 
\end{equation*}
where
\begin{equation*}
    N = \frac{C n \epsilon_{n, \np}^2}{\log(n  \epsilon_{n, \np}^2)}, \quad  \epsilon^2 = \frac{CN \log n}{n}, \quad a = (n \epsilon^2)^{1/a_1}, \quad \sigma^2 = (n \epsilon^2)^{-1/a_2}. 
\end{equation*}

By \citet[Lemma~9.15]{Ghosal2017} and the definitions of $N,\sigma^2,a$, we obtain the entropy bound for the sieve:
\begin{equation}\label{entropy-statement1}
\log N\!\left(\epsilon_{n,\np}/3,\ \cM_{n,\np},\ d_H\right)\lesssim n\epsilon_{n,\np}^2.
\end{equation}
Moreover, our assumption on $\Pi_{\param}$ implies $\E{\param}{1 - \rmP_\theta\left([-a, a]^d \right)} \lesssim \exp(- C_1 a^{a_1})$. 
By part~(ii) of \citet[Lemma~9.15]{Ghosal2017} and the assumptions on the prior $G$, we further have
\begin{equation}\label{entropy-statement2}
\begin{aligned}
    \Pi_{\pert}\!\left(\cM_{\np} \backslash \cM_{n, \np}\right) 
    &\leq \left(\frac{2 e \log \epsilon}{N} \right)^N + N \E{\param}{1 - \rmP_\theta\left([-a, a]^d \right)}  + G \left(  \lambda_1 (\Sigma) \geq \sigma^2 (1 + \epsilon^2)^n\right) + G \left(  \lambda_d (\Sigma) \leq \sigma^2\right)\\
    & \lesssim \left(\frac{2 e \log \epsilon}{N} \right)^N  + N \exp(- C_1 a^{a_1}) + \exp( - C_2 \sigma^{-2 a_2}) + \sigma^{-2 a_3} (1 + \epsilon_{n, \np}^2)^{-a_3 n} \\
    &\lesssim \exp\left(- C'' n \epsilon_{n, \np}^2 \right). 
\end{aligned}    
\end{equation}
for some constant $C''>0$. Combining \Cref{entropy-statement1} and \Cref{entropy-statement2} verifies \Cref{assumption:entropy}. 
\end{proof}

\begin{proof}[Proof of \Cref{thm-NPP}]
First note that the bounded Lipschitz metric is convex in its first argument. For $\eta \in [0,1]$ and probability measures $\rmP_0, \rmP_1, \rmQ$, 
\begin{equation}
\begin{aligned}
& d_{BL}\left(\eta \rmP_0 + (1 - \eta) \rmP_1, \rmQ\right) \\
&= \sup_{\|f\|_{BL} \leq 1} \left| \int f \, d(\eta \rmP_0 + (1 - \eta) \rmP_1 - \rmQ) \right| \\
&\leq \eta  \sup_{\|f\|_{BL} \leq 1} \left| \int f \, d(\rmP_0 - \rmQ) \right|  + (1 - \eta)  \sup_{\|f\|_{BL} \leq 1} \left| \int f \, d(\rmP_1 - \rmQ) \right| \\
&= \eta d_{BL} \left(\rmP_0, \rmQ\right) +  (1 -\eta) d_{BL} \left(\rmP_1, \rmQ\right).    
\end{aligned}
\end{equation}

\Cref{assumption-AN}(b) implies the total variation convergence 
\begin{equation}
    d_{TV}\left( \Pi_{\param}\left(d \theta \mid x_{1:n} \right), \cN \left(\hat \theta_{\mle}, \left( n V_{\theta_0} \right)^{-1}\right) \right) = O_{\rmP_0}(n^{-1/2}). 
\end{equation}
If we treat $\theta \mapsto \sqrt{n}\chi_\theta$ as a pushforward map, then by the triangle inequality, 
\begin{equation}
\begin{aligned}
&d_{BL}\left( \Pi_{\param}\left(\tilde \psi_n(\rmp_\theta) \mid x_{1:n} \right), \cN \left(0, \dot \chi_{\theta_0}^\top V_{\theta_0}^{-1} \dot \chi_{\theta_0} \right) \right)    \\
&\leq \underbrace{d_{BL}\left( \Pi_{\param}\left(\tilde \psi_n(\rmp_\theta) \mid x_{1:n} \right), (\sqrt{n}\chi)_\# \cN \left(\hat \theta_{\mle}, \left( n V_{\theta_0} \right)^{-1} \right) \right)}_{A_n} \\
&+ \underbrace{d_{BL} \left((\sqrt{n}\chi)_\# \cN \left(\hat \theta_{\mle},  \left( n V_{\theta_0} \right)^{-1} \right), \cN \left(0,  \dot \chi_{\theta_0}^\top V_{\theta_0}^{-1} \dot \chi_{\theta_0} \right) \right)}_{B_n}. 
\end{aligned}
\end{equation}
Let $\tilde \Pi_n$ be the law of $\sqrt{n}(\theta - \hat \theta_{\mle})$ where $\theta \sim \Pi_{\param}(\cdot \mid x_{1:n})$. We upper bound $A_n$ with the total variation bound in \Cref{assumption-AN}(b). 
\begin{equation}
\begin{aligned}
A_n 
&\leq d_{TV} \left((\sqrt{n}\chi)_\# \Pi_{\param}\left(\cdot \mid x_{1:n} \right),(\sqrt{n}\chi)_\# \cN \left(\hat \theta_{\mle}, \left( n V_{\theta_0} \right)^{-1}\right) \right) \\
&\leq d_{TV}\left(\tilde \Pi_n, \cN \left( 0, V_{\theta_0}^{-1}\right)  \right)  =o_{\rmP_0}(1). 
\end{aligned}
\end{equation}

Now we bound $B_n$. For $\theta \sim \cN \left(\hat \theta_{\mle},  \left( n V_{\theta_0} \right)^{-1} \right)$, we can reparametrize $\theta = \hat \theta_{\mle} +\left( n V_{\theta_0} \right)^{-1/2} Z$ for $Z \sim \cN(0, I_d)$. By a Taylor expansion, there exists $\tilde \theta_n$ in an $O_{\rmP_0}(n^{-1/2})$-neighbourhood of $\hat \theta_{\mle}$ such that
\begin{equation}
\chi_\theta = \chi(\hat \theta_{\mle}) + \dot \chi_{\hat \theta_{\mle}}^\top \left( n V_{\theta_0} \right)^{-1/2}  Z + n^{-1} \|V_{\theta_0} \|_2  Z^\top \ddot \chi_{\tilde \theta_n} Z . 
\end{equation}
We note that $\chi(\hat \theta_{\mle}) = 0$. 
Since $\hat \theta_{\mle} \overset{\rmP_0}{\to } \theta_0$, we also have $\tilde \theta_n \overset{\rmP_0}{\to } \theta_0$. Thus,
\begin{equation}
\sqrt{n}\chi_\theta  =\dot \chi_{\hat \theta_{\mle}}^\top V_{\theta_0}^{-1/2}  Z + O_{\rmP_0}(n^{-1/2})  \wto  \cN \left(0, \dot \chi_{\theta_0}^\top V_{\theta_0}^{-1} \dot \chi_{\theta_0} \right). 
\end{equation}
Finally, we use the equivalence between $d_{BL}$ and weak convergence to obtain $B_n = o_{\rmP_0}(1)$. 

For $\rmp$ drawn from the NPP model, we have
\begin{equation}
d_{BL} \left(\Pi\left(\tilde \psi_n(\rmp) \mid x_{1:n} \right), \cN \left(0, \dot \chi_{\theta_0}^\top V_{\theta_0}^{-1} \dot \chi_{\theta_0} \right) \right) \leq \eta_n   d_{BL} \left(\Pi_{\param}\left(\tilde \psi_n(\rmp_\theta) \mid x_{1:n} \right), \cN \left(0, \dot \chi_{\theta_0}^\top V_{\theta_0}^{-1} \dot \chi_{\theta_0} \right) \right) + (1 - \eta_n). 
\end{equation}

When $\rmp_0 \in \cM_{\param}$, $\eta_n \to 1$ by \Cref{thm:lambda-consistency}, and hence
\[
d_{BL} \left(\Pi\left(\tilde \psi_n(\rmp) \mid x_{1:n} \right), \cN\left(0, \dot \chi_{\theta_0}^\top V_{\theta_0}^{-1} \dot \chi_{\theta_0} \right) \right) \overset{\rmP_0}{\to} 0.
\]

When $\rmp_0 \notin \cM_{\param}$, $\eta_n \to 0$ by \Cref{thm:lambda-consistency}. Then
\begin{equation}
\begin{aligned}
&d_{BL}\left( \Pi\left(\psi(\rmp) \mid x_{1:n} \right), \delta_{\psi(\rmp_0)}\right) \\
&\leq \eta_n d_{BL}\left( \Pi_{\param}\left(\psi(\rmp_\theta) \mid x_{1:n} \right), \delta_{\psi(\rmp_0)}\right) + (1- \eta_n) d_{BL}\left( \Pi_{\pert} \left(\psi(\rmp) \mid x_{1:n} \right), \delta_{\psi(\rmp_0)}\right) \\
&\leq \eta_n + (1 - \eta_n)  d_{BL}\left( \Pi_{\pert} \left(\psi(\rmp) \mid x_{1:n} \right), \delta_{\psi(\rmp_0)}\right) \overset{\rmP_0}{\to} 0. 
\end{aligned}
\end{equation}
\end{proof}

\subsection{Proofs of \Cref{sect-theory-gnpp}}
The main goal of this section is to prove \Cref{thm-gNPP}, which shows that gNPP approximations are efficient and robust. This builds directly on \Cref{thm:gbf} and its proof, which establish consistency of the generalized Bayes factor $\hat \eta_n$ for model selection. Our first auxiliary result shows that the expected divergence $\E{\param}{\rho(\rmp_\theta, \rmp_{\theta_0}) \mid x_{1:n}}$ vanishes at the rate $n^{-1}$. 

\begin{lemma} \label{lemma:d-bias}
Let \Cref{assumption-AN,assumption-semimetric} be satisfied. Then $\E{\param}{\rho(\rmp_\theta, \rmp_{\theta_0}) \mid x_{1:n}} =   O_{\rmP_0}(n^{-1 })$. 
\end{lemma}

\begin{proof}
Let $B_{M_n n^{-1/2}} (\theta_0)$ be the neighbourhood defined in \Cref{assumption-semimetric}(b). Let $E_n$ be the event that $\hat \theta_{\mle} \in B_{M_n n^{-1/2}} (\theta_0)$. By \Cref{assumption-AN}(a), $\rmP_0(E_n) \to 1$ as $n \to \infty$. For the rest of the proof, we establish the result conditional on $E_n$.

Let $S_\theta := \nabla_\theta \rho(\rmp_\theta, \rmp_{\theta_0})$ and $H_\theta := \nabla_\theta^2 \rho(\rmp_\theta, \rmp_{\theta_0})$. By Taylor expansion, there exists a vector $\tilde \theta \in B\left( \theta_0, \|\hat \theta_{\mle} - \theta_0\|_2 \right)$ such that
\begin{equation}
\rho(\rmp_{\hat \theta_{\mle}}, \rmp_{\theta_0}) \leq S_{\theta_0}^\top \left(\hat \theta_{\mle} - \theta_0 \right) + \frac{1}{2} \left\|H_{\tilde \theta}^{1/2}\left( \hat \theta_{\mle} - \theta_0 \right) \right\|_2^2. 
\end{equation}
We note that $\nabla_\theta \rho(\rmp_\theta, \rmp_{\theta_0})|_{\theta = \theta_0} = 0$ since $\theta_0$ minimises $\rho(\rmp_\theta, \rmp_{\theta_0})$ on the open set $\Theta$. By the bounded Hessian assumption (\Cref{assumption-semimetric}(b)), there exists a constant $C_{\theta_0}$ such that 
\begin{equation}
\rho(\rmp_{\hat \theta_{\mle}}, \rmp_{\theta_0}) \leq C_{\theta_0} \left\| \hat \theta_{\mle} - \theta_0  \right\|_2^2 = O_{\rmP_0}(n^{-1}). 
\end{equation}

Similarly, there exists $\tilde \theta' \in B\left( \theta_0,  \|\hat \theta_{\mle} - \theta_0\|_2 \right)$ such that
\begin{equation}
\left\|\nabla_\theta \rho(\rmp_\theta, \rmp_{\theta_0})|_{\theta = \hat \theta_{\mle}} \right\|_2 = \left\|H_{\tilde \theta'} (\hat \theta_{\mle} - \theta_0) \right\|_2 \leq C_{\theta_0} \|\hat \theta_{\mle} - \theta_0 \|_2. 
\end{equation}

By Taylor expansion, we have
\begin{equation}
\begin{aligned}
\E{\param}{\rho(\rmp_\theta, \rmp_{\theta_0}) \mid x_{1:n}}  
&= \rho(\rmp_{\hat \theta_{\mle}}, \rmp_{\theta_0}) + \nabla_\theta \rho(\rmp_\theta, \rmp_{\theta_0})|_{\theta = \hat \theta_{\mle}}^\top \E{\param}{\theta - \hat \theta_{\mle} \mid x_{1:n}} \\
&\quad + \E{\param}{ \frac{1}{2}(\theta-\hat\theta_{\mle})^{\top}
\left\{\int_{0}^{1}(1-t)\,H_{\hat\theta_{\mle}+t(\theta-\hat\theta_{\mle})}\,dt\right\}
(\theta-\hat\theta_{\mle}) \mid x_{1:n}}. 
\end{aligned}
\end{equation}
Applying the uniform bound on $\|H_{\theta}\|_2$ in the neighbourhood of $\theta_0$ yields
\begin{equation*}
\begin{aligned}
\E{\param}{\rho(\rmp_\theta,\rmp_{\theta_0})\mid x_{1:n}}
&= \rho(\rmp_{\hat\theta_{\mle}},\rmp_{\theta_0})
+ \dot\rho_{\hat\theta_{\mle}}^{\top}\E{\param}{\theta-\hat\theta_{\mle}\mid x_{1:n} }
+ O_{\rmP_0}\!\left(\E{\param}{\|\theta-\hat\theta_{\mle}\|_2^2\mid x_{1:n}}\right).
\end{aligned}
\end{equation*}
Given the Bernstein--von Mises theorem for $\Pi_{\param}\left(\cdot \mid x_{1:n} \right)$ (\Cref{assumption-AN}), we have
\begin{equation}
\E{\param}{\|\theta - \hat \theta_{\mle}\|_2^2 \mid x_{1:n}} = O_{\rmP_0} \left(\frac{d}{n}\right), \quad \text{and} \quad \left\| \E{\param}{\theta \mid x_{1:n}} - \hat \theta_{\mle}\right\|_2 = O_{\rmP_0} \left(\sqrt{\frac{d}{n}}\right). 
\end{equation}
Finally, we combine the above inequalities to conclude that
\begin{equation}
\E{\param}{\rho(\rmp_\theta, \rmp_{\theta_0}) \mid x_{1:n}}  = \rho(\rmp_{\hat \theta_{\mle}}, \rmp_{\theta_0}) + O_{\rmP_0} \left(\frac{d}{n}\right) = O_{\rmP_0} \left(\frac{d}{n}\right). 
\end{equation}
\end{proof}

\begin{proof}[Proof of \Cref{contraction-div}]
By the triangle inequality, we have
\begin{equation}
\begin{aligned}
& \E{\param}{\diverge_{m,n}(\rmp_\theta, \rmp_0) \mid x_{1:n}}  - \diverge \left( \rmp_0, \rmp_{\theta_0}\right) \\
\leq\;&\E{\param}{\diverge_{m,n}(\rmp_\theta, \rmp_0) \mid x_{1:n}} - \E{\param}{\diverge \left(\rmp_\theta, \rmp_0 \right) \mid x_{1:n}} \\
&+ \E{\param}{\diverge \left(\rmp_\theta, \rmp_0 \right) \mid x_{1:n}}  - \diverge \left( \rmp_{\theta_0}, \rmp_0\right). 
\end{aligned}
\end{equation}
If $d(\cdot, \cdot)$ is a semimetric, then we can directly bound the second term by \Cref{lemma:d-bias}: 
\begin{equation}
\left|\E{\param}{\diverge \left(\rmp_\theta, \rmp_0 \right) \mid x_{1:n}}  - \diverge \left( \rmp_{\theta_0}, \rmp_0\right)\right|  \leq \E{\param}{\diverge \left(\rmp_\theta, \rmp_{\theta_0} \right) \mid x_{1:n}} =  O_{\rmP_0}\left(n^{-1} \right). 
\end{equation}
If $d(\cdot,\cdot)^{1/k}$ is a semimetric, then we apply the binomial expansion $x^k - y^k = (x - y) \sum_{j = 0}^{k-1} \binom{k}{j}(x - y)^{k -1- j} y^j$ to obtain
\begin{equation}
\begin{aligned}
&\left|\E{\param}{\diverge \left(\rmp_\theta, \rmp_0 \right) \mid x_{1:n}}  - \diverge \left( \rmp_{\theta_0}, \rmp_0\right)\right| \\
&= \left|\E{\param}{\left(\rho \left(\rmp_\theta, \rmp_0 \right) -  \rho \left( \rmp_{\theta_0}, \rmp_0\right) \right) \sum_{j = 0}^{k- 1} \binom{k}{j}\left(\rho \left(\rmp_\theta, \rmp_0 \right) -   \rho \left( \rmp_{\theta_0}, \rmp_0\right) \right)^{k -1- j} \rho \left( \rmp_{\theta_0}, \rmp_0\right)^j \mid x_{1:n}}  \right| \\
&\leq C \E{\param}{\left| \rho \left(\rmp_\theta, \rmp_0 \right) -  \rho \left( \rmp_{\theta_0}, \rmp_0\right) \right| \mid x_{1:n}} \\
&\leq C \E{\param}{\rho \left(\rmp_\theta, \rmp_{\theta_0} \right) \mid x_{1:n}} =  O_{\rmP_0}\left(n^{-1} \right),    
\end{aligned}
\end{equation}
where the inequality uses the uniform boundedness of $\rho(\cdot, \cdot)$. 

Let $M_n$ be a sequence such that $M_n \to \infty$ and $M_n = o(n^{1/2})$. 
To control the fluctuation of $\E{\param}{\diverge_{m,n}\left(\rmp_\theta, \rmp_0 \right) \mid x_{1:n}}$ around $\diverge( \rmp_\theta, \rmp_0)$, we decompose it into two terms inside and outside a ball of radius $M_n n^{-1/2}$:
\begin{equation}
\begin{aligned}
&\E{\param}{\diverge_{m,n}\left(\rmp_\theta, \rmp_0 \right) - \diverge( \rmp_\theta, \rmp_0) \mid x_{1:n}} \\
&=  \underbrace{\E{\param}{\left(\diverge_{m,n}\left(\rmp_\theta, \rmp_0 \right) - \diverge( \rmp_\theta, \rmp_0)\right) I_{B_{M_n n^{-1/2}}^c(\theta_0)} \mid x_{1:n}}}_{A_{m,n}} \\
&\quad + \underbrace{\E{\param}{\left( \diverge_{m,n}\left(\rmp_\theta, \rmp_0 \right) - \diverge( \rmp_\theta, \rmp_0)\right) I_{B_{ M_n n^{-1/2}}(\theta_0)} \mid x_{1:n}}}_{B_{m,n}}. 
\end{aligned}
\end{equation}
Since we can write $\diverge_{m,n}\left(\rmp_\theta, \rmp_0 \right) = \diverge\left(\rmp_\theta^m, \rmp_0^n \right)$, the quantity $\left| \diverge_{m,n}\left(\rmp_\theta, \rmp_0 \right) - \diverge( \rmp_\theta, \rmp_0)\right|$ is uniformly bounded. Then applying \Cref{assumption-AN} yields
\begin{equation}
\begin{aligned}
A_{m,n} 
&\leq C \Pi_{\param}\left(\| \theta - \theta_0 \|_2 > M_n n^{-1/2} \mid x_{1:n}\right) \\
&\leq  C \Pi_{\param}\left(\sqrt{n}\| \theta - \hat \theta_\mle \|_2 > M_n  - \sqrt{n} \| \theta_0 - \hat \theta_\mle \|_2 \mid x_{1:n}\right) \\
&\leq C \left(M_n  - \sqrt{n} \| \theta_0 - \hat \theta_\mle \|_2\right)^{-2}\E{\param}{n\| \theta - \hat \theta_\mle \|_2^2 \mid x_{1:n}}  \\
&= O_{\rmP_0}\left((M_n +  O_{\rmP_0}\left(1 \right))^{-2} \right)  O_{\rmP_0}\left(1 \right) = O_{\rmP_0}\left(M_n^{-2} \right),
\end{aligned}    
\end{equation}
for some constant $C>0$. The third line uses Chebyshev's inequality. The last line uses \Cref{assumption-AN}, which implies that the second moment
$\E{\param}{n\| \theta - \hat \theta_\mle \|_2^2 \mid x_{1:n}}$ 
converges in $\rmP_0$-probability to $\tr(V_{\theta_0}^{-1})$, hence
$\E{\param}{n\| \theta - \hat \theta_\mle \|_2^2 \mid x_{1:n}} =  O_{\rmP_0}\left(1\right)$.
Moreover, asymptotic normality of the MLE implies that the sequence
$\{\sqrt{n}\|\theta_0-\hat\theta_\mle\|_2\}_{n\ge 1}$ is bounded in probability, so
$\sqrt{n}\|\theta_0-\hat\theta_\mle\|_2=O_{\rmP_0}(1)$. 

To see that $A_{m,n}=O_{\rmP_0}(n^{-1})$, suppose to the contrary that there exists a fixed $m$ such that
$n|A_{m,n}|$ is unbounded in $\rmP_0$-probability. Then there exists a subsequence $\{n_k\}$ for which
$n_k|A_{m,n_k}|\to\infty$ in $\rmP_0$-probability. Define
\[
M_n := \begin{cases}
\sqrt{n_k}\,(n_k|A_{m,n_k}|)^{-1/4}, & n=n_k,\\
1, & \text{otherwise.}
\end{cases}
\]
Then $M_n=o(\sqrt{n})$ since $M_n/\sqrt{n}=1/\sqrt{n}\to 0$ for $n \neq n_k$ and
$M_{n_k}/\sqrt{n_k}=(n_k|A_{m,n_k}|)^{-1/4}\to 0$ for $n = n_k$. But for $n=n_k$, we also have $M_{n_k}^{-2}=\frac{\sqrt{|A_{m,n_k}|}}{\sqrt{n_k}}$, where
\[
\frac{|A_{m,n_k}|}{M_{n_k}^{-2}}=\sqrt{n_k|A_{m,n_k}|}\to\infty,
\]
so $|A_{m,n_k}|$ is not $O(M_{n_k}^{-2})$. This contradicts the assumption that
$A_{m,n}=O_{\rmP_0}(M_n^{-2})$ for every $M_n=o(\sqrt{n})$.

To bound $B_{m,n}$, we take the standard approach of analysing its mean and variance with respect to the approximating samples. For the mean, we apply Fubini's theorem:
\begin{equation}
\begin{aligned}
\E{}{B_{m,n}}
&=\E{\param}{\E{}{ \diverge_{m,n}\left(\rmp_\theta, \rmp_0 \right) - \diverge( \rmp_\theta, \rmp_0)} I_{B_{ M_n n^{-1/2}}(\theta_0)} \mid x_{1:n}} \\
&\leq \sup_{\theta \in B_{ M_n n^{-1/2}}(\theta_0)} \E{}{\diverge_{m,n}\left(\rmp_\theta, \rmp_0 \right) - \diverge( \rmp_\theta, \rmp_0)}.  
\end{aligned}
\end{equation}
Take $\theta_n \in \argmax_{\theta \in \bar B_{ M_n n^{-1/2}}(\theta_0)} \E{}{\diverge_{m,n}\left(\rmp_\theta, \rmp_0 \right) - \diverge( \rmp_\theta, \rmp_0)}$. As $n \to \infty$, the sequence $\theta_n \to \theta_0$ since $\|\theta_n - \theta_0 \|_2 \leq M_n n^{-1/2}$. We then apply \Cref{assumption-rate}(a) to conclude that 
\[
\E{}{\diverge_{m,n}\left(\rmp_{\theta_n}, \rmp_0 \right) - \diverge( \rmp_{\theta_n}, \rmp_0)} = O(r_{m,n}). 
\]

Now we control the variance. First, we upper bound the variance using Jensen's inequality:
\begin{equation}
\begin{aligned}
\var \left(B_{m,n} \right)  
&\leq   \E{\param}{\var \left(  \diverge_{m,n}\left(\rmp_\theta, \rmp_0 \right)  \right) I_{B_{ M_n n^{-1/2}}(\theta_0)} \mid x_{1:n}}. 
\end{aligned}
\end{equation}
By \Cref{assumption-rate}, we refine the upper bound as follows:
\begin{equation}
\var \left(B_{m,n}\right)  \leq r_{m,n}^2  \sup_{\theta \in \bar B_{ M_n n^{-1/2}}(\theta_0)} \cV(\theta) . 
\end{equation}
As $n \to \infty$, $\sup_{\theta \in \bar{B}_{M_n n^{-1/2}}(\theta_0)} \cV(\theta) \to \cV(\theta_0)$ by the continuity of $\cV$ at $\theta_0$, and hence $\var \left(B_{m,n}\right) = O(r_{m,n}^2)$. By Chebyshev's inequality, we conclude that $B_{m,n} = O_{\rmP_0}(r_{m,n})$.

Putting the bounds together, we have
\begin{equation}
   \E{\param}{\diverge_{m,n}(\rmp_\theta, \rmp_0) \mid x_{1:n}}  - \diverge(\rmp_{\theta_0}, \rmp_0) =  O_{\rmP_0}\left(  r_{m,n} \lor n^{-1} \right). 
\end{equation}
\end{proof}

\begin{proof}[Proof of \Cref{thm:gbf}]
Recall that we defined the generalized Bayes factor as
\begin{equation}
    \gBF_n := \Xi\left(\frac{\mathbb{E}[\diverge_n(\rmp_\theta, \rmp_0)]}{\mathbb{E}[\diverge_n(\rmp_\theta, \rmp_0) \mid x_{1:n}]} (n+1)^{-r}\right) \frac{\eta}{1-\eta}.
\end{equation}
The assumptions imply $r_n (n+1)^r  = o(1)$ and $\E{\param}{\diverge_n(\rmp_\theta, \rmp_0)} = O_{\P_0}(1)$, where $r_n$ is the rate at which the expected empirical divergence converges to the true divergence (see \Cref{contraction-div} and the subsequent discussion). By the continuous mapping theorem, $\gBF_n$ scales like
\begin{equation}
      \gBF_n  = \Xi\left(\frac{O_{\P_0}(1) }{\diverge(\rmp_{\theta_0}, \rmp_0) (n+1)^r  + o(1)} \right) \frac{\eta}{1 - \eta}. 
\end{equation}
If $\rmp_0 \in \cM_{\param}$, then $\diverge(\rmp_{\theta_0}, \rmp_0) = 0$ and $\gBF_n \overset{\P_0}{\to} \infty$ by the continuous mapping theorem. Analogously, if $\rmp_0 \notin \cM_{\param}$, then $\diverge(\rmp_{\theta_0}, \rmp_0) > 0$ and $\gBF_n \overset{\P_0}{\to} 0$.

We conclude with the desired result by applying the limit of $\gBF_n$ to $\hat \eta_n$.
\end{proof}

\begin{proof}[Proof of \Cref{thm-gNPP}]
By \Cref{assumption-AN,assumption-semimetric,assumption-rate}, we have $\E{\param}{\diverge_n(\rmp_\theta, \rmp_0) \mid x_{1:n}} = \diverge(\rmp_{\theta_0}, \rmp_0) + r_n$.  

By the decomposition provided in \Cref{eqn:npp_decomp} and \Cref{gBF}, we have
\begin{equation}
d_{BL} \left(\hat \Pi\left(\tilde \psi_n(\rmp) \mid x_{1:n} \right), \cN \left(0, \dot \chi_{\theta_0}^\top V_{\theta_0}^{-1} \dot \chi_{\theta_0} \right) \right) \leq \hat \eta_n   \underbrace{d_{BL} \left(\Pi_{\param}\left(\tilde \psi_n(\rmp_\theta) \mid x_{1:n} \right), \cN \left(0, \dot \chi_{\theta_0}^\top V_{\theta_0}^{-1} \dot \chi_{\theta_0} \right) \right)}_{\cA_n} + (1 - \hat \eta_n). 
\end{equation}
In the proof of \Cref{thm-NPP}, we showed that the convergence $\cA_n = o_{\P_0}(1)$ under \Cref{assumption-AN,assumption-chi}.  When $\rmp_0 \in \cM_{\param}$, $\hat \eta_n \overset{\P_0}{\to} 1$ by \Cref{thm:gbf}, and $d_{BL} \left(\hat \Pi\left(\tilde \psi_n(\rmp) \mid x_{1:n} \right), \cN \left(0, \dot \chi_{\theta_0}^\top V_{\theta_0}^{-1} \dot \chi_{\theta_0} \right) \right) \overset{\P_0}{\to} 0$.

When $\rmp_0 \notin \cM_{\param}$, $\hat \eta_n \overset{\P_0}{\to} 0$ by \Cref{thm:gbf}. Then
\begin{equation}
\begin{aligned}
&d_{BL}\left(\hat \Pi\left(\psi(\rmp) \mid x_{1:n} \right), \delta_{\psi(\rmp_0)}\right) \\
&\leq \hat \eta_n d_{BL}\left( \Pi_{\param}\left(\psi(\rmp_\theta) \mid x_{1:n} \right), \delta_{\psi(\rmp_0)}\right) + (1- \hat \eta_n) d_{BL}\left( \hat \Pi_{\pert} \left(\psi(\rmp) \mid x_{1:n} \right), \delta_{\psi(\rmp_0)}\right) \\
&\leq \hat \eta_n + (1 - \hat \eta_n)  d_{BL}\left( \hat \Pi_{\pert} \left(\psi(\rmp) \mid x_{1:n} \right), \delta_{\psi(\rmp_0)}\right) \overset{\P_0}{\to} 0. 
\end{aligned}
\end{equation}

\end{proof}

\subsection{Empirical Divergences} \label{apx:empirical_divergences}
In this section, we establish the convergence rates for the empirical divergences based on Wasserstein, MMD and KSD. We introduce specific conditions for each divergence.

\subsubsection{Wasserstein} \label{sect-Wasserstein}
We establish the convergence rate of $\E{\param}{\W_p^p(\hat \rmp_\theta^m, \hat \rmp_0^n) \mid x_{1:n}}$ over a sample space $\X \subseteq \R^\kappa$. This implies that \Cref{thm-gNPP} holds for the Wasserstein divergence, by replacing \Cref{contraction-div} with \Cref{thm-Wp} in the proof of \Cref{thm-gNPP}.

\begin{theorem}[Posterior expected Wasserstein convergence rate] \label{thm-Wp}
Suppose $p \geq 1$. Let \Cref{assumption-AN,assumption-semimetric,assumption-wass} be satisfied. Then
\begin{equation}
\E{\param}{\W_p^p(\hat \rmp_\theta^m, \hat \rmp_0^n) \mid x_{1:n}}- \W_p^p \left(\rmp_{\theta_0}, \rmp_0\right) =  O_{\P_0}\left(n^{-2/(\kappa \lor 4)} + m^{-2/(\kappa \lor 4)} \right). 
\end{equation}
\end{theorem}

In \Cref{assumption-wass}, we assume that both the parametric family and the true distribution have compact support and finite moments up to order $2p$.

\begin{assumption} \label{assumption-wass}
The distributions $\{\rmp_\theta\}_{\theta \in \Theta}$ and $\rmp_0$ satisfy
\begin{itemize}
    \item (Support) The distributions are supported on a compact subset $\X \subseteq \R^\kappa$ and have positive densities in the interior of their respective supports.
    \item (Moments)  The distributions have finite moments up to order $2p$. Additionally, the mapping $\theta \mapsto \E{\rmp_\theta}{\|X\|_2^{2p}}$ is continuous at $\theta_0$. 
\end{itemize}
\end{assumption}

\paragraph{Proofs}
\begin{lemma} \label{lemma:Wp-variance}
Let $X_1, \cdots, X_n \iid \rmp$ and $Y_1, \cdots , Y_m \iid \rmq$ be two independent samples with finite moments up to order $2p$, specifically $\E{\rmp}{\|X\|_2^j} + \E{\rmq}{\|Y\|_2^j} =: M_j^j < \infty$ for $j \in [2p]$. For any $n,m>0$,
\begin{equation}
\var\left(\W_p^p \left(\hat \rmp^n,\hat \rmq^m\right) \right) \leq 2^{p -1} M_{2p}^{2p}\left(\frac{1}{n} + \frac{1}{m} \right)^2,
\end{equation}
where $\hat \rmp^n$ and $\hat \rmq^m$ are the empirical measures formed by the $X_i$'s and $Y_i$'s.
\end{lemma}

\begin{proof}[Proof of \Cref{lemma:Wp-variance}]
By the definition of $\W_p$, we have
\begin{equation}
\W_p^p \left(\hat \rmp^n,\hat \rmq^m\right) = \min_{\sum_{i = 1}^n  w_{ij} = \frac{1}{m}, \sum_{j = 1}^m  w_{ij} = \frac{1}{n}} \sum_{i = 1}^n \sum_{j = 1}^m w_{ij}\|X_i - Y_j \|_2^p \leq \frac{\sum_{i = 1}^n \sum_{j = 1}^m \|X_i - Y_j \|_2^p}{nm}.
\end{equation}

Define $Z_{ij} := \|X_i - Y_j\|_2^p$. Since each pair $(X_i, Y_j)$ is independent, the collection $\{Z_{ij}\}$ is i.i.d. Under the bounded moment condition, the variance satisfies
\begin{equation}
\var\left(Z_{ij}\right) \leq 2^p M_{2p}^{2p}< \infty.
\end{equation}

Aggregating the variance shows
\begin{equation}
\var\left(\W_p^p \left(\hat \rmp^n,\hat \rmq^m\right) \right) \leq \frac{nm 2^p M_{2p}^{2p}}{n^2 m^2} = \frac{2^p M_{2p}^{2p}}{nm } \leq 2^{p -1} M_{2p}^{2p}\left(\frac{1}{n} + \frac{1}{m} \right)^2.
\end{equation}
where the last inequality uses $(nm)^{-1/2} \leq \frac{1}{2}(n^{-1} + m^{-1})$.
\end{proof}

\begin{lemma} \label{lemma:Wp-bias}
If $\rmp, \rmq$ are probability densities on a convex compact set $\X \subseteq \R^\kappa$ with nonempty interior, then for $p \geq 1$, we have
\begin{equation}
\sup_{\rmp, \rmq \in \cP(\X)}\E{}{\left| \W_p^p \left(\hat \rmp^n, \hat \rmq^m \right) -  \W_p^p \left(\rmp, \rmq \right) \right|} = O\left(n^{-2/(\kappa \lor 4)} + m^{-2/(\kappa \lor 4)} \right). 
\end{equation}
for $X_1, \cdots, X_n \iid \rmp$ and $Y_1, \cdots, Y_m \iid \rmq$. 
\end{lemma}

\begin{proof}[Proof of \Cref{lemma:Wp-bias}]
Recall $\hat \rmp^n :=  \frac{1}{n} \sum_{i = 1}^n \delta_{X_i}$ and $\hat \rmq^m := \frac{1}{m} \sum_{j = 1}^m \delta_{Y_j}$. We note that the function $x \mapsto \|x\|_2^p$ is convex by an application of Hölder's inequality. Then, following \citet[Remark 1.13]{Villani2003}, there exist Kantorovich potentials $\phi_n, \psi_m: \X \to \R$ such that
\begin{equation}
\W_p^p \left(\hat \rmp^n, \hat \rmq^m \right)  = \E{\hat \rmp^n}{\phi_n(X)} + \E{\hat \rmq^m}{\psi_m(Y)}. 
\end{equation}
Let $\Phi(\rmp,\rmq)$ be the set of pairs $(\phi, \psi) \in L^1(\rmp) \times L^1(\rmq)$ such that $\phi(x) + \psi(y) \leq \|x - y  \|_2^p$ for all $x, y \in \X$.
Since $\rmp$ and $\rmq$ are compactly supported, we have $(\phi_n, \psi_m) \in \Phi(\rmp, \rmq)$. By Kantorovich duality, we have 
\begin{equation}
\begin{aligned}
\W_p^p \left( \rmp , \rmq \right) &\geq \E{\rmp}{\phi_n(X)} + \E{\rmq}{\psi_m(Y)} \\
&= \W_p^p \left( \hat \rmp^n, \hat \rmq^m \right) + \int \phi_n(x) (\rmp(x) - \hat \rmp^n(x)) \, d x + \int \psi_m(x) (\rmq(x) - \hat \rmq^m(x)) \, d x . 
\end{aligned}
\end{equation}
and 
\begin{equation}
\W_p^p \left(\hat \rmp^n, \hat \rmq^m \right) \leq \W_p^p \left( \rmp, \rmq \right)  + \int \phi_n(x) (\rmp(x) - \hat \rmp^n(x)) \, d x + \int \psi_m(x) (\rmq(x) - \hat \rmq^m(x)) \, d x . 
\end{equation}
Define $\tilde \phi_n(x) = \phi_n(x) - \frac{L}{4} \|x\|_2^2$ and $\tilde \psi_m(y) = \psi_m(y) - \frac{L}{4} \|y\|_2^2$.  By \citet[Lemma 5]{ManoleWeed2024EmpiricalWass}, $\tilde \phi_n$ and $\tilde \psi_m$ are concave, $L$-Lipschitz, and uniformly bounded by $L$. 

Let $\cF_{L, U}(K)$ be the set of $L$-Lipschitz convex functions over a convex set $K \subseteq \R^\kappa$, where $\|f(x)\|_\infty \leq U$. By \citet[Lemma 5]{ManoleWeed2024EmpiricalWass}, the functions $\tilde \phi_n$ and $\tilde \psi_m$ are concave, $L$-Lipschitz, and uniformly bounded by $L$, thus $-\tilde \phi_n/L$ and $-\tilde \psi_m/L$ belong to $\cF_{1,1}(\X)$. 

Since $\X$ is convex and compact, we define
\begin{equation}
\Delta_{n,m} := \sup_{f \in \cF_{1,1}(\X)} \int f(x) (\rmp(x) - \hat \rmp^n(x)) \, d x + \sup_{g \in \cF_{1,1}(\X)} \int g(x) (\rmq(x) - \hat \rmq^m(x)) \, d x. 
\end{equation}
Using the fact that $-\tilde \phi_n/L$ and $-\tilde \psi_m/L$ belong to $\cF_{1,1}(\X)$, we have
\begin{equation}
    \W_p^p \left(\hat \rmp^n, \hat \rmq^m \right) - \W_p^p \left( \rmp , \rmq \right) \leq L \Delta_{n,m} + \frac{L}{4} \int \|x\|_2^2 \bigl((\hat \rmp^n(x) - \rmp(x)) + (\hat \rmq^m(x) - \rmq(x))\bigr) \, d x. 
\end{equation}
For the lower bound, there exists a pair of Kantorovich potentials $(\phi_{\rmp}, \psi_{\rmq}) \in \Phi(\rmp, \rmq)$ such that $\|\phi_{\rmp}\|_\infty \lor \|\psi_{\rmq}\|_\infty \leq 1$ and  $\W_p^p \left( \rmp , \rmq \right) = \E{\rmp}{\phi_{\rmp}(X)} + \E{\rmq}{\psi_{\rmq}(Y)}$. Thus, 
\begin{equation}
 \W_p^p \left(\hat \rmp^n, \hat \rmq^m \right) - \W_p^p \left( \rmp , \rmq \right) \geq  \int \phi_{\rmp}(x) (\rmp(x) - \hat \rmp^n(x)) \, d x + \int \psi_{\rmq}(x) (\rmq(x) - \hat \rmq^m(x)) \, d x . 
\end{equation}
Combining the above displays, we have
\begin{equation}
\begin{aligned}
&\E{}{\left|  \W_p^p \left(\hat \rmp^n, \hat \rmq^m \right) - \W_p^p \left( \rmp , \rmq \right)   \right|} \\
&\leq L \E{\rmp \otimes \rmq}{\Delta_{n,m}} + \frac{L}{4} \int \|x\|_2^2 \bigl((\hat \rmp^n(x) - \rmp(x)) + (\hat \rmq^m(x) - \rmq(x))\bigr) \, d x \\
&+ \E{\rmp}{\left| \int \phi_{\rmp}(x) (\rmp(x) - \hat \rmp^n(x)) \, d x\right|} + \E{}{\left| \int \psi_{\rmq}(x) (\rmq(x) - \hat \rmq^m(x)) \, d x \right|}. 
\end{aligned}
\end{equation}
Since $\|\phi_{\rmp}\|_\infty \lor \|\psi_{\rmq}\|_\infty \leq 1$, the functional $\rmp \mapsto  \E{\rmp}{\phi_{\rmp}(X)}$ is a bounded linear functional. By Chebyshev's inequality, $\P\left(\left|\E{\hat \rmp^n}{\phi_{\rmp}(X)} - \E{\rmp}{\phi_{\rmp}(X)} \right| \geq t \right) \leq \frac{\var_{\rmp}\left(\phi_{\rmp}(X)\right)}{nt^2} < \frac{4}{nt^2}$. The variance bound is uniform over all distributions $\rmp$ on $\X$, thus we have
\begin{equation}
\sup_{\rmp}\E{\rmp}{\left| \int \phi_{\rmp}(x) (\rmp(x) - \hat \rmp^n(x)) \, d x\right|} = O\left( n^{-1/2} \right). 
\end{equation}
Similarly, we have
\begin{equation}
\sup_{\rmq} \E{\rmq}{\left| \int \psi_{\rmq}(x) (\rmq(x) - \hat \rmq^m(x)) \, d x \right|} = O\left( m^{-1/2} \right), 
\end{equation}
and 
\begin{equation}
\sup_{\rmp, \rmq} \int \|x\|_2^2 \bigl((\hat \rmp^n(x) - \rmp(x)) + (\hat \rmq^m(x) - \rmq(x))\bigr) \, d x = O\left( n^{-1/2} + m^{-1/2} \right). 
\end{equation}
As a result, we have
\begin{equation} \label{eqn1:lemma:Wp-bias}
\sup_{\rmp, \rmq} \E{\rmp \otimes \rmq}{\left|  \W_p^p \left(\hat \rmp^n, \hat \rmq^m \right) - \W_p^p \left( \rmp , \rmq \right)   \right|} = L \sup_{\rmp, \rmq} \E{\rmp \otimes \rmq}{\Delta_{n,m}} + O\left( n^{-1/2} + m^{-1/2} \right). 
\end{equation}
To upper bound $\E{\rmp \otimes \rmq}{\Delta_{n,m}}$, we note that it is a sum of expectation suprema of empirical processes indexed by convex Lipschitz functions. Thus, we can bound the expectation of suprema by applying Dudley's chaining technique in terms of metric entropy of the class $\cF_{1, 1}(\X)$. One such result is \citet[Theorem 16]{VonLuxburg2004}, which states
\begin{equation}
\E{\rmp}{\sup_{f \in \cF(\X)} \int f(x) (\rmp(x) - \hat \rmp^n(x)) \, d x } \leq 2 \tau + \frac{4 \sqrt{2}}{\sqrt{n}} \int_{\tau/4}^{\sup_{f,f' \in \cF}\|f-f'\|_{L^2\left(\hat \rmp^n\right)}} \sqrt{\log N(\epsilon, \cF, L^2\left(\hat \rmp^n\right))}\, d \epsilon. 
\end{equation}

Since the $L^2$-distance is strictly smaller than the $L^\infty$-distance,
\begin{equation}
N \left(\epsilon, \cF_{1,1}(\X), L^2(\hat \rmp^n) \right) \leq   N \left(\epsilon, \cF_{1,1}(\X), L^\infty \right) \leq N \left(\epsilon, \cF_{1,1}([-1,1]^\kappa), L^\infty \right), 
\end{equation}
where the last inequality uses the compactness of $\X$. 

By \citet[Theorem 1]{GuntuboyinaSen2013L1}, there exists $\epsilon_0 > 0$ such that when $\epsilon \leq  \epsilon_0(B + 2)$, 
\begin{equation}
\log N \left(\epsilon, \cF_{1,1}([-1,1]^\kappa), L^\infty \right)  \leq C \left(\frac{B + 2}{\epsilon} \right)^{\kappa/2}, 
\end{equation}
where $C$ is a universal constant. 

Take $B = 0 \lor (\frac{2}{\epsilon_0}  - 2)$. Combining the above three displays, we have
\begin{equation}
\begin{aligned}
\E{\rmp \otimes \rmq}{\Delta_{n,m}} &\leq 2 \tau_n + \frac{4 \sqrt{2}}{\sqrt{n}} \int_{\tau_n/4}^2 C \left(\frac{2}{\epsilon_0} \right)^{\kappa/4} \epsilon^{-\kappa/4} \, d \epsilon + 2 \tau_m + \frac{4 \sqrt{2}}{\sqrt{m}} \int_{\tau_m/4}^2 C\left(\frac{2}{\epsilon_0} \right)^{\kappa/4} \epsilon^{-\kappa/4} \, d \epsilon \\
& =2 \tau_n  + \frac{C'}{\sqrt{n}} \left(\frac{2}{\epsilon_0} \right)^{\kappa/4}\left|2^{1-\kappa/4} - (\tau_n/4)^{1-\kappa/4}  \right| +  2 \tau_m + \frac{C'}{\sqrt{m}}\left(\frac{2}{\epsilon_0} \right)^{\kappa/4} \left|2^{1-\kappa/4} - (\tau_m/4)^{1-\kappa/4}  \right|  \\
&\leq C'' \epsilon_0^{-\kappa/4} \left(n^{-1/2} + m^{-1/2} \right) + 2 \tau_n + 2 \tau_m + \epsilon_0^{-\kappa/4} \left(\frac{\tau_n^{1-\kappa/4}}{\sqrt{n}} + \frac{\tau_m^{1-\kappa/4}}{\sqrt{m}} \right). 
\end{aligned}
\end{equation}
Choose $\tau_n =  \epsilon_0 n^{-2/\kappa}$. Then for constants $ C_{\epsilon_0, \kappa}, C'_{\epsilon_0, \kappa} $, 
\begin{equation}
\E{\rmp \otimes \rmq}{\Delta_{n,m}}  \leq C_{\epsilon_0, \kappa} \left(n^{-1/2} + m^{-1/2} \right) + C'_{\epsilon_0, \kappa} \left(n^{-2/\kappa} + m^{-2/\kappa} \right). 
\end{equation}
The constants only depend on $\epsilon_0$ and $\kappa$, so the bound is preserved after we take the supremum over $\rmp$ and $\rmq$. Combining the bound on $ \sup_{\rmp, \rmq}\E{\rmp \otimes \rmq}{\Delta_{n,m}}$ with \Cref{eqn1:lemma:Wp-bias}, we get
\begin{equation}
\sup_{\rmp, \rmq} \E{\rmp \otimes \rmq}{\left|  \W_p^p \left(\hat \rmp^n, \hat \rmq^m \right) - \W_p^p \left( \rmp , \rmq \right) \right|} = O\left( n^{-1/2} + m^{-1/2} \right) + O\left(n^{-2/\kappa} + m^{-2/\kappa} \right).
\end{equation}
The final statement comes from the observation that the $O\left( n^{-1/2} + m^{-1/2} \right)$ term dominates when $\kappa < 4$ and the $ O\left(n^{-2/\kappa} + m^{-2/\kappa} \right)$ terms dominate when $\kappa \geq 4$. 
\end{proof}

\begin{lemma} \label{lemma:Wp-boundedness}
Assume that $\text{diam}(\cX) < \infty$. Then $\sup_{\rmp, \rmq \in \cP(\X)} \W_p^p(\rmp, \rmq)< \infty$. 
\end{lemma}

\begin{proof}[Proof of \Cref{lemma:Wp-boundedness}]
Under the definition of $\W_p$, we have $\sup_{\rmp, \rmq \in \P(\X)} \W_p^p(\rmp, \rmq) \leq \text{diam}(\cX)^p < \infty$. 
\end{proof}

We also establish the convergence rate for the sample approximation to the $\W_p$ distance under the degenerate and non-degenerate cases.

\begin{lemma}\label{lemma:W1-concentration}
Assume that the distributions $\rmp_\theta$ and $\rmp_0$ are supported on a compact subset $\X \subseteq \R^\kappa$ and have positive densities in the interior of their supports. Additionally, assume that the mapping $\theta \mapsto \E{\rmp_\theta}{\|X\|_2^{2p}}$ is continuous at $\theta_0$. Suppose that $\lim_{m,n \to \infty} \frac{m}{m + n} = c \in (0,1)$. Then, \Cref{assumption-rate} is satisfied for $\diverge_{m,n}(\rmp, \rmq) = \W_p^p(\hat \rmp^m, \hat \rmq^n)$ and $\diverge(\rmp, \rmq) = \W_p^p  (\rmp, \rmq)$ with $r_{m,n} = n^{-2/(\kappa \lor 4)} + m^{-2/(\kappa \lor 4)}$.
\end{lemma}

\begin{proof}[Proof of \Cref{lemma:W1-concentration}]
By \Cref{lemma:Wp-bias}, \Cref{assumption-rate}(a) is satisfied with $r_{m,n} = n^{-2/(\kappa \lor 4)} + m^{-2/(\kappa \lor 4)}$. Take $\cV(\theta) = 2^p \E{\rmp_\theta}{\|X\|_2^p} + 2^p \E{\rmp_0}{\|Y\|_2^p}$. From \Cref{lemma:Wp-variance}, for any $(m,n)$, we have
\begin{equation}
\var\left(\W_p^p \left(\hat \rmp^n, \hat \rmq^m \right) \right) \leq \left(\frac{1}{m} + \frac{1}{n} \right)^2 \cV(\theta).
\end{equation}
Since $\X$ is compact, $\cV(\theta_0)$ is finite and $\cV(\theta)$ is continuous at $\theta_0$ by assumption. Hence, \Cref{assumption-rate}(b) is verified with $r_{m,n} = m^{-1} + n^{-1}$. Taking $r_{m,n} = n^{-2/(\kappa \lor 4)} + m^{-2/(\kappa \lor 4)}$ provides an upper bound for both conditions.
\end{proof}

\begin{proof}[Proof of \Cref{thm-Wp}]
When $p \geq 1$, the $\W_p$ metric is a metric and continuous under the weak topology. \Cref{lemma:Wp-boundedness} implies the uniform boundedness in \Cref{assumption-semimetric}. \Cref{lemma:W1-concentration} verifies \Cref{assumption-rate} at the rate $r_{m,n} = n^{-2/(\kappa \lor 4)} + m^{-2/(\kappa \lor 4)}$. This allows us to conclude from \Cref{contraction-div} that
\begin{equation}
  \E{\param}{\W_p^p(\hat \rmp_\theta^m, \hat \rmp_0^n) \mid x_{1:n}}  -  \W_p^p \left( \rmp_{\theta_0}, \rmp_0\right) =  O_{\P_0}\left(n^{-2/(\kappa \lor 4)} + m^{-2/(\kappa \lor 4)} \right). 
\end{equation}
\end{proof}

\subsubsection{MMD} \label{sect-mmd}
We establish the convergence rate for $\E{\param}{\mmd_U^2 \left(\p_\theta^m, \p_0^n \right) \mid x_{1:n}}$, based on the following U-statistic approximation to MMD \citep{Gretton2012-do}:
\begin{equation}\label{emp-mmd}
\mmd_U^2 \left(\rmp^m, \rmq^n \right) := \frac{ \sum_{i = 1}^m \sum_{j \neq i} k(x_i, x_j)}{m(m-1)} + \frac{\sum_{i = 1}^n \sum_{j \neq i} k(y_i, y_j)}{n(n-1)} \quad - 2\frac{\sum_{i = 1}^m \sum_{j = 1}^n k(x_i, y_j)}{mn},
\end{equation}
with i.i.d samples $x_{1:m}$ drawn from $\rmp$ and $y_{1:n}$ drawn from $\rmq$. We then use this to prove \Cref{thm-gNPP} holds for the MMD, by replacing \Cref{contraction-div} with \Cref{thm-mmd} in the proof of \Cref{thm-gNPP}.
Note for large-scale applications, MMD is often preferred over Wasserstein because the computation is quadratic rather than cubic in $n$, and the choice of kernel offers additional flexibility and control. 

\begin{theorem}[Posterior expected MMD convergence rate]\label{thm-mmd}
Let \Cref{assumption-AN,assumption-semimetric,assumption-mmd,assumption-mmd2} be satisfied. Then 
\begin{equation}
\E{\param}{\mmd_U^2\left(\rmp_\theta^m, \rmp_0^n\right) \mid x_{1:n}} - \mmd^2 \left(\rmp_{\theta_0}, \rmp_0\right) = O_{\P_0}\left(n^{-1/2} + m^{-1/2}\right). 
\end{equation}
\end{theorem}
\noindent \Cref{assumption-mmd,assumption-mmd2} impose standard regularity conditions on the RKHS and the parametric model~\citep{Gretton2012-do}.

\begin{assumption} \label{assumption-mmd}
The RKHS kernel $k: \X \times \X \mapsto \R$ is a symmetric, positive semi-definite, characteristic kernel such that $\sup_{\theta \in \Theta} \E{X, X' \sim \rmp_\theta}{k^2(X, X')} < \infty$ and $\E{Y, Y' \sim \rmp_0}{k^2(Y, Y')} < \infty$. 
\end{assumption}
We require a characteristic kernel to make the MMD a valid statistical divergence \citep{Sriperumbudur2011RKHS}. The bounded second moments are sometimes referred to as the \textit{Hilbert-Schmidt condition}, which is a sufficient assumption for the eigendecomposition of the kernel operator. Examples of kernels that satisfy \Cref{assumption-mmd} include the Gaussian kernel $k(x, y) = \exp\left(- \|x - y\|_2^2/\gamma^2 \right) $ and the Laplace kernel $k(x, y) = \exp\left(- \|x - y\|_2/\gamma \right)$.

\begin{assumption} \label{assumption-mmd2}
The mapping $\theta \mapsto \E{X, X' \sim \rmp_\theta}{k^2(X, X')}$ is continuous at $\theta_0$.  
\end{assumption}

This assumption requires that the parametric model behaves smoothly around $\theta_0$, with respect to the RKHS geometry. In particular, it requires that the mapping $\theta \mapsto \int k^2(x,x')\,\rmp_\theta(x)\rmp_\theta(x')\,dx\,dx'$ is continuous at $\theta_0$. This holds, for example, for models with densities $\rmp_\theta$ that depend continuously on $\theta$, such as $\p_\theta=\cN(\theta,1)$.

\paragraph{Proofs}
\begin{lemma}[RKHS-Cauchy-Schwarz] \label{RKHS-CS}
    For any $x, y \in \X$, $k^2(x, y) \leq k(x, x) k(y,y)$. 
\end{lemma}
\begin{proof}
Under the RKHS formalism, 
\begin{equation}
    k^2(x, y)  = \langle k(x, \cdot), k(y, \cdot) \rangle^2 \leq  \langle k(x, \cdot), k(x, \cdot) \rangle^2 \langle k(y, \cdot), k(y, \cdot) \rangle^2 = k(x,x) k(y,y). 
\end{equation}
\end{proof}
The following lemma establishes the convergence rate of the empirical MMD, applying the results of \cite{Gretton2012-do}. 
\begin{lemma} \label{lemma:mmd-concentration}
Let \Cref{assumption-mmd,assumption-mmd2} be satisfied. Then \Cref{assumption-rate} is satisfied for $\diverge_{m,n}(\rmp, \rmq) = \mmd_U^2(\rmp^m, \rmq^n)$ and $\diverge(\rmp, \rmq) = \mmd^2(\rmp, \rmq)$ with $r_{m,n} = m^{-1/2} + n^{-1/2}$. 
\end{lemma}

\begin{proof} [Proof of \Cref{lemma:mmd-concentration}]
Since $\diverge_{m,n}(\rmp, \rmq)$ is an unbiased estimate of $\diverge(\rmp, \rmq)$, \Cref{assumption-rate}(a) is satisfied for any positive sequence $r_{m,n} \to 0$.    For \Cref{assumption-rate}(b), we need to bound the variance of the empirical MMD. First, by \Cref{emp-mmd}, we have
\begin{equation} \label{eqn-mmd-proof:1}
\begin{aligned}
\var\left(\mmd_U^2(\rmp^m, \rmq^n) \right) & \leq 2  \var_{X_i \iid \rmp}\left(\frac{ \sum_{i = 1}^m \sum_{j \neq i} k(X_i, X_j)}{m(m-1)} \right) + 2\var_{Y_i \iid \rmq}\left(\frac{\sum_{i = 1}^n \sum_{j \neq i} k(Y_i, Y_j)}{n(n-1)} \right) \\
&+ 8 \var_{X_i \iid \rmp, Y_i \iid \rmq}\left(\frac{\sum_{i = 1}^m \sum_{j = 1}^n k(X_i, Y_j)}{mn} \right). 
\end{aligned}
\end{equation}
The last term is simply the variance of a sum of i.i.d. variables, thus
\begin{equation}
  \var_{X_i \iid \rmp, Y_i \iid \rmq}\left(\frac{\sum_{i = 1}^m \sum_{j = 1}^n k(X_i, Y_j)}{mn} \right) \leq  \frac{\E{X \sim \rmp, Y \sim \rmq}{k^2(X, Y)}}{mn}. 
\end{equation}
By \Cref{RKHS-CS}, 
\begin{equation}
     \var_{X_i \iid \rmp, Y_i \iid \rmq}\left(\frac{\sum_{i = 1}^m \sum_{j = 1}^n k(X_i, Y_j)}{mn} \right) \leq \frac{\E{X \sim \rmp}{k(X, X)} \E{Y \sim \rmq}{k(Y,Y)}}{mn}. 
\end{equation}

The first two terms in \Cref{eqn-mmd-proof:1} are variances of one-sample U-statistics. By \citet[Section 5.2.1, Lemma A]{Serfling2009}, we have the following bound:
\begin{equation}
    \var_{X_i \iid \rmp}\left(\frac{ \sum_{i = 1}^m \sum_{j \neq i} k(X_i, X_j)}{m(m-1)} \right) \leq \frac{2 \var_{X, X' \sim \rmp}\left(k(X, X')\right)}{m} \leq \frac{2 \E{X, X' \sim \rmp}{k^2(X, X')}}{m}. 
\end{equation}
Analogously, we have
\begin{equation}
    \var_{Y_i \iid \rmq}\left(\frac{\sum_{i = 1}^n \sum_{j \neq i} k(Y_i, Y_j)}{n(n-1)} \right) \leq  2 \frac{\E{Y, Y' \sim \rmq}{k^2(Y, Y')}}{n}. 
\end{equation}
Putting the bounds together and applying Jensen's inequality with some algebra, we obtain a simple bound without the cross term,
\begin{equation} \label{eqn-mmd-proof:2}
    \var\left(\mmd_U^2(\rmp^m, \rmq^n) \right) \leq \frac{8 \E{X, X' \sim \rmp}{k^2(X, X')}}{m} + \frac{8 \E{Y, Y' \sim \rmq}{k^2(Y, Y')}}{n}. 
\end{equation}
Define 
\begin{equation}
    \cV(\theta) := 8 \left( \E{X, X' \sim \rmp_\theta}{k^2(X, X')} +  \E{Y, Y' \sim \rmp_0}{k^2(Y, Y')}\right). 
\end{equation}
The function satisfies $\cV(\theta_0) < \infty$ by \Cref{assumption-mmd} and $\cV(\theta)$ is continuous at $\theta_0$ by \Cref{def-mmd2}. Then \Cref{eqn-mmd-proof:2} implies the bound,
\begin{equation}
     \var\left(\mmd_U^2(\rmp_\theta^m, \rmp_0^n) \right)   \leq \left(\frac{1}{m} + \frac{1}{n} \right)\cV(\theta) \leq r_{m,n}^2 \cV(\theta). 
\end{equation}
for $r_{m,n} = m^{-1/2} + n^{-1/2}$. 
\end{proof}
\begin{lemma}\label{lemma:MMD-boundedness}
Let \Cref{assumption-mmd} be satisfied.  Then $ \sup_{\theta \in \Theta} \mmd_U^2(\rmp_\theta^m, \rmp_0^n) < \infty$.
\end{lemma}
\begin{proof}
This follows directly from the following representation for the MMD \citep{Gretton2012-do}:
\begin{equation} \label{def-mmd2}
\mmd^2 \left(\rmp, \rmq\right) =  \E{X, X' \sim \rmp}{k(X, X')} - 2\E{X \sim \rmp, Y \sim \rmq}{k(X, Y)}  + \E{Y, Y' \sim \rmq}{k(Y, Y')}. 
\end{equation}
By \Cref{RKHS-CS}, we have
\begin{equation}
    \sup_{\theta \in \Theta} \mmd_U^2(\rmp_\theta^m, \rmp_0^n) \leq 2 \sup_{\theta \in \Theta}  \E{X, X' \sim \rmp_\theta}{k(X, X')} +2 \E{Y, Y' \sim \rmp_0}{k(Y, Y')} < \infty. 
\end{equation} 
Since the expectation of $k^2(x, x')$ is uniformly bounded in each case, $ \mmd^2 \left( \cH_k, \rmp_\theta, \rmp_0 \right)$ is uniformly bounded by Jensen's inequality. 
\end{proof}
\begin{proof}[Proof of \Cref{thm-mmd}]
The divergence $\mmd_U^2 \left(\rmp_\theta^m, \rmp_0^n\right)$  is slightly different from the MMD between empirical distributions, $\mmd^2 \left(\rmp_\theta^m, \rmp_0^n\right)$. Only the latter applies in the setting of \Cref{contraction-div}. But the difference between them is negligible. By Lemma 2 of \cite{Briol2019}, however, this difference is bounded by a factor of $m^{-1} + n^{-1}$, 
\begin{equation} \label{Lemma-Barp2019MMD-2}
\sup_{\theta \in \Theta} \left| \mmd_U^2\left(\rmp_\theta^m, \rmp_0^n\right)   - \mmd \left(\rmp_\theta^m, \rmp_0^n\right) \right| \leq  2 (m^{-1} + n^{-1}) \sup_{x, x' \in \X} k(x, x') = O(m^{-1} + n^{-1}). 
\end{equation}

Using this fact, we have
\begin{equation}
\begin{aligned}
&\E{\param}{\mmd_U^2\left(\rmp_\theta^m, \rmp_0^n\right) \mid x_{1:n}} -  \E{\param}{\mmd^2\left(\rmp_\theta^m, \rmp_0^n\right) \mid x_{1:n}} \\
&\leq \sup_{\theta \in \Theta} \left|\mmd_U^2\left(\rmp_\theta^m, \rmp_0^n\right)  - \mmd^2\left(\rmp_\theta^m, \rmp_0^n \right) \right| = O(m^{-1} + n^{-1}). 
\end{aligned}
\end{equation}
$\mmd$ is a pseudometric and continuous under the weak topology. \Cref{lemma:MMD-boundedness} implies the uniform boundedness in \Cref{assumption-semimetric}. \Cref{lemma:W1-concentration} verifies \Cref{assumption-rate} with a rate of $r_{m,n} = n^{-1/2} + m^{-1/2}$. This allows us to conclude that
\begin{equation}
\E{\param}{\mmd_U^2\left(\rmp_\theta^m, \rmp_0^n \right) \mid x_{1:n}}- \mmd^2 \left( \rmp_{\theta_0}, \rmp_0\right) = O_{\P_0}\left(n^{-1/2} + m^{-1/2}\right). 
\end{equation}
Combining the two displays above yields the desired convergence rate.
\end{proof}

\begin{remark} \label{remark-mmd-synthetic}
Consider the model in \Cref{sec-synthetic} with the MMD induced by an inverse multiquadric (IMQ) kernel. The parametric model is the normal mean model $\P_\theta=\mathcal{N}(\theta,1)$. In this case, we can use \Cref{thm-mmd} after verifying \Cref{assumption-AN,assumption-semimetric,assumption-mmd,assumption-mmd2}. The MMD is uniformly bounded by a fixed constant, so \Cref{assumption-semimetric}(a) is satisfied. Taking the reference measure $U=\mathcal{N}(0,1)$, the transport map from $U$ to $\P_\theta$ is $T_\theta(u)=u+\theta$, 
which is linear in $\theta$ and therefore has uniformly bounded first and second derivatives. Together with the discussion right after \Cref{assumption-semimetric}, this verifies \Cref{assumption-semimetric}(b). \Cref{assumption-AN} holds because the normal mean model is asymptotically normal in the usual sense. For \Cref{assumption-rate}, this is verified in \Cref{lemma:mmd-concentration}: the IMQ kernel is bounded, symmetric, positive semidefinite, and characteristic, and the mapping $\theta \mapsto \int k^2(x,x')\,\rmp_\theta(x)\rmp_\theta(x')\,dx\,dx'$
is continuous in $\theta$. By \Cref{thm-mmd}, the MMD achieves the rate $O_{\P_0}(n^{-1/2}+m^{-1/2})$; hence, if $m=n$, any $r<1/2$ is admissible and the generalized Bayes factor based on the MMD achieves consistency.
\end{remark}
\subsubsection{KSD} \label{sect-KSD}
We establish the convergence rate for $\E{\param}{\ksd_U(\rmp_0^n, \rmp_\theta) \mid x_{1:n}}$, where $\ksd_U$ is a one-sample U-statistic \citep{Liu2016}: 
\begin{equation}\label{empirical-ksd}
\ksd_U \left(\rmp^n, \rmq \right) := \frac{1}{n(n-1)} \sum_{1 \le i \neq j \leq n} u_{\rmq, k}(x_i, x_j), 
\end{equation}
where $x_{1:n} \iid \rmp$ and $u_{\rmq, k}(x, x') := \nabla \log \rmq(x) k(x, x') \nabla \log \rmq(x')+ 2 \nabla \log \rmq(x)^\top \nabla_{x'} k(x,x')  + \tr(\nabla_{x,x'} k(x, x'))$. 
This implies that \Cref{thm-gNPP} holds for the KSD, by replacing \Cref{contraction-div} with \Cref{thm-ksd} in the proof of \Cref{thm-gNPP}.
\begin{theorem}[Posterior expected KSD convergence rate] \label{thm-ksd}
Let \Cref{assumption-AN,assumption-KSD,assumption-KSD2,assumption-KSD3} be satisfied. As $n \to \infty$, $\E{\param}{\ksd_U(\rmp_0^n, \rmp_\theta) \mid x_{1:n}}$ converges in $[\P_0^\infty]$-probability to $\ksd(\rmp_0, \rmp_{\theta_0}) $ at the rate of
\begin{equation}
\E{\param}{\ksd_U(\rmp_0^n, \rmp_\theta) \mid x_{1:n}} = \ksd(\rmp_0, \rmp_{\theta_0}) + O_{\P_0}(n^{-1/2}).
\end{equation}    
\end{theorem}
\Cref{assumption-KSD,assumption-KSD2,assumption-KSD3} involve regularity conditions on the kernel and the parametric model.
Recall that $\Delta_{\rmq, \rmp}(x) := \nabla_x \log \rmp(x) - \nabla_x \log \rmq(x)$
\begin{assumption} \label{assumption-KSD}
The kernel $k(x, x’)$ is symmetric, integrally positive definite, uniformly bounded, and belongs to the Stein class of all continuous densities $\rmp_0$ and $\rmp_\theta$. Additionally, $\sup_{\theta \in \Theta}\E{X \sim \rmp_0}{\left\|\Delta_{\rmp_0, \rmp_\theta}(X)\right\|_2^2} < \infty$.
\end{assumption}
\noindent The first part of \Cref{assumption-KSD} is satisfied when the tail of $\rmp_0$ decays exponentially and $k$ is positive definite, characteristic and uniformly bounded. The Stein class requirement ensures that KSD operates as a valid statistical divergence \citep{Liu2016}. 
For example, the squared-exponential kernel $k(x, y) = \exp\left(- \|x - y\|_2^2/\gamma^2 \right)$ belongs to the Stein class for smooth densities on $\R^d$. 

The second part of \Cref{assumption-KSD} requires that the relative Fisher information between $\rmp_0$ and $\rmp_\theta$ be uniformly bounded over $\theta\in\Theta$.  For example, if $\p_0=\mathcal{N}(\theta_0,1)$ and $\p_\theta=\mathcal{N}(\theta,1)$, then the relative Fisher information scales as $O(\|\theta-\theta_0\|_2^2)$ in $\theta$. In this case, the second part of \Cref{assumption-KSD} is satisfied when $\Theta$ is compact. 

We also need assumptions on the regularity of the KSD. 
\begin{assumption}\label{assumption-KSD2}
There exists an $M_n n^{-1/2}$ neighborhood of $\theta_0$ for some $M_n \to \infty$ such that
\begin{enumerate}
\item[(a)] the mapping  $\theta \mapsto \E{X, X' \sim \rmp_0}{u_{\rmp_\theta, k}(X, X')}$ is twice differentiable with bounded Hessian (in the $L_2$ sense).
\item[(b)] the mapping $\theta \mapsto \var_{X, X' \sim \rmp_0}\left[u_{\rmp_\theta, k}(X, X')\right]$ is bounded and continuous at $\theta_0$.
\end{enumerate}
\end{assumption}
\noindent When $\rmp_0$ and $\rmp_\theta$ are smooth densities, \Cref{assumption-KSD2}(a) is equivalent to stating that $\theta \mapsto \ksd(\rmp_0, \rmp_\theta)$ is twice differentiable with a bounded Hessian in a shrinking neighborhood of $\theta_0$.  \Cref{assumption-KSD2}(b) is satisfied when the parametric model is regular at $\theta_0$. 

To ensure a notion of boundedness for the empirical KSD, we require an additional assumption:
\begin{assumption} \label{assumption-KSD3}
We assume that $\sup_{x, x' \in \R}k(x, x') < \infty$ and $\ksd_U(\rmp_0^n, \rmp_\theta) \asto \ksd(\rmp_0, \rmp_\theta)$ uniformly in $\theta$. 
\end{assumption}
\noindent The first part of \Cref{assumption-KSD3} requires $k$ to be uniformly bounded, which holds for e.g. Gaussian or Laplace kernels. The second part of \Cref{assumption-KSD3} requires more effort to verify; it essentially involves establishing a uniform law of large numbers (ULLN) for $\ksd_U(\rmp_0^n, \rmp_\theta)$ across all $\theta \in \Theta$, as done in \citet{Barp2019MMD}.

Sufficient conditions for \Cref{assumption-KSD3} includes (1) the domination of $u_{\rmp_\theta, k}(x, x’)$ by an integrable and symmetric kernel $g(x, x’)$, and (2) the existence of a sequence of sets where the mappings $\theta \mapsto \E{X’ \sim \rmp_0}{u_{\rmp_\theta, k}(x, X’)}$ and $\theta \mapsto u_{\rmp_\theta, k}(x, x’)$ are equicontinuous for all $x \in \X$ and all $(x,x’) \in \X \times \X$, respectively \citep{Yeo2001}. Alternatively, it suffices that the function class $\{u_{\P_\theta, k}\}_{\theta \in \Theta}$ is $[\P_0 \times \P_0]$-Glivenko-Cantelli. 
\paragraph{Proofs}
\begin{lemma} \label{lemma:KSD-bias}
Let \Cref{assumption-AN,assumption-KSD,assumption-KSD2} be satisfied. Then $\E{\param}{\ksd \left(\rmp_0, \rmp_\theta \right)  \mid x_{1:n}} =\ksd \left(\rmp_0, \rmp_{\theta_0} \right) + O_{\P_0}(n^{-1})$. 
\end{lemma}
\begin{proof}[Proof of \Cref{lemma:KSD-bias}]
Let $B_{M_n n^{-1/2}} (\theta_0)$ be the neighborhood defined in \Cref{assumption-KSD2}. Let $E_n$ be the event that $\hat \theta_{\mle} \in B_{M_n n^{-1/2}} (\theta_0)$.  By \Cref{assumption-AN}(a), $\P_0(E_n) \to 1$ as $n \to \infty$. For the rest of the proof, we condition on $E_n$.

Let $S_\theta := \nabla_\theta \ksd(\rmp_\theta, \rmp_{\theta_0}), H_\theta := \nabla_\theta^2 \ksd(\rmp_\theta, \rmp_{\theta_0})$.  By \Cref{assumption-KSD2}, there exists $L > 0$ such that
\begin{equation}
    \sup_{\theta \in B_{M_n n^{-1/2}} (\theta_0)} \| H_\theta\|_2 \leq L. 
\end{equation}

By Taylor expanding $\ksd(\rmp_0, \rmp_\theta)$ around the MLE,  we have
\begin{equation}
    \ksd(\rmp_0, \rmp_\theta) = \ksd(\rmp_0, \rmp_{\hat \theta_{\mle}}) + S_{\hat \theta_{\mle}}^T (\theta - \hat \theta_{\mle}) + (\theta - \hat \theta_{\mle})^T H_{\hat \theta_{\mle}}(\theta - \hat \theta_{\mle}) + o(\|\theta - \hat \theta_{\mle}\|_2^2). 
\end{equation}
By the bonded Hessian condition, we have a sandwich inequality, 
\begin{equation}
\left| \ksd(\rmp_0, \rmp_\theta)   - \ksd(\rmp_0, \rmp_{\hat \theta_{\mle}}) \right| \leq  L \|\theta - \hat \theta_{\mle}\|_2^2 +  o(\|\theta - \hat \theta_{\mle}\|_2^2). 
\end{equation}
Applying \Cref{assumption-AN}(b) to the posterior expectation of the above display, we have
\begin{equation}
\E{\param}{\ksd(\rmp_0, \rmp_\theta)  \mid x_{1:n}} = \ksd(\rmp_0, \rmp_{\hat \theta_{\mle}})  + O_{\P_0}(n^{-1}). 
\end{equation}

For $ \ksd(\rmp_0, \rmp_{\hat \theta_{\mle}})$, a Taylor expansion around $\theta_0$ yields, 
\begin{equation}
\begin{aligned}
\ksd(\rmp_0, \rmp_{\hat \theta_{\mle}}) & =  \ksd(\rmp_0, \rmp_{\theta_0}) + S_{\tilde \theta}^T (\hat \theta_{\mle} - \theta_0) \\
&= \ksd(\rmp_0, \rmp_{\theta_0})  + L O_{\P_0}(\|\hat \theta_{\mle} - \theta_0\|_2^2) =  \ksd(\rmp_0 \rmp_{\theta_0}) +  O_{\P_0}(n^{-1}).  
\end{aligned}
\end{equation}
Combining the two displays above yields the desired result. 
\end{proof}
\begin{lemma}\label{lemma:KSD-bounded}
Let \Cref{assumption-KSD} be satisfied. Assume that $\sup_{x, x' \in \X}k(x,x') < \infty$.  Then $\sup_{\theta \in \Theta} \ksd \left( \rmp_0, \rmp_\theta\right)< \infty$. 
\end{lemma}
\begin{proof}[Proof of \Cref{lemma:KSD-bounded}]
From its definition, we can upper bound the KSD by Cauchy-Schwarz, 
\begin{equation}
    \ksd(\rmp_0, \rmp_\theta) = \E{X, X' \sim \rmp_0}{\Delta_{\rmp_0, \rmp_\theta}(X)^T k(X, X') \Delta_{\rmp_0, \rmp_\theta}(X')} \leq \sup_{x,x' \in \X}\left\| k(x, x') \right\|_2 \E{X \sim \rmp_0}{\left\|\Delta_{\rmp_0, \rmp_\theta}(X)\right\|_2^2}.  
\end{equation}
Since $k(a, b) < C$ for all $a, b \in \R$, for $x, x' \in \X$,  $\|k(x, x') \|_2 \leq \tr(k(x, x')) \leq \kappa C$ which uses the assumption that $k(x, x') $ is positive definite.  Since $\E{X \sim \rmp_0}{\left\|\Delta_{\rmp_0, \rmp_\theta}(X)\right\|_2^2}$ is uniformly bounded over $\Theta$ by \Cref{assumption-KSD}, $\ksd(\rmp_0, \rmp_\theta)$ is uniformly bounded. 
\end{proof}
\begin{lemma}\label{lemma:KSD-rate}
Let \Cref{assumption-KSD,assumption-KSD2} be satisfied. Then \Cref{assumption-rate} is satisfied for $\diverge_n(\rmp, \rmq) = \ksd_U(\rmp^n, \rmq)$ and $\diverge(\rmp, \rmq) = \ksd(\rmp, \rmq)$ with $r_n = n^{-1/2}$. 
\end{lemma}
\begin{proof}[Proof of \Cref{lemma:KSD-rate}]
\cite[Theorem 3.6]{Liu2016} shows that under \Cref{assumption-KSD}, $\ksd_U\left(\rmp_0^n, \rmp_\theta \right) $ is a valid U-statistic for $\ksd(\rmp_0, \rmp_\theta)$, 
\begin{equation}
    \E{}{\ksd_U\left(\rmp_0^n, \rmp_\theta \right)} = \ksd(\rmp_0, \rmp_\theta). 
\end{equation} 
Thus  \Cref{assumption-rate}(a) holds true for any $r_n$. 

To verify \Cref{assumption-rate}(b), we apply the variance bound for one-sample U-statistics \cite[Section 5.2.1, Lemma A]{Serfling2009} to obtain:
\begin{equation} \label{KSD-variance-bound}
\var\left(\ksd_U\left(\rmp_0^n, \rmp_\theta \right)  \right) \leq \frac{2 \var_{X, X' \sim \rmp_0}\left(u_{\rmp_\theta, k}(X, X') \right)}{n} \leq \frac{2 \E{X, X' \sim \rmp_0}{u_{\rmp_\theta, k}^2(X, X')}}{n}. 
\end{equation}
Set $\cV(\theta) = \E{X, X' \sim \rmp_0}{u_{\rmp_\theta, k}^2(X, X')}$  and $r_n = n^{-1/2}$ provides the desired result.  
\end{proof}

\begin{proof}[Proof of \Cref{thm-ksd}]
By the triangle inequality, we have
\begin{equation}
\begin{aligned}
& \E{\param}{\ksd_U(\rmp_0^n, \rmp_\theta) \mid x_{1:n}} - \ksd \left( \rmp_0, \rmp_{\theta_0}\right) \\
&\leq \underbrace{\E{\param}{\ksd_U\left(\rmp_0^n, \rmp_\theta \right) - \ksd(\rmp_0, \rmp_\theta) \mid x_{1:n}} }_{A_n} + \underbrace{\E{\param}{\ksd \left(\rmp_0, \rmp_\theta \right) \mid x_{1:n}}  - \ksd \left(\rmp_0, \rmp_{\theta_0} \right)}_{R_n}. 
\end{aligned}
\end{equation}
\Cref{lemma:KSD-bias} proved that $R_n = O_{\P_0}(n^{-1})$. To bound $A_n$, we bound its expectation and variance with respect to the randomness in $\ksd_U(\rmp_0^n,\rmp_\theta)$

By Fubini's theorem, we exchange the integrals and apply the unbiasedness of U-statistics:
\begin{equation}
 \E{}{A_n} = \E{\param}{\E{}{\ksd_U\left(\rmp_0^n, \rmp_\theta \right) - \ksd(\rmp_0, \rmp_\theta)}  \mid x_{1:n}} = 0. 
\end{equation}
For the variance, we have
\begin{equation}
    \var\left[A_n \right]    = \E{\param}{\var\left[\ksd_U\left(\rmp_0^n, \rmp_\theta \right) \right] \mid x_{1:n}} \leq 2n^{-1}  \E{\param}{\var_{X, X' \sim \rmp_0}\left[u_{\rmp_\theta, k}(X, X') \right]  \mid x_{1:n}}. 
\end{equation}
By \Cref{assumption-KSD2}, the posterior expectation $\E{\param}{\var_{X, X' \sim \rmp_0}\left[u_{\rmp_\theta, k}(X, X') \right]  \mid x_{1:n}}$ is bounded in probability.  Thus, by Chebyshev's bound, we conclude that $A_n = O_{\P_0}(n^{-1/2})$. 

Let $M_n$ be a sequence such that $M_n \to \infty$ and  $M_n = o(n^{1/2})$. 
To control the fluctuation of $\E{\param}{\ksd_U\left(\rmp_0^n, \rmp_\theta \right) \mid x_{1:n}}$ around $\ksd(\rmp_0, \rmp_\theta)$, we decompose it into two terms inside and outside of a ball of radius $M_n n^{-1/2}$:
\begin{equation}
\begin{aligned}
&\E{\param}{\ksd_U\left(\rmp_0^n, \rmp_\theta \right) - \ksd(\rmp_0, \rmp_\theta) \mid x_{1:n}} =  \underbrace{\E{\param}{\left(\ksd_U\left(\rmp_0^n, \rmp_\theta \right) - \ksd(\rmp_0, \rmp_\theta) \right) I_{B_{M_n n^{-1/2}}^c(\theta_0)} \mid x_{1:n}}}_{A_n}    \\
&+ \underbrace{\E{\param}{\left( \ksd_U\left(\rmp_0^n, \rmp_\theta \right) - \ksd(\rmp_0, \rmp_\theta) \right) I_{B_{ M_n n^{-1/2}}(\theta_0)} \mid x_{1:n}}}_{B_n}. 
\end{aligned}
\end{equation}
To bound $A_n$, we use the ULLN for $\ksd_U\left(\rmp_0^n, \rmp_\theta \right)$. By Holder's inequality, we have
\begin{equation}
\begin{aligned}
A_n &\leq \sup_{\theta \in \Theta} \left|\ksd_U\left(\rmp_0^n, \rmp_\theta \right) - \ksd(\rmp_0, \rmp_\theta)  \right| \Pi_{\param}\left(\| \theta - \theta_0 \|_2 > M_n n^{-1/2} \mid x_{1:n}\right) \\
&= o_{\P_0}(1) \Pi_{\param}\left(\| \theta - \theta_0 \|_2 > M_n n^{-1/2} \mid x_{1:n}\right).   
\end{aligned}
\end{equation}
Then we bound the posterior probability using \Cref{assumption-AN}.  
\begin{equation}
\begin{aligned}
& \Pi_{\param}\left(\| \theta - \theta_0 \|_2 > M_n n^{-1/2} \mid x_{1:n}\right) \\
&\leq  \Pi_{\param}\left(\sqrt{n}\| \theta - \hat \theta_\mle \|_2 > M_n  - \sqrt{n} \| \theta_0 - \hat \theta_\mle \|_2 \mid x_{1:n}\right) \\
&\leq\inf_\eta \exp\left( -\eta (M_n  - \sqrt{n} \| \theta_0 - \hat \theta_\mle \|_2 )\right)\E{\param}{ \exp\left( \eta \sqrt{n}\| \theta - \hat \theta_\mle \|_2  \right) \mid x_{1:n}} \\
&= O_{\P_0}\left(\exp \left(- C’ M_n \right) \right),
\end{aligned}
\end{equation}
for some constants $C$ and $C'$. The third line applies Hoeffding's inequality. The fourth line uses the fact that weak convergence implies the convergence of characteristic functions (to a sub-Gaussian limit). Thus, if we take $M_n = n^{1/3}$, we get $A_n =   O_{\P_0}\left(n^{-1} \right)$.

To bound $B_n$, we analyze its mean and variance with respect to the randomness in $\ksd_U \left(\rmp^n, \rmq \right)$. To control the mean, we apply Fubini's theorem
\begin{equation}
\begin{aligned}
\E{}{B_n} &=\E{\param}{\E{}{ \ksd_U\left(\rmp_0^n, \rmp_\theta \right) - \ksd(\rmp_0, \rmp_\theta)} I_{B_{ M_n n^{-1/2}}(\theta_0)} \mid x_{1:n}} =  0. 
\end{aligned}
\end{equation}

Now we control the variance. Some calculation shows
\begin{equation}
\begin{aligned}
\var \left(B_n \right)  =   \E{\param}{\var \left( \ksd_U\left(\rmp_0^n, \rmp_\theta \right)  \right) I_{B_{ M_n n^{-1/2}}(\theta_0)} \mid x_{1:n}}. 
\end{aligned}
\end{equation}
Let $\cV(\theta)= \var_{X, X' \sim \rmp_0}\left[u_{\rmp_\theta, k}(X, X') \right]$. 
By \Cref{KSD-variance-bound}, we refine the upper bound as follows
\begin{equation}
\var \left(B_n\right)  \leq 2 n^{-1}  \sup_{\theta \in \bar B_{ M_n n^{-1/2}}(\theta_0)} \cV(\theta). 
\end{equation}
As $n \to \infty$, $\sup_{\theta \in \bar{B}_{M_n n^{-1/2}}(\theta_0)} \cV(\theta) \to \cV(\theta_0)$ by the continuity of $\cV$ at $\theta_0$, thus $B_n = O(n^{-1/2})$ by Chebyshev's inequality.

Combining the bounds, we obtain
\begin{equation}
\E{\param}{\ksd_U(\rmp_0^n, \rmp_\theta) \mid x_{1:n}} = \ksd(\rmp_0, \rmp_{\theta_0}) + O_{\P_0}(n^{-1/2}).
\end{equation}
\end{proof}

\section{Details on Empirical Studies} \label{sec:empirical_apx}

\subsection{gNPP}

%Another alternative is to estimate the response surface directly. However, this approach is primarily effective when the treatment is continuous. In our case, since the treatment is count-valued, modeling a response surface with count-valued inputs becomes challenging to interpret and compare. This complexity makes it preferable to have a single scalar summary of the effect. 

\paragraph{Semiparametric model.} The semiparametric model in the gNPP follows the model proposed by \citet{Hahn2020} for causal inference, with an additional transformation step to account for non-normality. The model is,
\begin{equation} \label{BART}
    y_i = T_\lambda^{-1}(\mu(w_i, \hat{a}(w_i)) + \tau a_i + \epsilon_i), \quad \epsilon_i \sim \mathcal{N}(0, \sigma^2)
\end{equation}
where:
\begin{itemize}
    \item The function $T_\lambda(\cdot)$ is the Yeo-Johnson transformation~\citep{Yeo2000}. It corrects for non-normality in the outcome expression level.
    \item The function $\mu(\cdot, \cdot)$ is a sum of piecewise constant regression trees. We place a BART prior on $\mu$, following previous work in Bayesian causal inference~\citep{Hill2011}.
    \item The function $\hat{a}(w_i)$ is a propensity model, obtained by training a five-layer neural network to predict $a_i$ from $w_i$ under a mean-squared loss. 
    \citet{Hahn2020} show that including $\hat{a}(\cdot)$ in Bayesian causal inference reduces estimator bias.
    \item The coefficient $\tau$ determines the effect of the treatment. We place an improper flat prior on $\tau$. For simplicity we assume the (transformed) outcome depends linearly on $a_i$ even though it may depend nonlinearly on $w_i$.
    \item We place a half-Normal prior on the variance $\sigma$.
\end{itemize}
The Yeo-Johnson transformation, $T_\lambda$, is a monotonic function used to reduce skewness and approximate normality \citep{Yeo2000}. The parameter $\lambda$ is fit via maximum likelihood. The transformation is defined as:
\begin{equation*}
    T_\lambda(y) = 
\begin{cases} 
\frac{((y + 1)^\lambda - 1)}{\lambda}, & \text{if } y \geq 0, \lambda \neq 0, \\
\log(y + 1), & \text{if } y \geq 0, \lambda = 0, \\
\frac{-((-y + 1)^{2 - \lambda} - 1)}{2 - \lambda}, & \text{if } y < 0, \lambda \neq 2, \\
-\log(-y + 1), & \text{if } y < 0, \lambda = 2.
\end{cases}
\end{equation*}

In the model, we assign a BART prior to $\mu$. The BART function is represented as a sum of piecewise constant binary regression trees. Each tree $T_l$ partitions the covariate space $\cA \times \X$, with each partition element $A_b$ assigned a parameter $m_{lb}$. The function $g_l(x)$ takes the value $m_{lb}$ if $x \in A_b$ and $0$ otherwise. The overall function is then $\mu(x) = \sum_{l=1}^L g_l(x)$.

We use 50 trees, each constrained by a prior that favors small trees and leaf parameters near zero, making them “weak learners.” The prior specification follows that of \cite{Chipman2010}, where the probability of a node splitting at depth $h$ is $\eta(1 + h)^{-\beta}$ with $\eta \in (0, 1)$ and $\beta \in [0, \infty)$. The splitting variable and cut-point are chosen uniformly at random. Large trees have low prior probability, with typical values $\eta = 0.95$ and $\beta = 2$. Leaf parameters $m_{lb}$ follow independent priors $\cN(0, \sigma_m^2)$, where $\sigma_m = \sigma_0 / \sqrt{L}$. The induced marginal prior for $\mu(a, w)$ is centered at zero, with $95\%$ of the prior mass within $\pm 2\sigma_0$. 

\begin{figure}[h]
\centering
\includegraphics[width=0.95\textwidth]{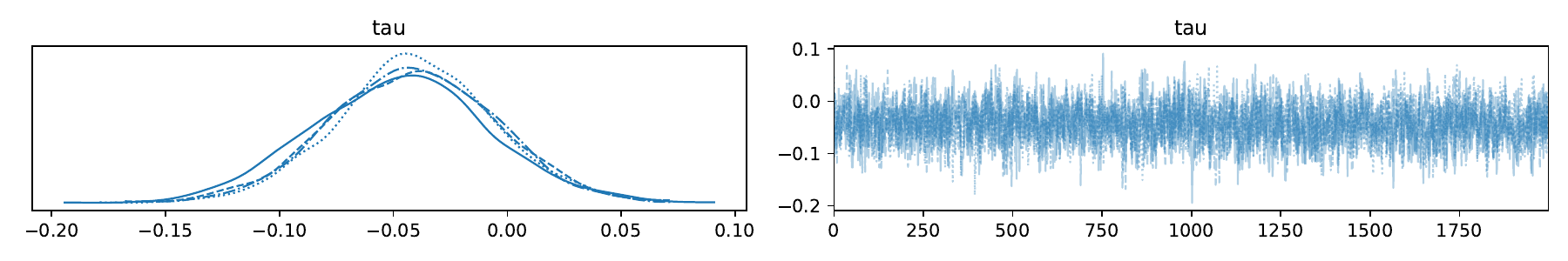}
\caption{Traceplots of the BART model MCMC inference for the effect of FOXP3 (treatment) on GZMH (outcome), corresponding to the parameter $\tau$. Different lines correspond to separate chains (4 total). } \label{fig:traceplot}
\end{figure}

We sample from the BART posterior via MCMC, using the PyMC-BART package~\citep{quiroga2023bayesian}. In general we found the chains were well-mixed, after a burn-in period of 1000 steps (\Cref{fig:traceplot}). Our reported results pool samples from four separately initialized chains.

To account for posterior uncertainty in $w_{1:n}$, we introduce a Bayesian bootstrap model for the distribution of $w_{1:n}$ \citep{Rubin1981}. The posterior of $\ATE(\p)$ is obtained as the pushforward distribution under the product posterior of the conditional distribution $\p$ and the covariate distribution $\p_w$:
\begin{equation}\label{ATE-post}
\p \sim \Pi_\textsc{BART}\left(\p \mid x_{1:n} \right), \quad \p_w \sim \text{BB}\left(\frac{1}{n} \sum_{i = 1}^n \delta_{w_i} \right),
\end{equation}
where $\p$ implicitly contains the parameters for the conditional distribution, and $\p_w$ is the distribution of $w$. We sample from the posterior of the ATE by simulating from the posterior predictive of \Cref{BART}. 

Let $\Pi_{\textsc{BART}}(\ATE\left(\p\right) \mid x_{1:n})$ be the posterior distribution of the ATE under the BART model. The gNPP posterior is then given by
\begin{equation*}
\hat \Pi\left(\ATE\left(\p\right) \mid x_{1:n} \right) = \Pi_{\param}\left(\ATE\left(\p_\theta\right) \mid x_{1:n} \right) \hat \eta_n + \Pi_{\textsc{BART}}(\ATE\left(\p\right) \mid x_{1:n}) (1 -  \hat \eta_n), 
\end{equation*}
where $\hat \eta_n$ is the generalized mixing weight based on the MMD.   

\begin{figure}[h]
\centering
\includegraphics[width=0.95\textwidth]{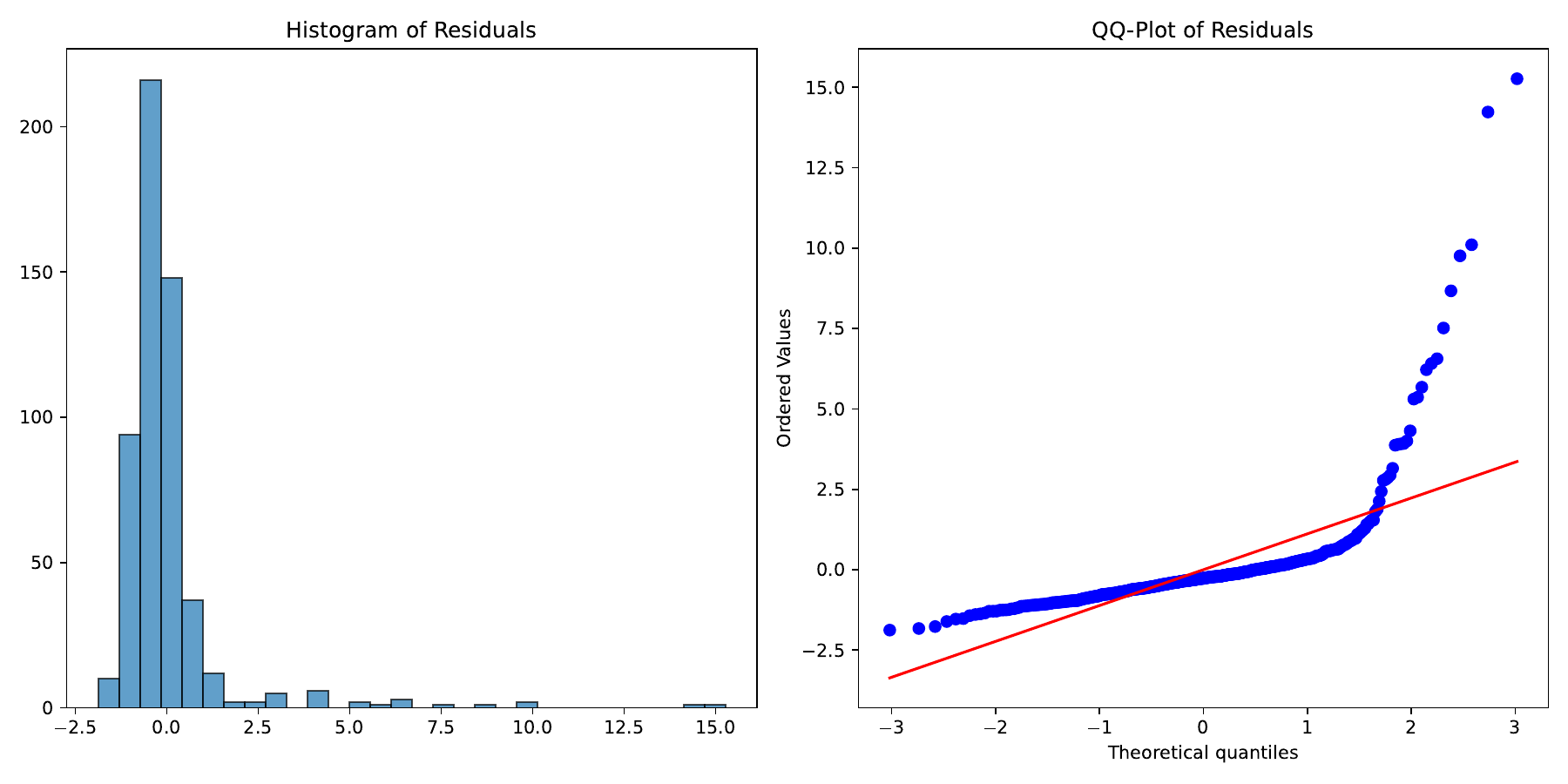}
\caption{Diagnostic plots of the causal linear model for the effect of FOXP3 (treatment) on GZMH (outcome). } \label{fig:lm-diagnostic}
\end{figure}

\paragraph{Parametric model.} The parametric model assumes a linear relationship between the target gene expression $(y)$, the treatment gene $(a)$, and the cell-type/state representation $(z)$. The parametric model is specified as:
\begin{equation} \label{app-parametric}
    \p_\theta(y \mid a, w) = \cN \left(c + \tau a + \gamma^T z, \sigma^2 \right),
\end{equation}
where $c$ is the intercept $\tau$ is the coefficient of the treatment gene $a$, $\gamma$ is a vector of coefficients for the confounding genes in $w$, and $\sigma^2$ is the noise variance.

Let $\theta \equiv [c, \tau, \gamma]$. Assuming a flat prior $\p(\theta) \propto 1$ and that the variables $c, \tau, \gamma$  are apriori independent from $a$ and $x$, the posterior distribution is given by:
\begin{equation} \label{app-pm-post-theta}
    \p \left(\theta \mid x_{1:n} \right) \propto \p_\theta(y_{1:n} \mid a_{1:n}, z_{1:n}) = \cN \left(\hat \theta, \hat V \sigma^2 \right), 
\end{equation}
where 
\begin{align}  \label{app-pm-post-notations}
\begin{bmatrix}
\hat c \\
\hat \tau \\
\hat \gamma
\end{bmatrix} &= \hat \theta \equiv (\tilde X^T \tilde X)^{-1} \tilde X^T y, \quad \hat V \equiv (\tilde X^T \tilde X)^{-1}, \quad \text{for }\tilde X \equiv \begin{bmatrix}
1 & a_1 & z_1^T \\
1 & a_2 & z_2^T \\
\vdots & \vdots & \vdots \\
1 & a_n & z_n^T
\end{bmatrix}.
\end{align}
Under the parametric model~\eqref{app-parametric}, the CATE is constant across $x$ so the posterior of $\ATE(\p)$ simplifies to 
\begin{equation*}
 \ATE\left(\p_\theta \right) = \tau (q_{98}(a) - q_0(a)), \quad \tau \sim \Pi\left(\tau \mid x_{1:n} \right). 
\end{equation*}

\Cref{app-pm-post-theta} shows that the joint posterior of $(c, \tau, \gamma)$ follows a multivariate normal distribution. After marginalization, we obtain the posterior for ATE, denoted $\Pi_{\param}\left( \ATE\left(\p_\theta \right) \mid x_{1:n} \right)$. 
\begin{equation} \label{ATE-pm-post}
\Pi_{\param}\left( \ATE\left(\p_\theta \right) \mid x_{1:n} \right) = \cN \left(\hat \tau (q_{98}(a) - q_0(a)), \hat V_{22} (q_{98}(a) - q_0(a))^2 \right), 
\end{equation}
where $\hat V_{22}$ is the second diagonal entry of $\hat V$ defined in \Cref{app-pm-post-notations}.

\subsection{Effect of TCF7 on SELL} \label{apx:tcf7_sell}

\begin{figure}[h]
\centering
\begin{subfigure}{0.49\textwidth}
    \centering
    \includegraphics[width=0.85\textwidth]{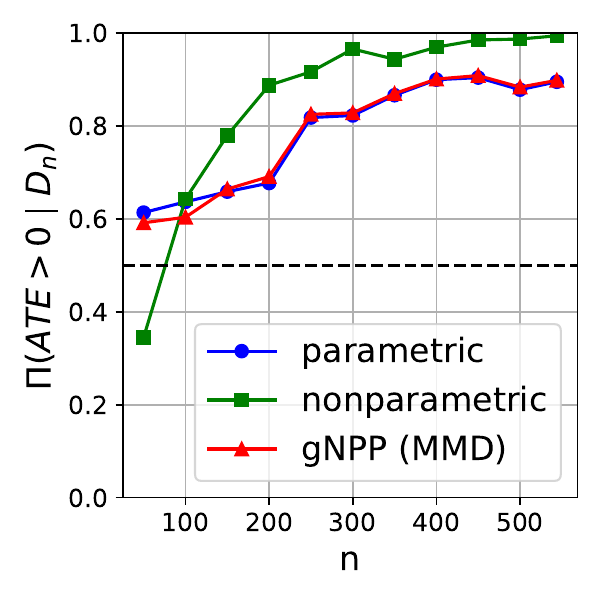}
    \caption{}
\end{subfigure}
\hfill
\begin{subfigure}{0.49\textwidth}
    \centering
    \includegraphics[width=0.85\textwidth]{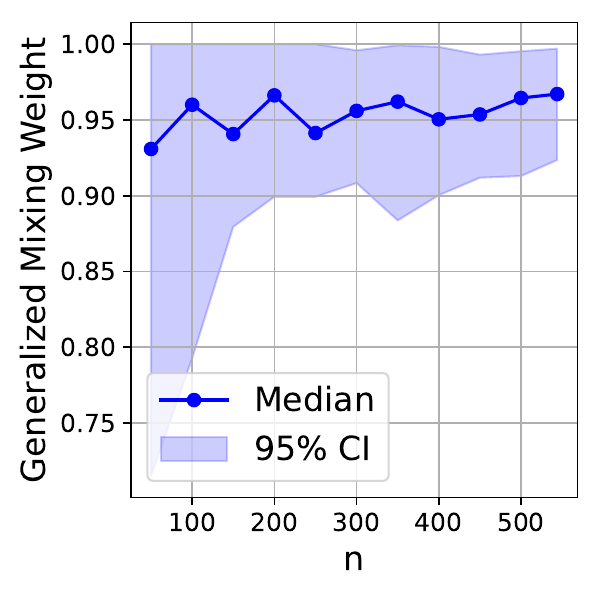}
    \caption{}
\end{subfigure}
\caption{\textbf{Effect of TCF7 on SELL.} a. Posterior probability of the ATE being positive under the parametric, nonparametric, and gNPP models. $n$ denotes the size of the (subsampled) dataset. Values are the median across 10 independent data subsamples and model samples. b. Generalized mixing weights, $\hat{\eta}_n$. The estimated confidence interval (CI) is across independent data subsamples and model samples.  }
\label{fig:summary_TCF7_SELL}
\end{figure}

Besides considering interventions on FOXP3, we also considered the effect of interventions on \textit{TCF7} (Transcription Factor 7) on \textit{SELL} (L-selectin). Elevated levels of SELL are associated with favorable survival outcomes in breast cancer~\citep{Kumari2021-ph}.
The gNPP posterior suggests that increasing the expression of TCF7 is likely to increase the expression of SELL (\Cref{fig:summary_TCF7_SELL}a). In this case, the generalized mixing weight places strong weight on the parametric model (\Cref{fig:summary_TCF7_SELL}b).

\end{document}